\documentclass[fleqn]{aa}
\usepackage{graphicx}
\usepackage{txfonts}
\usepackage{amssymb}
\usepackage{amsmath}
\usepackage{natbib}
\usepackage{mathrsfs}
\usepackage{booktabs}
\usepackage{multicol}
\usepackage{cases}
\usepackage{color}
\usepackage{dsfont}
\usepackage{hyperref}
\usepackage{placeins}
\hypersetup{
colorlinks=true,
linkcolor=blue,
citecolor=blue,
}
\usepackage[switch]{lineno}

\newcommand{\mesa}{{\textsc{MESA}}}
\newcommand{\kepler}{\textit{Kepler}}

\newcommand{\ind}[1]{_{\rm #1}}

\def\m2s2{\,m$^{2}$\,s$^{-2}$} 

\def\aov{\alpha\ind{ov}}

\newcommand{\vaisala}{Brunt-V\"ais\"al\"a}

\newcommand{\numax}{\nu_{\rm max}}
\newcommand{\deltanu}{\Delta\nu}

\newcommand{\rint}{r_{\rm i}}
\newcommand{\rext}{r_{\rm o}}

\newcommand{\dpun}{\Delta\Pi_1}
\newcommand{\dn}{\Delta\nu}
\newcommand{\Brcarre}{\langle B_r^2 \rangle}

\newcommand{\starA}{KIC~3326392}
\newcommand{\starB}{KIC~5700274}
\newcommand{\starC}{KIC~7458743}
\newcommand{\starD}{KIC~9227589}
\newcommand{\starE}{KIC~9697268}
\newcommand{\starF}{KIC~10613561}
\newcommand{\starG}{KIC~10976412}
\newcommand{\starH}{KIC~11408970}
\newcommand{\nstar}{71}

\newenvironment{itemize*}%
  {\begin{itemize}%
    \setlength{\itemsep}{1pt}%
    \setlength{\parskip}{1pt}}%
  {\end{itemize}}

\newcommand{\omg}{\langle\Omega\rangle_{\rm g}}

\newcommand\T{\rule{0pt}{2.6ex}}
\newcommand\B{\rule[-1.2ex]{0pt}{0pt}}

\begin{document}
\title{Near-critical magnetic fields in \kepler\ red giants}
\titlerunning{Near-critical magnetic fields in \kepler\ red giants}
\author{
S. Deheuvels\inst{1,2},
J.Ballot\inst{1},
F. Ligni\`eres\inst{1},
G. Li\inst{3},
M. Villate\inst{1}
}

\institute{IRAP, Universit\'e de Toulouse, CNRS, CNES, UPS, 31400 Toulouse, France
\and Institut Universitaire de France (IUF)
\and Centre for Astrophysics, University of Southern Queensland, Toowoomba, QLD 4350, Australia
}

\offprints{S. Deheuvels\\ \email{sebastien.deheuvels@irap.omp.eu}
}


\abstract
{The recent seismic detection of magnetic fields in red giants cores has given the opportunity to characterize these fields, potentially giving information about their origin and their role in the internal transport of angular momentum. }
{We detect strong deviations from the regular pattern of g-mode periods in eight \kepler\ red giants showing $l=1$ doublets. In three of these stars, the modes show partial suppression. We investigate the magnetic origin of these features and determine the characteristics of the core fields that can produce such signatures (strength, topology).}
{We need to invoke strong, near-critical fields. Assessing the effects of such fields on the mixed mode frequencies requires a non-perturbative approach. We use and adapt a formalism that was recently proposed following a similar development as the traditional approximation for rotation (TAR). We then compute asymptotic expressions of mixed mode frequencies including magnetic effects and attempt to reproduce the observed oscillation spectra.}
{We show that for near-critical fields, information can be obtained about the radial profile of the radial field $B_r$, as opposed to weaker fields for which only a weighted average of $B_r^2$ can be measured. 
For the eight targets, we find that the $l=1$ doublets cannot be identified as the $m=\pm1$ components. Instead, we show that very good fits to all the observations can be obtained by identifying the two components as $m=0$ and $m=1$. These solutions correspond to fields with intensities ranging from 100 to 700\,kG that are confined well below the H-burning shell. Our best-fit models for the eight stars have low masses (1.1-1.2\,$M_\odot$) and the maximal size of their convective core during the main sequence approximately corresponds to the radial extent of the measured magnetic fields. The detected fields could thus have been generated by dynamo action in the main-sequence convective core.}
{}


\maketitle
\nolinenumbers

\section{Introduction \label{sect_intro}}

Magnetic fields can be found at all stages of stellar evolution, during the pre-main sequence (\citealt{villebrun19}), the main sequence (\citealt{donati09}, \citealt{vidotto14}), the red giant phase (\citealt{auriere15}, \citealt{cantiello16}), and in compact stars such as white dwarfs (\citealt{ferrario15}) or neutron stars (e.g., \citealt{thompson93}, \citealt{raynaud20}). They are expected to impact stellar evolution by redistributing angular momentum and they are considered as one of the most promising candidates to solve the long-standing problem of angular momentum transport in stars.

The seismology of subgiants and red giants recently provided invaluable constraints to make progress on this question. These stars harbor oscillation modes with a mixed character, behaving both as pressure modes (p modes) in the envelope and as gravity modes (g modes) in the core. Mixed modes led to the measurement of internal rotation all along the post-main-sequence evolution: for subgiants (\citealp{deheuvels12, deheuvels14, deheuvels20}), red-giant-branch stars (\citealt{beck12}, \citealt{mosser12b} \citealt{gehan18}, \citealt{li24}, \citealt{hatt24}), and core-He burning giants (\citealt{deheuvels15}, \citealt{mosser24}). At all stages, these observations showed that the cores of these stars are spinning much slower than predicted by purely hydrodynamical processes of angular momentum transport (\citealt{ceillier13}, \citealt{marques13}), demonstrating the need for an additional mechanism redistributing angular momentum. Magnetic fields could efficiently transport angular momentum, either directly by applying a magnetic torque on the stellar core (\citealt{maeder14}), or by turbulence associated with magnetohydrodynamic (MHD) instabilities, such as the Tayler instability (\citealt{tayler73}, \citealt{spruit02}, \citealt{fuller19}, \citealt{petitdemange23}, \citealt{barrere23}), or the magneto-rotational instability (\citealt{rudiger15}, \citealt{jouve15}, \citealt{meduri24}). The lack of observational constraints on internal magnetic fields has been a major obstacle to explore the hypothesis of a magnetic transport of angular momentum.

Mixed modes were recently used to obtain the first observational constraints on the strength and topology of internal magnetic fields (\citealt{li22}). The signature of magnetic fields on mixed mode frequencies had been first derived using a perturbative approach in the special case of a dipolar field with various radial profiles, either aligned on the rotation axis (\citealt{gomes20}, \citealt{mathis21}, \citealt{bugnet21}) or inclined (\citealt{loi21}). \cite{li22} obtained a more general expression of magnetic frequency shifts for magnetic fields with an arbitrary configuration, and they detected these signatures in three \kepler\ red giants. Overall, magnetic fields have now been detected in \nstar\ red giants using \kepler\ data (\citealp{li22, li23}, \citealt{deheuvels23}, \citealt{hatt24}, \citealt{villate26}). The detected fields have intensities ranging between about 10 and 600\,kG in the deep core, and they show a wide diversity of field geometries. {We also note the recent detection of a mainly-toroidal magnetic field in the deep interior of a main-sequence F star (\citealt{takata26}).} Complementarily, the existence of a population of red giants with suppressed dipole and quadrupole mixed modes (\citealt{mosser12c}, \citealt{garcia14}, \citealt{stello16}) has been interpreted as the consequence of strong core magnetic fields that exceed the critical strength $B_{\rm c}$ above which magneto-gravity waves can no longer propagate (\citealt{fuller15}). When these waves reach regions with super-critical fields, they are thought to be refracted upward into outgoing slow magnetic waves (\citealt{lecoanet17}, \citealt{loi17}, \citealt{rui23}, \citealt{david26}) and eventually dissipated, which would cause mode suppression. However, the detection of partially suppressed dipole modes that retain a g-like character (\citealt{mosser17}) seems incompatible with this interpretation and remains to be explained.

Magnetic frequency shifts strongly depend on the frequency $\omega$ of the mode (they vary as $\omega^{-3}$), so that strong fields significantly modify the regular period spacing of high-order g modes at low frequency (\citealt{li22}, \citealt{bugnet22}). Such departures from regular g-mode period spacing were detected by \citet{deheuvels23} in 11 \kepler\ red giants, and attributed to strong, near-critical core magnetic fields. All these stars show one single component per dipole triplet, identified as the $m=0$ component. 
In this work, we identify eight \kepler\ red giants whose dipole modes show a distortion of the g-mode pattern that is very similar to the stars studied in \cite{deheuvels23}, but these stars show $l=1$ doublets. We here investigate the magnetic origin of these features.

In Sect. \ref{sect_analysis}, we extract mode frequencies from the oscillation spectra of the eight red giants. We then show in Sect. \ref{sect_mode_ID} that the most obvious identification of the detected components as $m=\pm1$ yields poor agreement with the observations, and we explore the identification as $m=0$ and $m=1$ components. To produce the observed distortion in the g-mode pattern, the magnetic fields need to be near-critical, so that their effects on mode frequencies cannot be treated with a perturbative approach. \cite{rui23} and \cite{rui24} developed a formalism inspired from the traditional approximation of rotation in order to take magnetic effects into account in a non-perturbative manner. In Sect. \ref{sect_nonpert}, we adapt this formalism to our case. In Sect. \ref{sect_field_profile}, we show that measuring the seismic signature of near-critical fields can yield constraints on the radial profile of the radial magnetic field $B_r$. We apply this formalism to the targets of our sample in Sect. \ref{sect_fitdata} and we discuss the results in Sect. \ref{sect_discussion}.

\section{Red giants with strongly distorted doublets \label{sect_analysis}}

In \cite{li24}, we measured g-mode properties and internal rotation rates for red giants in the catalog of \cite{yu18}. In the process of this work, we identified a subsample of red giants showing strongly distorted g-mode patterns.

\subsection{Stretched \'echelle diagrams \label{sect_stretched_ED}}

\begin{figure}
\begin{center}
\includegraphics[width=0.9\linewidth]{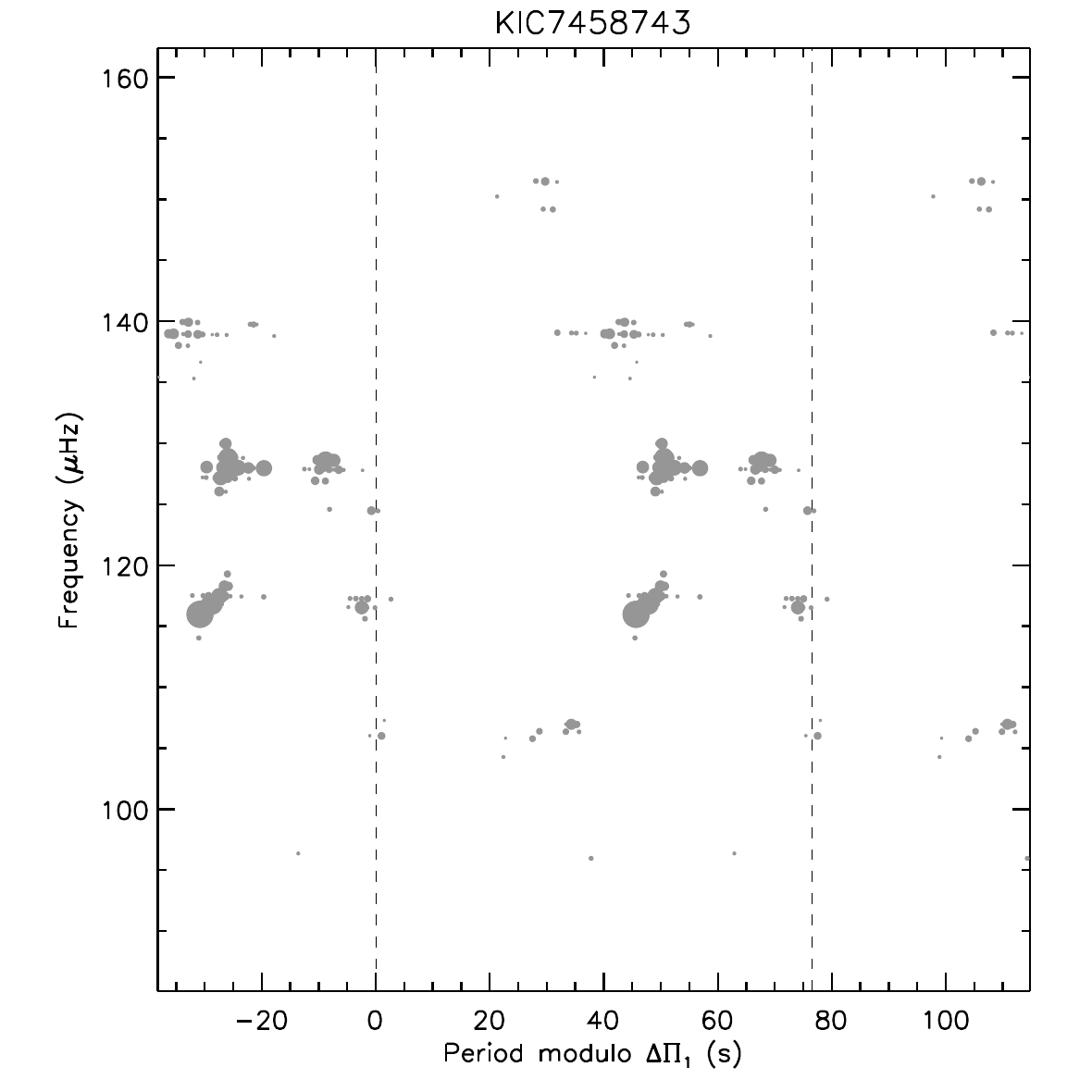}
\end{center}
\caption{Stretched period \'echelle diagram for KIC\,7458743, obtained from 4-yr \kepler\ time series. Only peaks with a SNR above eight are shown, the size of the symbols being proportional to the peak amplitude. For clarity, peaks in the vicinity of $l=0$ and $l=2$ modes are omitted. The \'echelle diagram is folded with an apparent period spacing of $\dpun^{\rm meas} = 76.5$\,s, which produces the best visual alignment of the ridges.
\label{fig_echelle_745}}
\end{figure}

While high-radial-order g modes are nearly regularly spaced in period, mixed modes are not. \cite{mosser15} proposed to apply a ``stretching'' to the mixed mode periods in order to make them equally spaced by $\dpun$. For this purpose, stretched periods $\tau$ are computed by solving the differential equation  $\hbox{d}\tau = \hbox{d}P/\zeta$, where $\zeta$ is defined as the fraction of the mode kinetic energy that is enclosed in the g-mode cavity ($\zeta$ tends to 1 for pure g modes, and 0 for pure p modes). In the non-magnetic case, we expect stretched periods of same $l$ to align on a vertical ridge when plotting them in an \'echelle diagram folded with the period spacing $\dpun$. For eight \kepler\ red giants, the resulting stretched \'echelle diagrams show two strongly distorted ridges, as shown in Fig. \ref{fig_echelle_745}. At this stage, we have not yet found evidence for such distortion in red giants that show more than 2 components per $l=1$ mode. 

\subsection{Extraction of mode frequencies \label{sect_MLE}}

We performed a full seismic analysis of the eight \kepler\ red giants of our sample. The light curves were taken from Kepler Light Curves Optimized For Asteroseismology (KEPSEISMIC) and downloaded from Mikulski Archive for Space Telescopes (MAST). KEPSEISMIC light curves were corrected from outliers, jumps, and drifts through the implementation of the Kepler Asteroseismic Data Analysis and Calibration Software (KADACS, \citealt{garcia11}). The gaps shorter than 20 days were filled following in-painting techniques described in \cite{garcia14} and \cite{pires15}. The power density spectra (PSD) were then obtained using the Lomb-Scargle periodogram (\citealt{lomb76}; \citealt{scargle82}). The background of the PSD is modeled as a sum of the contributions from the granulation (two Harvey-like profiles, \citealt{harvey85}, with distinct characteristic timescales) and the photon noise (white noise). To obtain an estimate of the background, we fit this model completed with a contribution from the oscillations (modeled as a Gaussian function) to the observed PSD with a maximum-likelihood estimation (MLE) method. 

\begin{figure*}
\begin{center}
\includegraphics[width=0.24\linewidth]{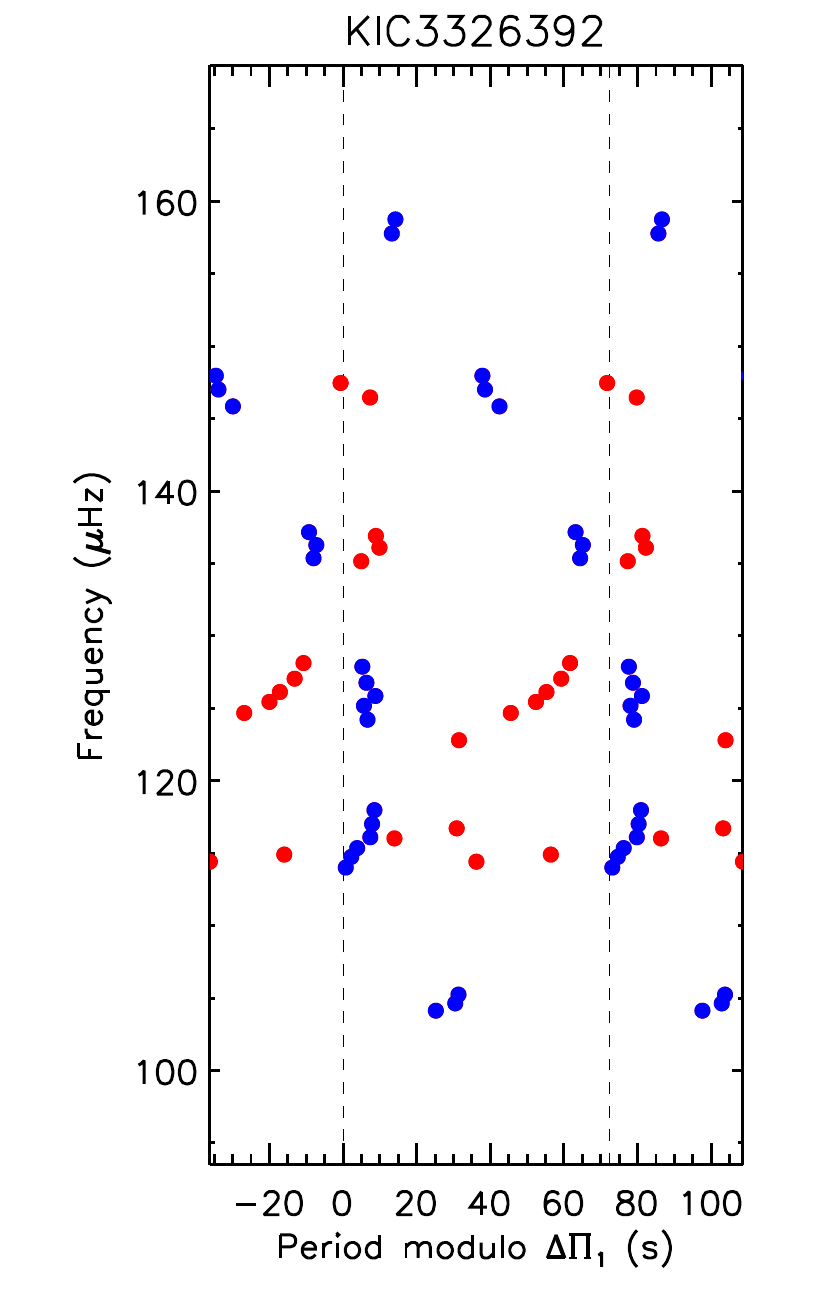}
\includegraphics[width=0.24\linewidth]{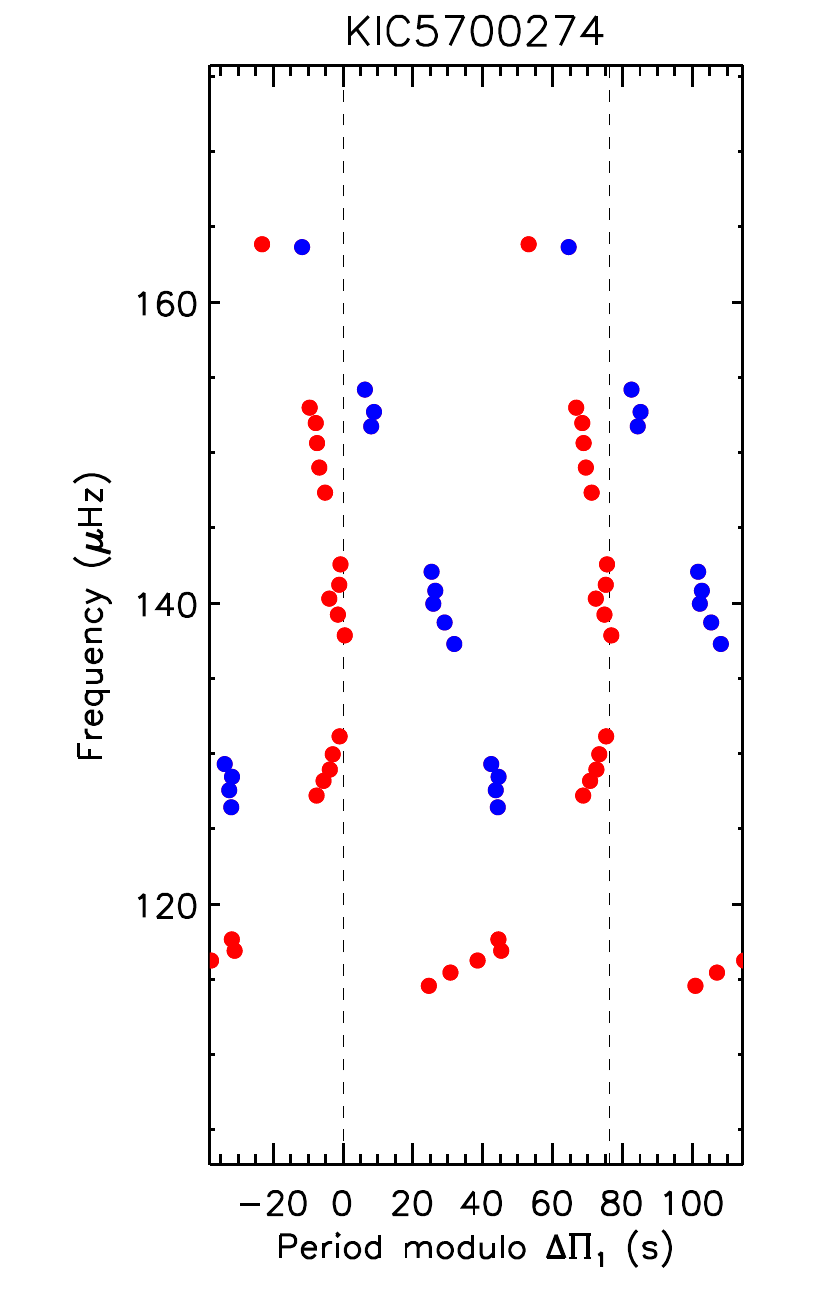}
\includegraphics[width=0.24\linewidth]{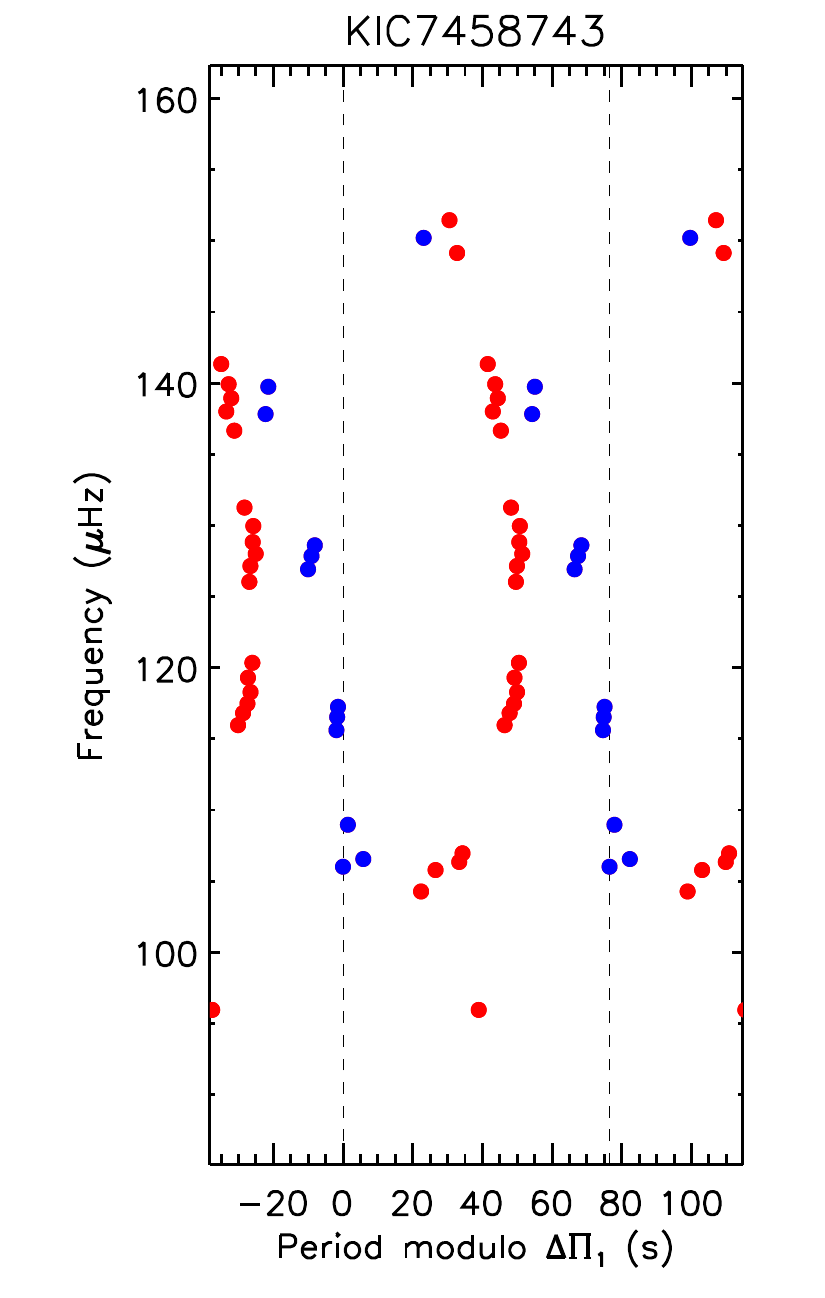}
\includegraphics[width=0.24\linewidth]{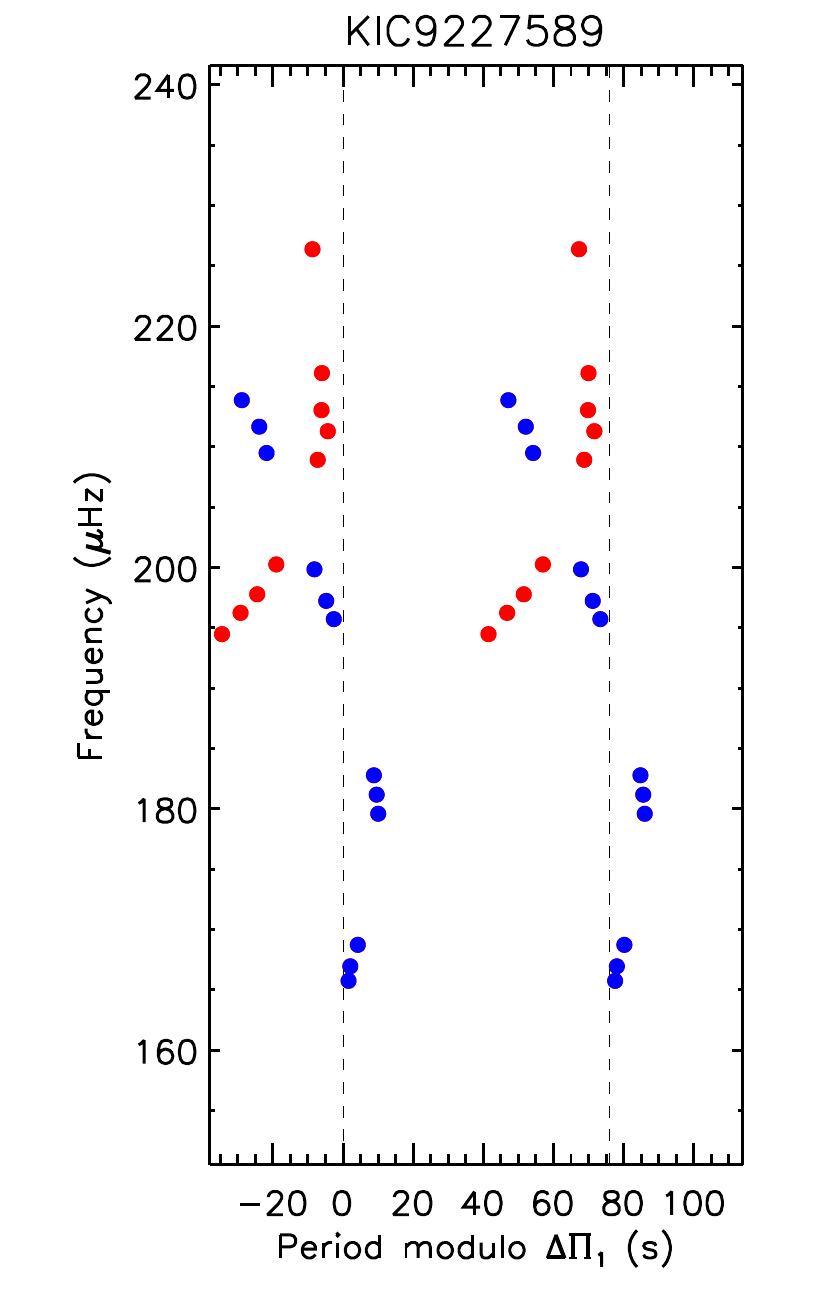}
\includegraphics[width=0.24\linewidth]{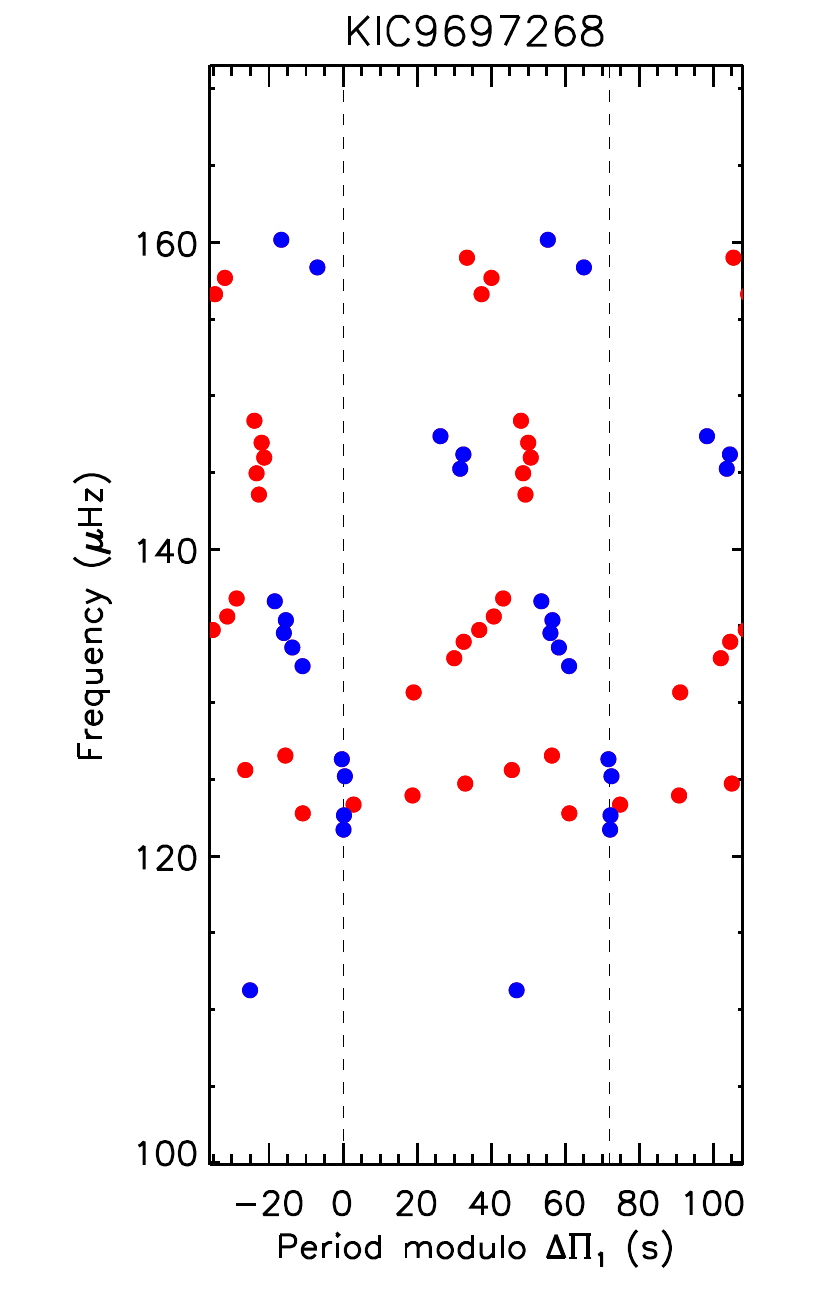}
\includegraphics[width=0.24\linewidth]{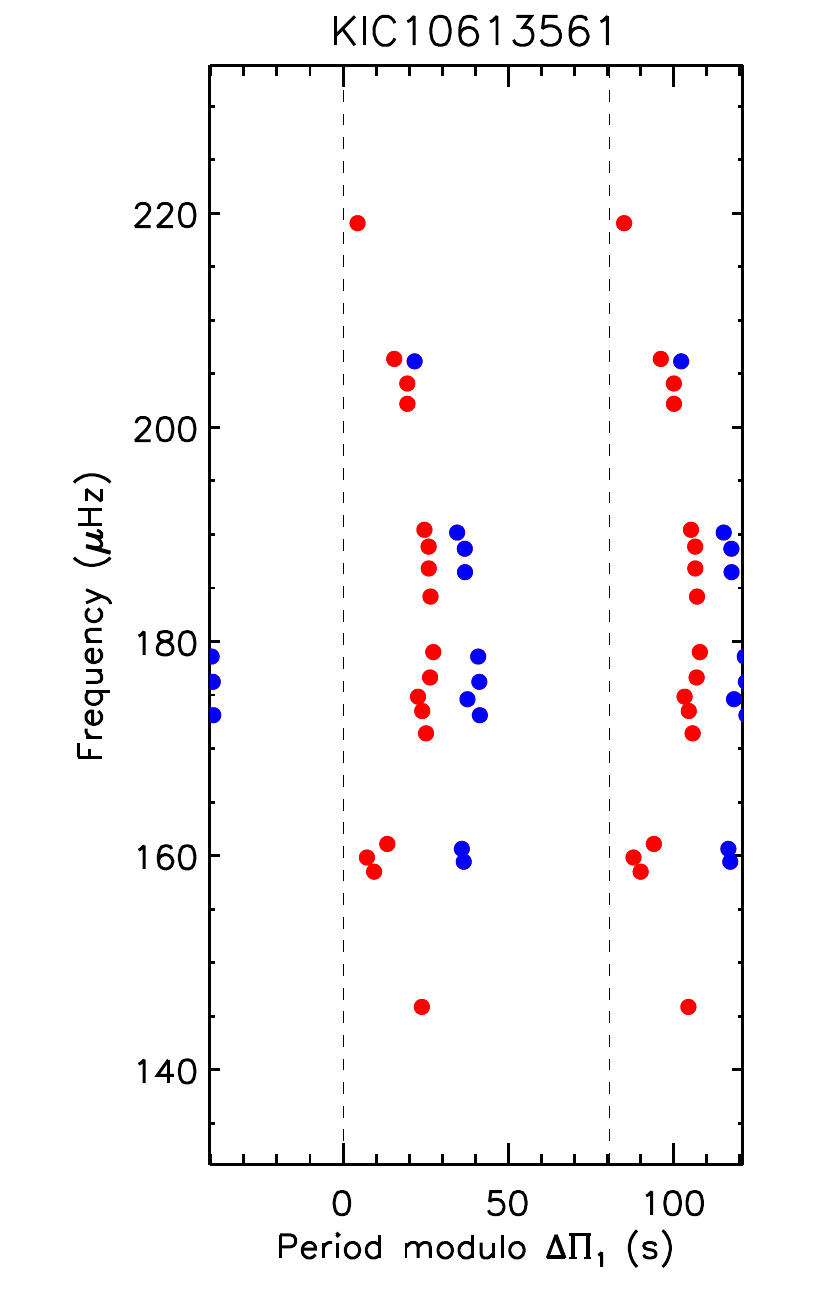}
\includegraphics[width=0.24\linewidth]{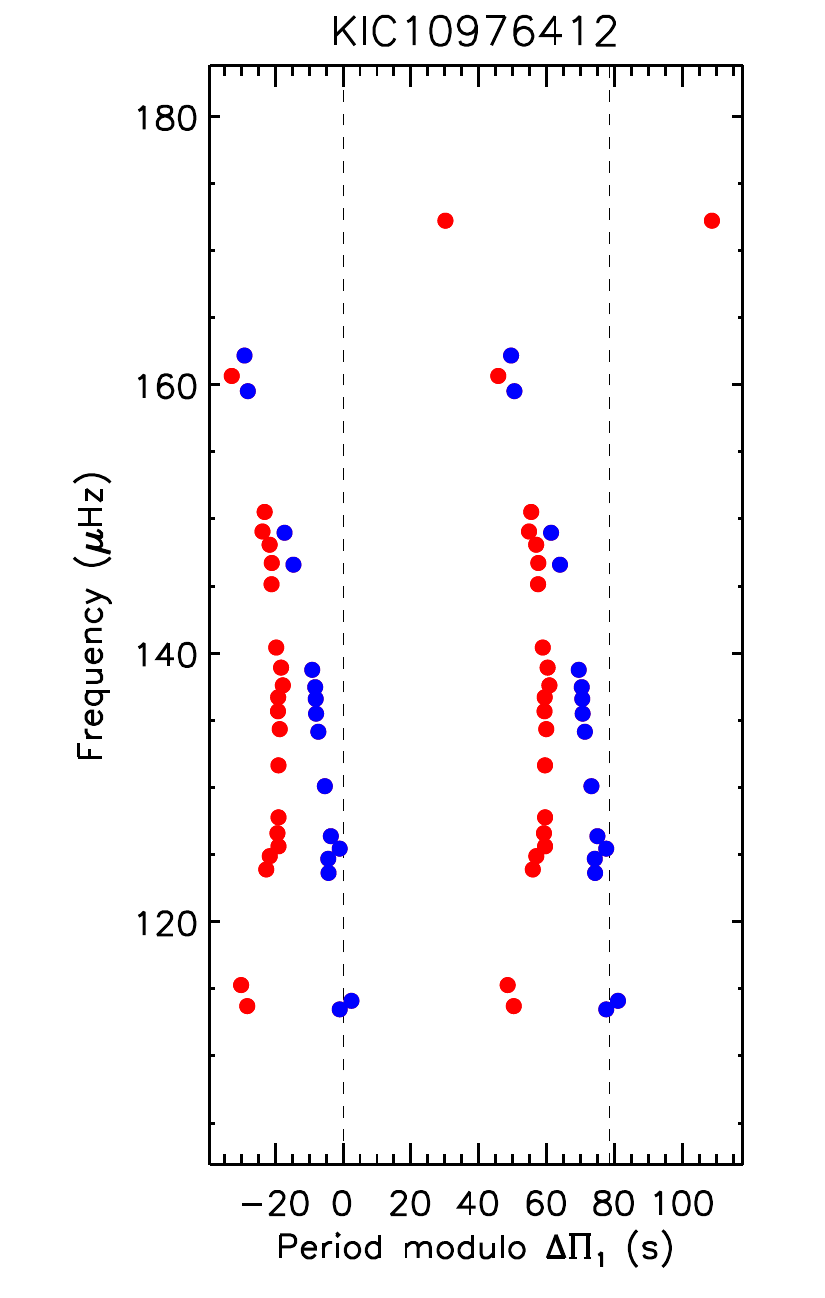}
\includegraphics[width=0.24\linewidth]{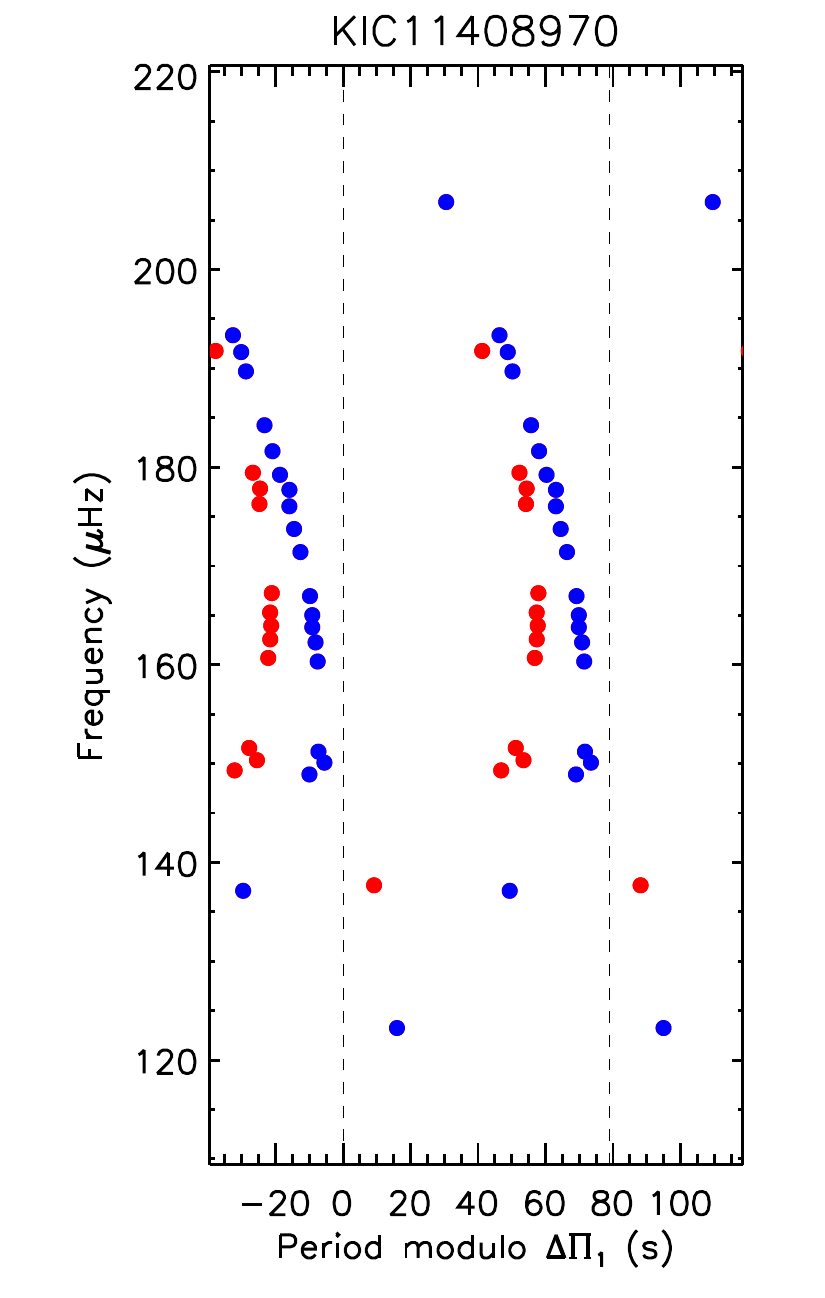}
\end{center}
\caption{Stretched period \'echelle diagrams of the eight \kepler\ red giants of our sample. The \'echelle diagrams were folded using apparent period spacings $\dpun^{\rm meas}$ that offer good visual alignment of the ridges. To guide the eye, the ridges were highlighted with different colors (red for the most distorted ridge, blue for the other).
\label{fig_stretch_echelle}}
\end{figure*}

To identify significant peaks in the PSD, we searched for peaks that exceed the threshold $x_p$ corresponding to a false-alarm probability of $p = 0.05$ over the whole frequency range where oscillation modes are expected ($\numax \pm 5 \,\deltanu$). These peaks can be used to plot stretched \'echelle diagrams (see Fig. \ref{fig_echelle_745}).The list of significant modes was then extended to include additional peaks from the PSD that align on the two clear ridges identified in the stretched \'echelle diagram, while having a power exceeding eight times the background level (see \citealt{mosser15}). Using the identified significant peaks, we obtained a list of guessed frequencies for dipole mixed modes. We then proceeded to fit a model of the oscillation spectrum to the observed PSD. Each mode was modeled as a Lorentzian function with free central frequency $\nu_{n,l,m}$, line width $\Gamma_{n,l,m}$ and height $H_{n,l,m}$. We used an MLE approach to fit this model to the PSD. The list of measured frequencies are given in Appendix \ref{app_freq} for all eight stars. The results are shown in the shape of stretched \'echelle diagrams in Fig. \ref{fig_stretch_echelle}. As mentioned above, the ridges are not straight, so that we can only determine an apparent value of the period spacing $\dpun^{\rm meas}$ that provides a rough visual alignment of the ridges. The values of $\dpun^{\rm meas}$ for each star are given in Table \ref{tab_stars}. 

\begin{table}
\begin{center}
\caption{Characteristics of \kepler\ red giants showing strongly distorted $l=1$ doublets. 
\label{tab_stars}}
\begin{tabular}{l c c c c}
\hline \hline
\T  KIC Id & $M$ & $\dn$ & $\dpun^{\rm meas}$ & z \\
\B & $(M_\odot)$ & ($\mu$Hz) & (s) & (s)  \\
\hline
\T 3326392 & $1.15 \pm 0.05$ & $10.64 \pm 0.01$ & $72.4$ & $-6.0$ \\
5700274 & $1.20 \pm 0.07$ & $11.89 \pm 0.03$ & $76.3$ & $-4.3$ \\
7458743 & $1.13 \pm 0.07$ & $10.89 \pm 0.02$ & $76.5$ & $-2.4$ \\
9227589 & $1.23 \pm 0.06$ & $15.25 \pm 0.02$ & $76.0$ & $-9.8$ \\
9697268 & $1.43 \pm 0.09$ & $11.38 \pm 0.02$ & $72.0$ & $-7.7$ \\
10613561 & $1.24 \pm 0.08$ & $ 14.76 \pm 0.10$ & $80.6$ & $-4.5$ \\
10976412 & $1.13 \pm 0.07$ & $ 11.58 \pm 0.02$ & $78.5$ & $-1.6$ \\
\B 11408970 & $1.12 \pm 0.07$ & $13.85 \pm 0.01$ & $79.0$ & $-4.7$ \\
\hline
\end{tabular}
\tablefoot{The columns give the KIC number of the stars in our catalog, the mass derived from seismic scaling relations (\citealt{yu18}),  the average large separation of p modes $\deltanu$, the apparent period spacing of g modes $\dpun^{\rm meas}$ (see Sect. \ref{sect_MLE}), and the distance from the degeneracy sequence $z$ (see Sect. \ref{sect_distorted_doublets}). }
\end{center}
\end{table}

\subsection{Distorted doublets \label{sect_distorted_doublets}}

Figure \ref{fig_stretch_echelle} confirms that the eight stars under study all have $l=1$ doublets that show strong distortion in the stretched \'echelle diagram. For clarity, two distinct colors have been used to highlight the ridges, but at this stage this is only to guide the eye. In the following sections, no assumption has been made about modes belonging to one ridge or the other. Several observations can be made about the stretched \'echelle diagrams: 
\begin{itemize}
\item In all cases the two ridges either cross or get closer to each other in the high-frequency part of the \'echelle diagram. 
\item The intensity of the curvature seems different for the two ridges (in Fig. \ref{fig_stretch_echelle}, we represent the ridge showing the strongest curvature in red).
\item The three stars for which the ridges are the most curved (KIC\,3326392, KIC\,9227589, and KIC\,9697268) show hints of mode suppression, the most distorted ridge (the red ridge) vanishing at low frequencies, while the less distorted one (the blue one) remains. 
\item There seems to be a difference in mode height between the two ridges. This can be seen by eye for KIC\,7458743 in Fig. \ref{fig_echelle_745}. Our MLE fits confirm that the oscillation modes of the red ridge have 
heights that are on average 1.9 times larger than those of the blue ridge for this star.
\end{itemize}

\begin{figure}[!htp]
\begin{center}
\includegraphics[width=0.99\linewidth]{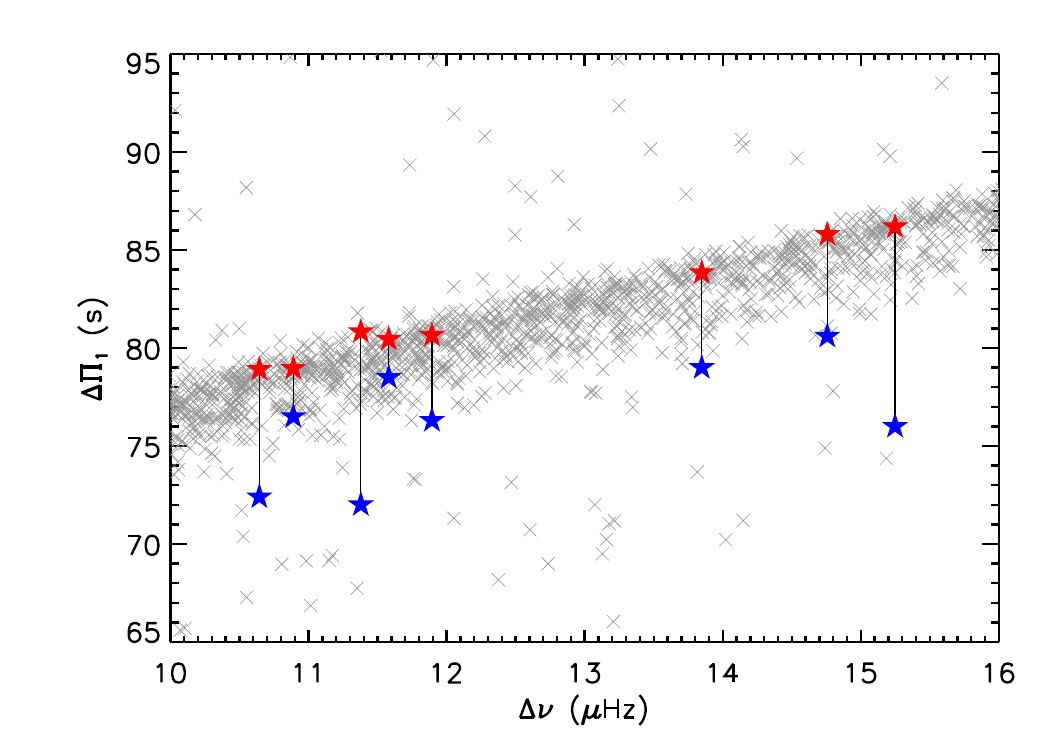}
\end{center}
\caption{Location of the stars under study in the $(\deltanu$-$\dpun)$ plane. The apparent period spacings of g modes $\dpun^{\rm meas}$ are shown as blue star symbols and the red star symbols indicate the asymptotic period spacing $\dpun$ measured by accounting for magnetic fields in a non-perturbative manner (Sect. \ref{sect_fitdata}). Gray crosses correspond to all red giants studied in \cite{li24}.
\label{fig_dn_dp}}
\end{figure}

\subsection{Location in the $\deltanu$-$\dpun$ plane  \label{sect_dn_dp}}

\cite{deheuvels22} showed that red giants collapse on a clear sequence in the $\deltanu$-$\dpun$ plane (the \textit{degeneracy sequence}) when the electron gas becomes degenerate in the core. The strongly magnetic stars of \cite{deheuvels23} had a measured period spacing $\dpun^{\rm meas}$ that was too low, so that these stars appeared \textit{below} the degeneracy sequence in the $\deltanu$-$\dpun$ plane, where no stars are expected. Accounting for the magnetic field enabled to determine the true period spacing $\dpun$, which placed these stars back on the degeneracy sequence, as expected (\citealt{deheuvels23}). In Fig. \ref{fig_dn_dp}, we place our stars in the $\deltanu$-$\dpun$ plane (blue star-shaped symbols) using the apparent period spacing $\dpun^{\rm meas}$, and their position can be compared to the degeneracy sequence (well identified with the gray crosses). 
The degeneracy sequence is mass-dependent, higher-mass stars having a lower $\dpun$ for a given $\deltanu$. For each star of the sample, we determined the equation of the degeneracy sequence corresponding to its stellar mass (taken from seismic scaling relations, see \citealt{yu18}). For this purpose, we selected stars among the catalog of \cite{li24} that have the same mass within 1-$\sigma$ errors. We fitted a third-degree polynomial to the relation $\dpun(\deltanu)$ for these stars, giving the degeneracy sequence for the target mass. The difference $z$ between the measured period spacing $\dpun^{\rm meas}$ and the one predicted from the degeneracy sequence is given for each star in Table \ref{tab_stars}. We find that all our targets are located from 1.6\,s to 9.8\,s below the degeneracy sequence. The typical scatter around the degeneracy sequence does not exceed 4~s. We thus conclude that six out of eight of our targets can be considered significantly below the degeneracy sequence, confirming the similarity between our targets and those of \cite{deheuvels23}.

\subsection{Origin of ridge distortion}

In \cite{deheuvels23}, we identified two potential sources for the ridge distortions: buoyancy glitches (corresponding to sharp variations in the profile of the \vaisala\ frequency, with length scales shorter than the mode wavelength), or internal magnetic fields. We ruled out buoyancy glitches, which would produce very different modulation in the ridges (see Appendix A of \citealt{deheuvels23}) and fail to account for the abnormally-low measured period spacings. These arguments also hold in the present case. Moreover, with buoyancy glitches, we would expect the two ridges of the doublet to show similar deviations. Indeed, to first order, the modes of each doublet have the same eigenfunctions, so that they should be affected by a buoyancy glitch in the same way. The fact that we observe different curvatures for the two ridges in all our stars makes is even less likely that buoyancy glitches are causing the observed ridge distortion. In the following sections we explore the magnetic hypothesis.

\section{Mode identification \label{sect_mode_ID}}

If the magnetic field is not axisymmetric (that is, if it does not have an axis of symmetry that coincides with the rotation axis), it splits every $l=1$ mode into three components, each of which being also split in three components by the effect of rotation. In principle, a non-axisymmetric field can thus produce nine components per multiplet (\citealt{gough90}, \citealt{loi21}). Although some of these components are likely to have much smaller amplitudes than the others, we still expect to have more than three components per multiplet if the effects of non-axisymmetry of the field are non-negligible. The fact that only two components per multiplet are observed suggests that the non-axisymmetry effects are indeed weak. We thus discuss the mode identification {under this assumption}.

\subsection{Identification as $m=\pm1$ components \label{sect_ID11}}

\begin{figure}[!htp]
\begin{center}
\includegraphics[width=0.8\linewidth]{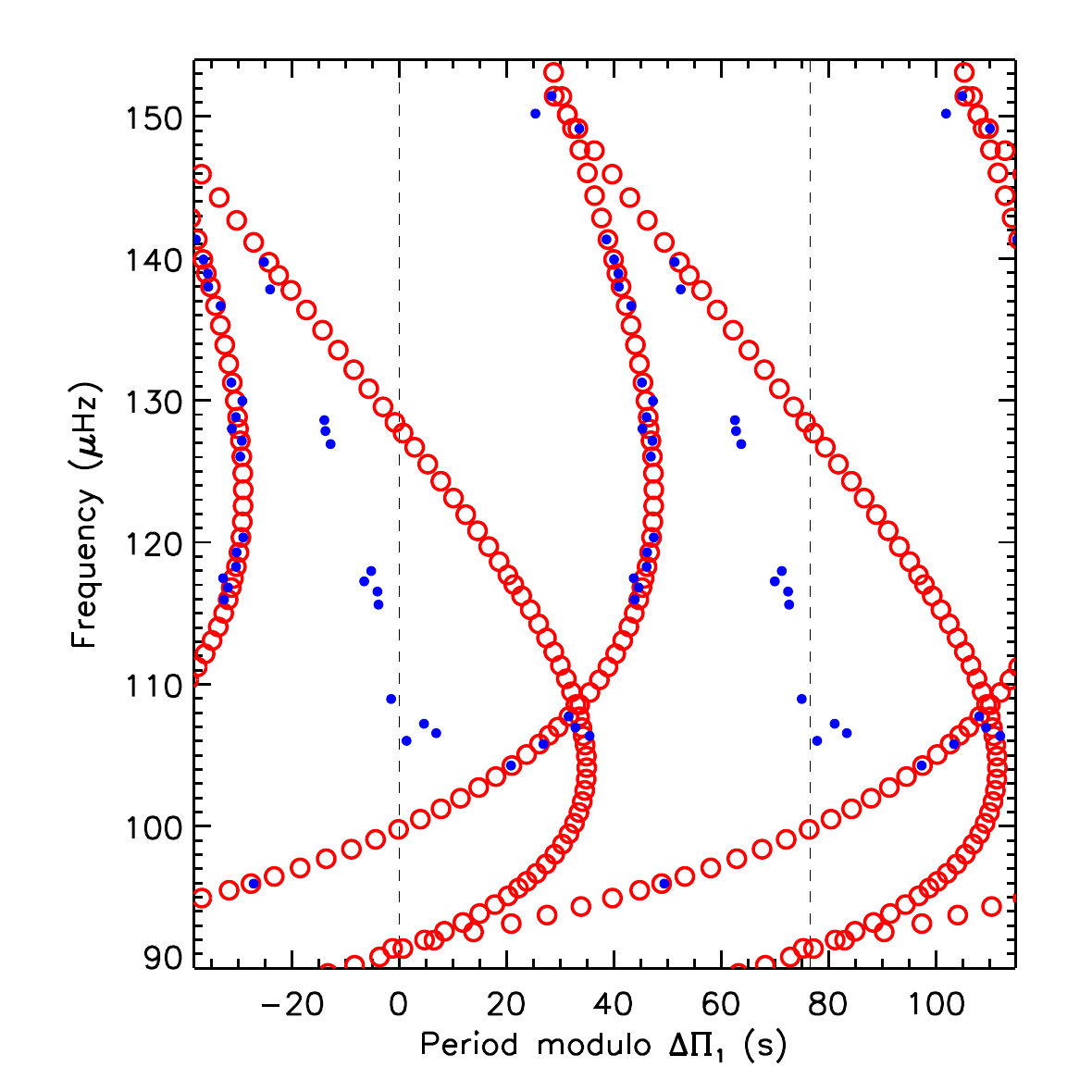}
\end{center}
\caption{Stretched \'echelle diagram of KIC\,7458743. Filled blue circles show the observed mode frequencies. Open red circles show asymptotic mixed mode frequencies including rotation and magnetic field effects, assuming that the observed ridges correspond to $m=\pm1$ components. The core rotation was determined to ensure a crossing of the ridges at $\nu_{\rm cross} = 150\,\mu$Hz and magnetic field properties were adjusted to reproduce the $m=-1$ ridge (see Sect. \ref{sect_ID11}).
\label{fig_745_ID11}}
\end{figure}

The most natural mode identification for an $l=1$ doublet is that we are seeing the $m=\pm1$ components. This occurs when the inclination angle between the rotation axis and the line of sight is close to $90^\circ$, that is, the star is seen equator-on. In this case, the $m=0$ component geometrically cancels. {For a predominantly radial magnetic field, the} mode frequencies depends on $|m|$, so that the $m=-1$ and $m=+1$ are affected in the same way by the magnetic field. Only rotational effects can thus produce different curvatures for the two ridges. If the core rotation is slow, the rotational splitting is small and the $m=\pm1$ ridges are nearly parallel, contrary to the observations. A faster core rotation can make the two ridges cross (\citealt{gehan18}), as we observe in most of our stars. {However, for all eight stars, we found it impossible to satisfactorily fit the shape of the two ridges and the frequency at which they cross. In Fig. \ref{fig_745_ID11}, we show the example of \starC, for which} the two ridges cross around a frequency of $\nu_{\rm cross} \approx 150\,\mu$Hz. The core rotation needed to produce a crossing of the two ridges at $\nu_{\rm cross}$ is expressed as $\omg/(2\pi) = \nu_{\rm cross}^2 \dpun$ (\citealt{gehan18}, {the presence of a magnetic field is expected to modify this expression, but its effect is negligible for \starC).} This gives a core rotation of $\omg/(2\pi) \approx 1.78\,\mu$Hz (which would make this star one of the fastest rotators in the \kepler\ catalog, \citealt{li24}). Using the method that is described in Sect. \ref{sect_fitdata}, we fit an asymptotic expression of mixed mode frequencies including the effects of rotation and magnetic fields in a non-perturbative way to one of the observed ridges, taking $\omg/(2\pi) = 1.78\,\mu$Hz. The results are shown in Fig.~\ref{fig_745_ID11} in a stretched \'echelle diagram. As can be seen, one of the ridges is very well reproduced, but the second one is completely off. The difference of slopes between the two ridges at the frequency of the crossing is very different in the model and in the observations. {If we fit the two ridges simultaneously, considering the core rotation rate and the magnetic field properties as free parameters, the best solution is obtained when the ridges cross around 130\,$\mu$Hz and the quality of the fit is poor.}
We reach the same conclusion for all eight stars of the sample, we find it impossible to reproduce the observations if we identify the ridges as the $m=\pm1$ components.

Moreover, the identification as $m=\pm1$ components fails to account for the difference in mode amplitudes between the two ridges, as observed for KIC~7458743 (the $m=\pm1$ components have the same average over the observed disk, so they are expected to have the same amplitude). It does not explain either the suppression of only one of the two ridges, as observed in three stars of the sample (since the magnetic field affects $m=\pm1$ components in the same way, they should be identically suppressed). For all these reasons, we rule out the identification of the ridges as $m=\pm1$ components.

\subsection{Identification as $m=0$ and $m=1$ components \label{sect_ID01}}

The alternative identification is that one of the two ridges is the $m=0$ component. This offers much more flexibility to fit the observed ridges because in general the magnetic field has a different impact on $m=0$ and $m=\pm1$ components. {The inclination angle between the stellar rotation axis and the line of sight would need to be intermediate, so that both $m=0$ and $m=1$ components are visible. This configuration could account for the fact that the heights of the two ridges differ in some cases, for example for \starC.} However, it raises the question of why we do not observe three ridges. Two hypotheses can be made to potentially account for this.

First, the core rotation could be too slow to separate the $m=\pm1$ components. This would occur if the core rotation rate is smaller than the typical uncertainties to which mode frequencies can be estimated in the \kepler\ data. Typically, these uncertainties reach values of about 10~nHz for the most g-like modes near $\numax$, which is close to the frequency resolution with 4 years of data ($\nu_{\rm res} = 7.9$~nHz). In this case, these stars would be exceptionally-slow rotators compared to the bulk of \kepler\ red giants. Indeed, among the 2,006 red giants in the catalog of \cite{li24}, 90\% of stars have core rotation rates between 380~nHz and 1170~nHz, and only 8 stars have core rotation rates slower than 100~nHz. {This catalog also comprises 489 red giants for which no significant rotational splitting was detected, which can be due to a low inclination angle, but also possibly to a very slow core rotation.} One possible explanation for a slow core rotation in our targets would be that the strong internal field produces a very efficient transport of angular momentum that enforces solid-body rotation in the star. {We note that near solid-body rotation was found by \cite{li24} for six red giants with slow-spinning cores.} In \cite{li24}, envelope rotation rates {significantly above zero} were measured for only 12\% of the catalog (243 stars out of 2,006), the other stars having envelope rotation rates compatible with zero. The slowest measured envelope rotation rate is 16~nHz, barely above the frequency resolution. This means that 88\% of the stars in the catalog by \cite{li24} have envelope rotation below 16~nHz. If a mechanism enforces solid-body rotation in these stars, the core could rotate at rates that are undetectable even with 4 years of \kepler\ data.

Another possible explanation would be that through a yet unknown mechanism, when the magnetic field becomes near-critical ($\sim B_{\rm c}$), mode suppression could affect only the $m=+1$ component or only the $m=-1$ component. In the following sections, we work under the hypothesis that the two detected ridges correspond to $m=0$ and $m=1$ components, which is the only way we found of reproducing the observations.

\section{Non-perturbative approach \label{sect_nonpert}}

The magnetic fields causing the strong deviation from classical g-mode pattern in our stars are near-critical, so that the effects of magnetic fields on the oscillations need to be addressed with a non-perturbative approach (\citealt{rui24}). Before presenting the details of our non-perturbative approach, adapted from \cite{rui23} (Sect. \ref{sect_nonpert_assumptions} to \ref{sect_example}), we briefly recall a few aspects of the perturbative treatment of magnetic effects on oscillation mode frequencies in Sect. \ref{sect_pert}. This will be useful to compare with the results of the non-perturbative treatment in the next sections. 

\subsection{Perturbative approach \label{sect_pert}}

A perturbative approach was used early on to estimate the effects of a magnetic field on solar p modes (\citealt{gough90}), on g modes in SPB stars (\citealt{hasan05}), and later on mixed modes in red giants, in the special case of a dipolar field with various radial profiles (\citealt{gomes20}, \citealt{mathis21}, \citealt{bugnet21}, \citealt{loi21}). \cite{li22} generalized the expression of magnetic frequency shifts to magnetic fields with an arbitrary configuration. This study only assumes that the azimuthal field $B_\varphi$ does not completely dominate over the radial field $B_r$ (it is valid provided $B_\varphi/B_r \ll N/\omega$, where $N$ is the \vaisala\ frequency and $\omega$ is the mode angular frequency, $N/\omega$ having typical values of the order of $10^2$ in red giants).

\cite{li22} showed that the magnetic frequency shifts depend on only two parameters: 
\begin{itemize}
\item The average frequency shift $\omega_{\rm B}$ of all the components within the multiplet, which is given by the expression
\begin{equation}
\omega_{\rm B} = \frac{\mathcal{I}}{\mu_0 \omega^3} \Brcarre,
\label{eq_omegaB}
\end{equation}
where $\mathcal{I}$ is a term depending on the internal structure of the reference model. The term $\Brcarre$ is an average of $B_r^2$, expressed as
\begin{equation}
\Brcarre = \frac{1}{4\pi} \int_{\rint}^{\rext} K(r) \int_{0}^{2\pi} \int_{0}^{\pi} B_r^2 \sin\theta \, \hbox{d}\theta \, \hbox{d}\varphi \, \hbox{d}r,
\label{eq_pert_Br}
\end{equation}
where $\rint$ and $\rext$ are the inner and outer turning points of the g-mode cavity. The weight function $K(r)$ probes the g-mode cavity and sharply peaks in the H-burning shell. We stress that the average frequency shift $\omega_{\rm B}$ strongly increases with decreasing mode frequency (it varies as $\omega^{-3}$).
\item The ``asymmetry'' parameter $a_{\rm asym}$, which is a dimensionless quantity that depends on a horizontal average of $B_r^2$ weighted by the second-order Legendre polynomial $P_2(\cos\theta)$ and expressed as 
\begin{equation}
a_{\rm asym} = \displaystyle \frac{\int_{\rint}^{\rext} K(r) \iint B_r^2 P_2(\cos\theta) \sin\theta \,\hbox{d}\theta  \,\hbox{d}\varphi \,\hbox{d}r }{\int_{\rint}^{\rext} K(r) \iint B_r^2 \sin\theta \,\hbox{d}\theta  \,\hbox{d}\varphi \,\hbox{d}\varphi \,\hbox{d}r}
\label{eq_pert_asym}
\end{equation} 
\end{itemize}
Essentially, the parameter $\omega_{\rm B}$ gives a measure of the field strength and $a_{\rm asym}$ provides information on the field topology. The average quantity $ a_{\rm asym}$ alone can be produced by an infinity of field geometries (\citealt{mathis23}) but its value gives broad constraints on the shape of the field. The parameter $a_{\rm asym}$ indeed ranges between $-1/2$ (for a field completely concentrated on the equator) and $1$ (for a field concentrated at the poles). For a purely dipolar field, $a_{\rm asym}$ takes values between $-0.2$ and $0.4$. \cite{das24} showed that having access to magnetic frequency shifts for $l=2$ modes could help characterize the field geometry.

If the effects of non-axisymmetry of the field are negligible (that is, if $B_r^2$ is axisymmetric or if the magnetic splitting $\omega_{\rm B}$ is smaller than the rotational splitting, see \citealt{li22}), then the magnetic field produces an additional frequency shift to the $m=0$ and $m=\pm1$ components, expressed as
\begin{align}
& \delta\omega_{\rm g} (m=0) = (1-a_{\rm asym}) \, \omega_{\rm B} \label{eq_shift_pert_m0} \\
& \delta\omega_{\rm g} (m=\pm1)  = \left(1+\frac{a_{\rm asym}}{2} \right) \omega_{\rm B} 
\label{eq_shift_pert_m1}
\end{align}
In previous works (\citealt{li22}, \citealt{deheuvels23}), $\omega_{\rm B}$ was expressed as $\omega_{\rm B} = \delta\omega_0 (\omega_{\rm max}/\omega)^3$, where $\delta\omega_0$ is the average magnetic frequency shift at the frequency of maximum power of the oscillations $\omega_{\rm max}$.

We attempted to fit expressions given by Eq. \ref{eq_shift_pert_m0} and \ref{eq_shift_pert_m1} to the observed mode frequencies of the stars in our sample following the method that was used in \cite{deheuvels23}, assuming that we detect the $m=0$ and $m=1$ components with negligible core rotation. For the four stars that show the most strongly curved ridges (\starA, \starB, \starD, and \starE), no satisfactory solution could be found. We managed to fit either one ridge or the other but not both of them simultaneously. For the other four stars, with less strongly distorted ridges, reasonable fits could be obtained using the perturbative approach. However, the magnetic frequency shift that was measured indicates magnetic fields that are close to the critical field strength, so that a non-perturbative approach is necessary (\citealt{rui24}). 

\subsection{Non-perturbative approach: assumptions and analytical problem \label{sect_nonpert_assumptions}}

We adapted the approach introduced by \cite{rui23}. The authors include the Lorentz force in the oscillation equations using the Boussinesq approximation, which means that the effects of compressibility are taken into account only in the momentum equation. They assume that the oscillations are adiabatic and they make the Cowling approximation, which consists in neglecting the perturbation to the gravitational potential. We make the same assumptions here, also neglecting the effects of rotation in the oscillation equations (even if rotation were to be considered, it could be treated a posteriori as a perturbation because red giants are slow rotators). The magnetic field is assumed to be axisymmetric, which simplifies a lot the equations (the solutions vary as $e^{im\varphi}$, with $m$ the azimuthal order). We assume that $B_\varphi/B_r \ll N/\omega$, so that horizontal field components can be neglected. Finally, the horizontal geometry of $B_r(r,\theta)$ is assumed to be the same in the whole g-mode cavity, so that it is expressed as $B_r(r,\theta) = B_0 f(\theta) g(r)$.

For gravity modes, the radial wavenumber is much larger than the horizontal wavenumber ($k_r / k_h  \sim N/\omega \sim 10^2$ in red giant cores), so that their displacements are nearly horizontal. Since we consider high-order g modes, the radial wavenumber is also much larger than the length scale of variations of structural quantities. For this reason, \cite{rui23} adopt the JWKB approximation in the radial direction (but not in the horizontal direction). In the oscillation equations, the eigenfunctions can then be written as the product of a plane-like wave function $E(r)$ and a slowly varying envelope function, yielding for instance for the pressure perturbation
\begin{equation}
p'(r,\theta) = \hat{p}(r,\theta) E(r).
\end{equation}
The phase function can be expressed as
\begin{equation}
E(r) = \exp \left( - i \int k_r \,\hbox{d}r \right).
\label{eq_phase_function}
\end{equation}
In the oscillation equations, the radial derivative $\partial / \partial r$ can then be replaced by $-ik_r$, which is considered to be much larger than the radial gradients of envelope functions ($\partial_r \hat{p}$, $\partial_r \widehat{\xi_r}$...). 

Upon making the change of variable $\mu = \cos \theta$ and inserting the equation of motion into the continuity equation (see details in Appendix \ref{app_nonpert}), {one finds that the pressure perturbation is solution of the eigenvalue problem}
\begin{equation}
\mathcal{L}\hat{p} + \lambda \hat{p} = 0,
\label{eq_tarm}
\end{equation}
where
\begin{equation}
\mathcal{L}  \hat{p} = \frac{\partial}{\partial\mu} \left[ \frac{1-\mu^2}{1-b^2f^2} \frac{\partial \hat{p}}{\partial\mu} \right] - \frac{m^2}{(1-\mu^2)(1-b^2f^2)} \hat{p},
\end{equation}
\begin{equation}
\lambda = \left( \frac{r\omega k_r}{N} \right)^2 
\label{eq_lambda}
\end{equation}
and 
\begin{equation}
b = \frac{k_r B_0 g(r)}{\omega\sqrt{4\pi\rho_0}}.
\end{equation}
In the general case, the coefficients of operator $\mathcal{L}$ depend on both $r$ and on $\mu$. The problem is thus not separable in a radial part and a horizontal part. However, with the JWKB approximation, the eigenfunctions of Eq. \ref{eq_tarm} depend only parametrically on $r$, as explained in \cite{ogilvie04} (see also \citealt{mathis09}), so that Eq. \ref{eq_tarm} can be solved in $\mu$ for each value of $r$. It is important to stress that the eigenvalue $\lambda$ also depends parametrically on $r$. In the non-magnetic case ($b = 0$), Eq. \ref{eq_tarm} reduces to the general Legendre equation. We then recover $\lambda = l(l+1)$ and the horizontal eigenfunctions correspond to the spherical harmonics $Y_l^m(\theta,\varphi)$.

The quantity $b$ in Eq. \ref{eq_tarm} has been introduced by \cite{rui23} as $b = k_r v_{\textrm{A}, r} / \omega $, where $\overrightarrow{v_{\rm A}} = \overrightarrow{B_0} /\sqrt{4\pi\rho_0}$ is the Alfv\`en velocity. Because $k_r \gg k_h$, $b$ approximately corresponds to the ratio $\omega_{\rm A}/\omega$, where $\omega_{\rm A} = \overrightarrow{k} \cdot \overrightarrow{v_{\rm A}}$ is the Alfv\`en frequency. The eigenvalue $\lambda$ can also be written as $\lambda = (b/a)^2$, where
\begin{equation}
a = \frac{N v_{\rm A}}{r \omega^2}.
\label{eq_a}
\end{equation}
The quantity $a$ is proportional to the ratio $\omega_{\rm supp}^2 / \omega^2$, where $\omega_{\rm supp}$ is the frequency below which the magnetic field is expected to prevent the propagation of gravity waves (\citealt{rui23}). 
Although the details of how magnetogravity waves are suppressed by the magnetic field remain to be fully understood, it is thought that when they reach the critical frequency $\omega_{\rm supp}$, these waves refract into outgoing slow magnetic waves that approach infinite wavenumbers and dissipate (\citealt{lecoanet17}, \citealt{rui23}, \citealt{david26}). Alternately, magnetogravity waves can be damped out by the phase-mixing process when they become resonant with Alfv\`en waves ($b>1$).

\subsection{Profile of the magnetic field \label{sect_field}}

In \cite{rui23} and \cite{rui24}, the authors assumed that the field has a dipolar geometry ($f(\theta) = \cos\theta$). For the radial dependence $g(r)$, they adopted the Prendergast magnetic field geometry (\citealt{prendergast56}). However, it should be noted that \cite{rui24} built their Prendergast field by imposing a vanishing surface field. The resulting field slowly decreases as a function of the radius, and it is nearly uniform over the g-mode cavity. Thus, the results presented by the authors approximately correspond to the case $g(r) =$ constant. In this study, we explore different latitudinal and radial dependences for the field. In order to produce different levels of confinement of $B_r$ in the core, we assumed a Gaussian-shape profile for $g(r) = \exp\left[-r^2/(2{\sigma_{\rm B}}^2)\right]$.
The parameter $\sigma_{\rm B}$ 
measures the radial extension of the core field and is considered as a free parameter in our following analysis.

The latitudinal dependence $f(\theta)$ of $B_r$ is a priori not known. In the perturbative treatment, it intervenes only in the calculation of the asymmetry parameter $a_{\rm asym}$ (Eq. \ref{eq_pert_asym}). In this study, we chose to build latitudinal profiles $f(\theta)$ that correspond to given values of $a_{\rm asym}$. Even though the asymmetry parameter does not enter the calculation of magnetic frequency shifts in the non-perturbative approach, it is still interesting to relate $f(\theta)$ to $a_{\rm asym}$ because in the high-frequency part of the spectrum, the effects of the magnetic fields on the oscillations are less strong, so that we expect to approximately recover the solution of the perturbative approach. More importantly, we found that latitudinal profiles corresponding to the same value of $a_{\rm asym}$ produce similar results for all the purposes of our study (see Sect. \ref{sect_latitudinal_field} below). Naturally, an infinity of latitudinal profiles produce a given value of $a_{\rm asym}$. 
{We here chose $f(\theta)$ of the type
\begin{equation}
f(\theta) = P_1(\cos\theta) e^{\delta P_2(\cos\theta)}
\label{eq_latitudinal_field}
\end{equation}
where $\delta\in\mathds{R}$.}
{This is a large-scale field that is antisymmetric with respect to the equator. It has the advantage that it can be made to concentrate near the poles or near the equator depending on the value of $\delta$. Indeed, it can be shown that $a_{\rm asym} \rightarrow 1$ for large positive values of $\delta$, and $a_{\rm asym} \rightarrow -1/2$ for large negative values of $\delta$ (despite the fact that the profile vanishes at the equator itself).}
When plugging the chosen expression of $f(\theta)$ in Eq. \ref{eq_pert_asym}, we find that the function $a_{\rm asym}(\delta)$ 
is continuous and strictly increasing over the interval $]-1/2,1[$. Thus, for any target $a_{\rm asym}$, we can find a unique value of $\delta$ so that the ratio in Eq. \ref{eq_pert_asym} is equal to $a_{\rm asym}$ (see Appendix \ref{app_latitudinal_profile}). We also imposed that the maximum value of $|f|$ is 1. This way, we ensure that the mode becomes suppressed if either $a$ of $b$ exceeds unity somewhere in the star. 
{The multiplication by $P_1(\cos\theta)$ in the expression of $f(\theta)$ ensures that $\int_{0}^\pi f(\theta)\sin(\theta) \,\hbox{d}\theta$ vanishes. In these conditions,}
the chosen field can be made divergence-free, as expected from Maxwell's equations, by adding an adequate $\theta$-component $B_\theta(r,\theta)$ (see Appendix \ref{app_divergence_free}).

\subsection{Numerical solution of the horizontal problem \label{sect_horizontal}}

For each radius $r$, Eq. \ref{eq_tarm} must be solved numerically. For this purpose, we use a relaxation method. Following \cite{rui23}, we introduce variables $\mathcal{P} = p'/\rho_0\omega^2r^2$ and $\mathcal{Z}_\theta = \sqrt{1-\mu^2} \xi_\theta/r$, and the system of equations is then written as
\begin{align}
\frac{\partial \mathcal{P}}{\partial \mu} & = - \frac{1-b^2 f^2(\mu)}{1-\mu^2} \mathcal{Z}_\theta \\
\frac{\partial \mathcal{Z}_\theta}{\partial \mu} & = \left( \frac{b^2}{a^2} - \frac{m^2}{(1-\mu^2)\left[ 1-b^2 f^2(\mu) \right]} \right) \mathcal{P}
\end{align}
We start from $b=0$ (for which the eigenfunctions correspond to spherical harmonics) and we gradually increase $b$, using the solution from the previous iteration as an initial guess for the next. For each value of $b$, this procedure provides the eigenfunctions ($\mathcal{P}$ and $\mathcal{Z}_\theta$) and the parameter $a$. For a dipole field geometry, \cite{rui23} showed that $a$ increases with $b$ until the internal gravity wave branch connects with a slow magnetic wave branch, at which point $a$ reaches a maximum value $a_{\rm max}$. For larger values of $b$, $a$ then decreases, which shows that the wave propagates back to larger radii inside the star and does not reach the deep core. For an axisymmetric dipolar field ($a_{\rm asym} = 0.4$), \cite{rui23} found that in the case $(l=1,m=0)$, the maximum value of $a$ is reached when $b\rightarrow\infty$ so that the wave is not refracted upwards, but reaches infinite wavenumber and dissipates in the layer where $a$ is maximal. In this study, we integrate the equations until either $b$ reaches 1 or $a(b)$ reaches a maximum. 

\begin{figure}
\begin{center}
\includegraphics[width=0.9\linewidth]{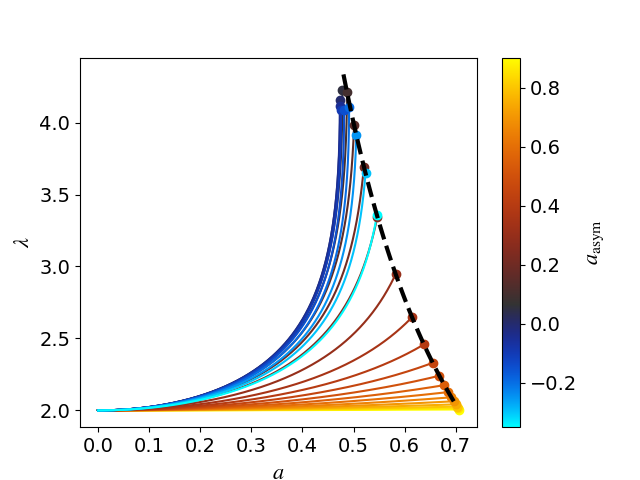}
\includegraphics[width=0.9\linewidth]{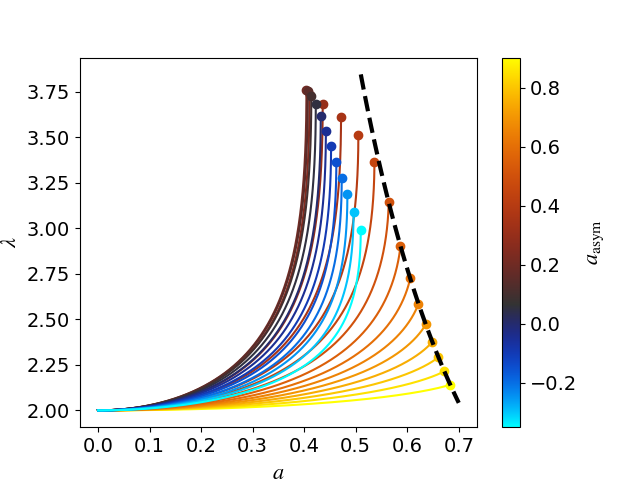}
\end{center}
\caption{Numerical solution of the horizontal problem for different values of the asymmetry parameter $a_{\rm asym}$. We show $\lambda$ as a function of $a$ for $(l=1, m=0)$ modes (top panel) and for $m=\pm1$ modes (bottom panel). The colored dots indicate that mode suppression is reached, and the black dashed line corresponds to $b=1$. 
\label{fig_lambda}}
\end{figure}

In Fig. \ref{fig_lambda}, we plot the numerical solutions to the horizontal problem obtained with the latitudinal profile given in Eq. \ref{eq_latitudinal_field} and various values of $a_{\rm asym}$. We plot $\lambda=(b/a)^2$ as a function of $a$. For the case $a_{\rm asym} = 0.4$ corresponding to an axisymmetric dipolar field, we recover the same $a(b)$ relation as \cite{rui23}. For negative asymmetry parameters, we find that $\lambda$ increases with increasing $a_{\rm asym}$ for a given $a$, while the situation is reversed for $a_{\rm asym} > 0$. The cause of mode suppression depends on the value of $a_{\rm asym}$. The modes become suppressed because of resonance with Alfv\`en waves ($b$ exceeds unity) for 
{$a_{\rm asym}\lesssim-0.204$ or $a_{\rm asym}\gtrsim0.086$}
with $m=0$ modes, and for 
{$a_{\rm asym}\gtrsim0.464$}
with $m=\pm1$ modes. For other values of $a_{\rm asym}$, mode suppression occurs because $a$ reaches $a_{\rm max}$.

\subsection{Numerical solution of the radial problem}

The JWKB approximation in the radial direction yields the relation
\begin{equation}
\omega = \frac{1}{\pi (n_{\rm g} + \varepsilon_{\rm g})} \int_{\rint}^{\rext} \frac{N}{r} \sqrt{\lambda} \, \hbox{d}r
\label{eq_asympt_mag}
\end{equation} 
where the expression of $k_r$ was taken from Eq.~\ref{eq_lambda}. As mentioned above, $\lambda$ depends parametrically on $r$ and its value needs to be determined for each layer of the g-mode cavity. For a given field geometry, Eq.~\ref{eq_tarm} can be solved as explained in Sect. \ref{sect_horizontal} to obtain a relation between $a$ and $b$, and thus between $a$ and $\lambda = (b/a)^2$. This means that $\lambda(r)$ is known if we can determine $a(r)$ for all radii within the g-mode cavity. The parameter $a$ depends on equilibrium quantities -- $N(r)$, $\rho(r)$ -- and on the mode frequency $\omega$ (see Eq. \ref{eq_a}). To calculate $a$, we thus need a stellar model. The search for a stellar model representative of the studied star is addressed in Sect. \ref{sect_stellar_model}, and here, we assume that we have access to such a model. The dependence of $a$ on $\omega$ 
{is an additional complication because the mode frequency is not known at this stage. We thus proceed iteratively, initializing $\omega$ to}
its non-magnetic asymptotic frequency (which is obtained by setting $\lambda=2$ in Eq. \ref{eq_asympt_mag}) {to obtain initial values for $a(r)$}. We interpolate the $\lambda(a)$ relation to determine the value of $\lambda(r)$ at each radius in the cavity. Equation~\ref{eq_asympt_mag} is then solved to obtain the asymptotic mode frequencies including the effects of the strong field. {This process is iterated, the obtained mode frequencies being used to compute updated values of $a(r)$, until convergence.}
If for some $r$ in the cavity $a(r)$ exceeds $a_{\rm max}$ or $b(r)$ reaches unity, then we consider that the mode is suppressed and it is ignored. 

\subsection{Example \label{sect_example}}

\begin{figure}
\begin{center}
\includegraphics[width=0.49\linewidth]{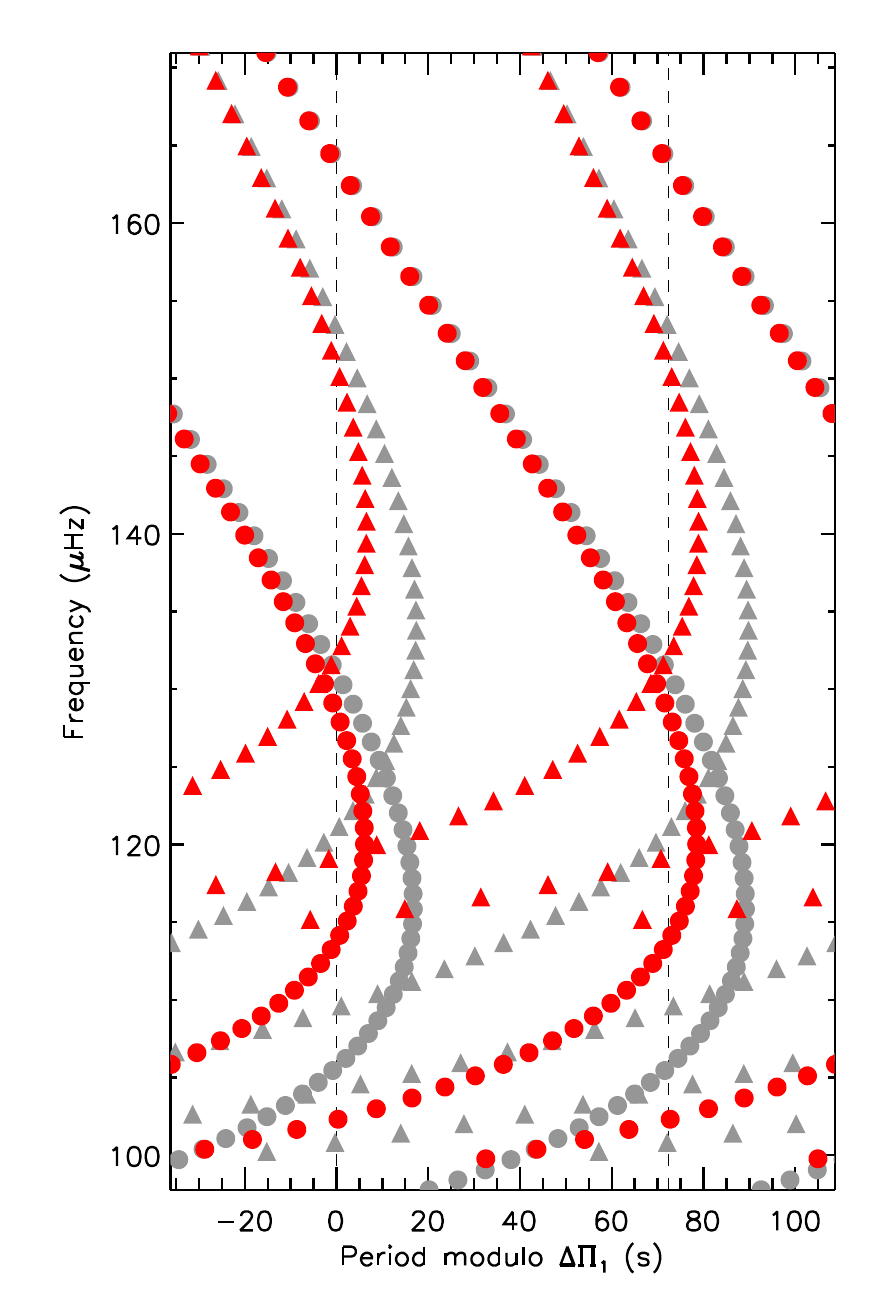}
\includegraphics[width=0.49\linewidth]{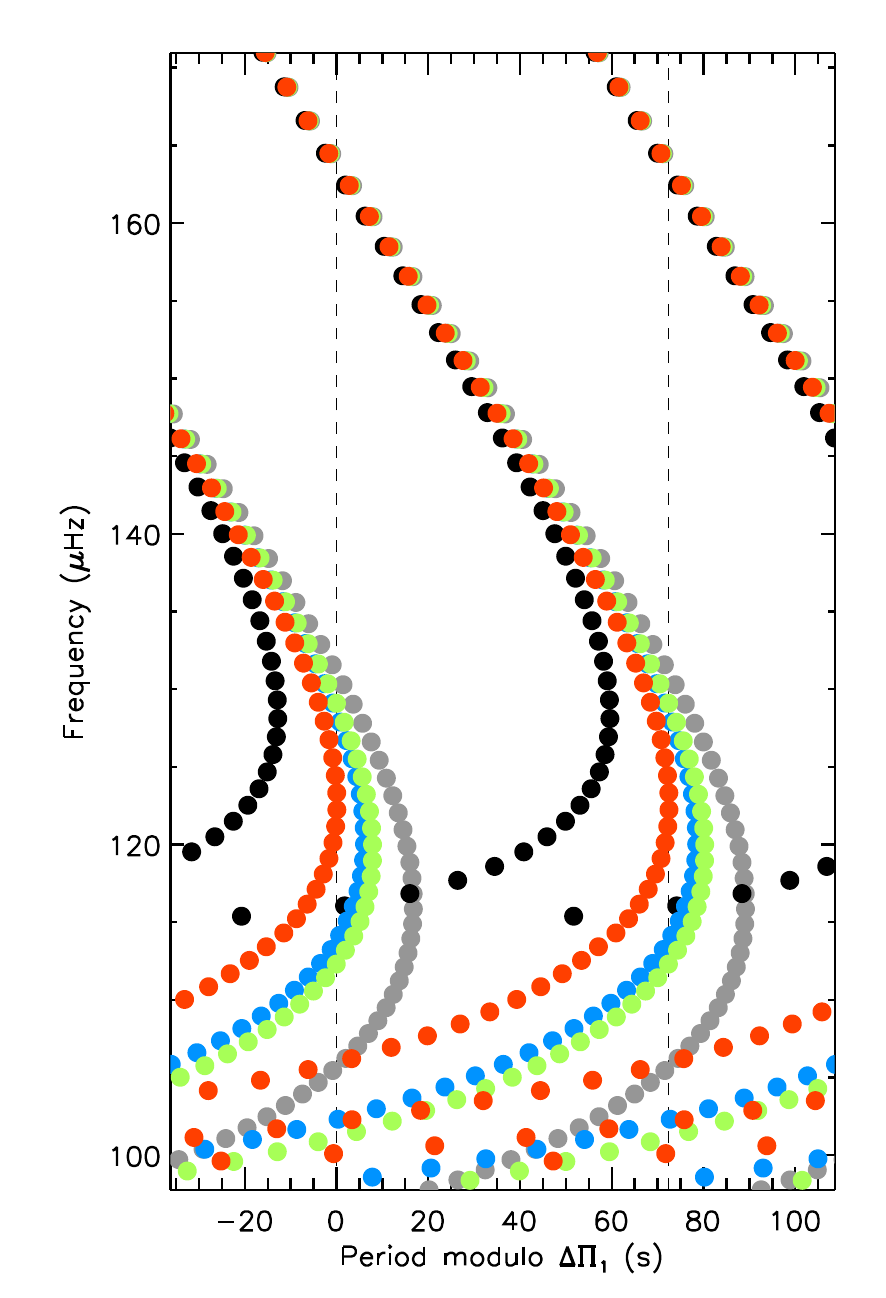}
\end{center}
\caption{Period \'echelle diagrams of $l=1$ g modes for the example case described in Sect. \ref{sect_example}. \textit{Left:} Frequencies of $m=0$ (circles) and $m=1$ (triangles) modes computed for a field with a radial extension $\sigma_{\rm B} = 0.003\,R_\star$. Magnetic effects are treated either with a non-perturbative approach (red symbols), or a perturbative approach (gray symbols). \textit{Right:} Frequencies of $m=0$ modes for values of $\sigma = 0.0015$ (red), $0.002$ (green), $0.003$ (blue) and for a constant field (black).
\label{fig_echelle_nonpert}}
\end{figure}

\begin{figure*}
\begin{center}
\includegraphics[width=0.32\linewidth]{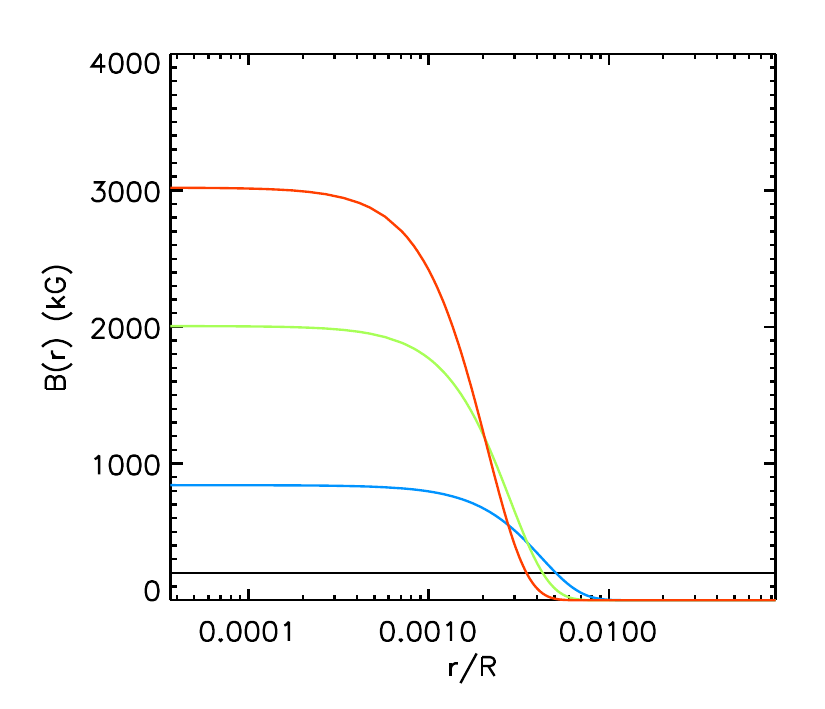}
\includegraphics[width=0.32\linewidth]{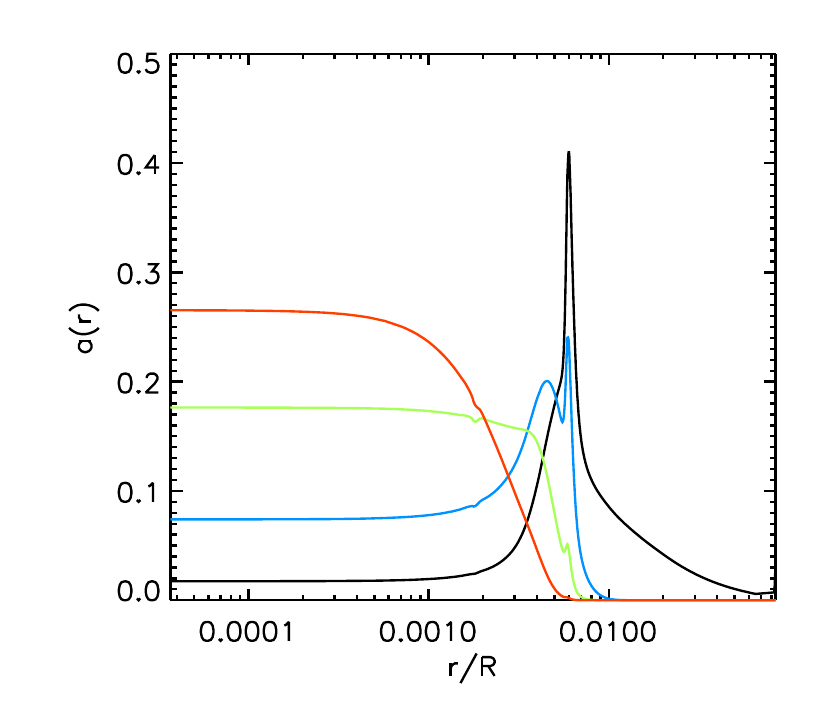}
\includegraphics[width=0.32\linewidth]{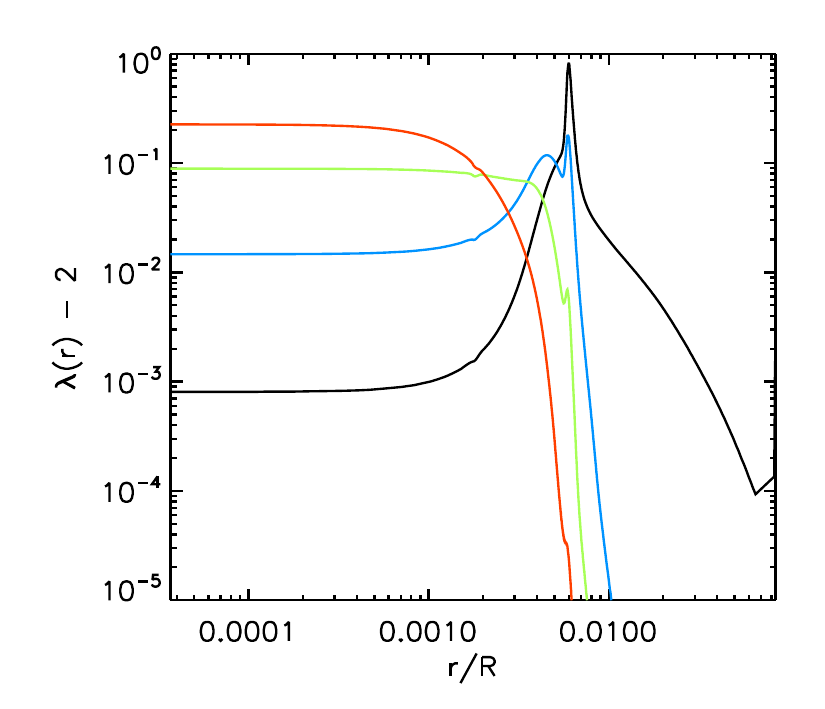}
\end{center}
\caption{Impact of the radial profile of $B_r$ on the mode frequencies. We considered fields with radial extensions of $\sigma_{\rm B} = 0.0015\,R_\star$ (red lines) , $0.002\,R_\star$ (green lines), $0.003\,R_\star$ (blue lines) and a constant field (black lines). \textit{Left:} Radial profile of $B_r$. \textit{Middle:} Profile of the parameter $a(r)$ in the g-mode cavity for the mode closest to $\numax$ (Eq. \ref{eq_a}). \textit{Right:} Profile of $\lambda(r)$ deduced from the interpolation of the $a(\lambda)$ relation for the same mode (see text). We show the departure from the value $\lambda = 2$ corresponding to the perturbative approach.
\label{fig_nonpert_radial}}
\end{figure*}

We here show an example of the oscillation spectrum obtained in the presence of a strong magnetic field. We consider a 1.1-$M_\odot$ stellar model evolved with the evolution code \mesa\ (\citealt{paxton11}) until it reaches a radius of $R_\star = 5.6\,R_\odot$. We assume a magnetic field $B_r(r,\theta) = B_0 g(r) f(\theta)$ as explained in Sect. \ref{sect_field}, setting $\sigma_{\rm B} = 0.003\,R_\star$ for the radial profile $g(r)$. We choose an asymmetry parameter $a_{\rm asym} = 0.34$ and build the corresponding function $f(\theta)$ (see Appendix \ref{app_latitudinal_profile}). In order to easily compare the results from the perturbative and non-perturbative approaches, it is convenient to relate the field intensity $B_0$ to the magnetic perturbative magnetic frequency shift $\delta\omega_0$. In this example, we considered a value of $\delta\omega_0/(2\pi) = 2.2\,\mu$Hz, which corresponds to $\Brcarre^{0.5}= 200$~kG. Using Eq. \ref{eq_omegaB}, $B_0$ is then obtained as 
\begin{equation}
B_0 = \left[ \Brcarre \frac{2}{\int_{\rint}^{\rext} K(r) g^2(r) \, \hbox{d}r \, \int_0^\pi f^2(\theta) \sin\theta \, \hbox{d}\theta} \right] ^{0.5}.
\label{eq_B0}
\end{equation}
We then calculate the frequencies of $(l=1,m=0)$ and $(l=1,m=1)$ modes by solving Eq. \ref{eq_asympt_mag} ($m=\pm1$ modes have the same frequencies because we ignored rotation here). The results are shown in the shape of a period \'echelle diagram in the left panel of Fig. \ref{fig_echelle_nonpert} (red symbols).

For comparison, we also computed mode frequencies using a perturbative approach as explained in Sect. \ref{sect_pert} with the same field configuration. The knowledge of $\delta\omega_0/(2\pi) = 2.2\,\mu$Hz and $a_{\rm asym} = 0.34$ uniquely determines the magnetic shifts, given by Eq. \ref{eq_shift_pert_m0} and \ref{eq_shift_pert_m1}. The results are overplotted in Fig. \ref{fig_echelle_nonpert} (gray symbols). As expected, for higher-frequency modes, the magnetic effects are weaker and the mode frequencies from the non-perturbative approach nearly overlap those obtained with the perturbative approach. At lower frequencies, the two sets of frequencies clearly separate and the perturbative approach is no longer valid. {We also see evidence for suppression of $m=1$ modes at low frequencies}.

\section{Impact of the profile of $B_r$ on mode frequencies \label{sect_field_profile}}

For weak fields, the perturbative approach holds, and the magnetic frequency shifts depend only on weighted averages of the radial field $B_r$ in the radial and latitudinal directions (Eq. \ref{eq_pert_Br} and \ref{eq_pert_asym}). Two different fields producing the same weighted averages are thus expected to have the same impact on mode frequencies. In the non-perturbative approach, the solutions of Eq. \ref{eq_tarm} depend on the full radial profile $g(r)$ and latitudinal profile $f(\theta)$ of $B_r$. This makes the situation more complex, but offers the potential opportunity to constrain the shape of the magnetic field.

\subsection{Radial profile of $B_r$\label{sect_radial_field} }

Using the same example case as described in Sect. \ref{sect_example} (stellar model and magnetic field configuration), we computed magnetically-distorted mode frequencies with different radial extensions of the core field. We considered values of $\sigma_{\rm B}$ of $0.0015\,R_\star$, $0.002\,R_\star$, and $0.003\,R_\star$. For our stellar model, the H-burning shell is located at approximately $6\times 10^{-3}\,R_\star$ so that the amplitudes of these three fields in the H-burning shell corresponds to 0.04\%, 1.3\%, and 14.5\% of the central amplitude, respectively. Finally, we considered a field that is constant over the whole cavity, similar to the field assumed by \cite{rui23}. For each field, $B_0$ was chosen to produce the same value of $\Brcarre$ (Eq. \ref{eq_B0}). The left panel of Fig. \ref{fig_nonpert_radial} shows the considered fields. Since the weight function $K(r)$ peaks in the H-burning shell, fields that are confined below this shell have much larger values of $B_0$. The period \'echelle diagrams of $m=0$ modes for these different fields are shown in the right panel of Fig. \ref{fig_echelle_nonpert}. At higher frequencies, the modes approximately correspond to the perturbative solutions, as expected, but at lower frequencies, it is striking to see that the considered fields produce very different spectra. 

The departure from the perturbative solution (gray symbols) is maximal for the constant field. It decreases as $\sigma_{\rm B}$ decreases to reach a minimum for $\sigma_{\rm B} \approx 2\times10^{-3}\,R_\star$ and then increases again as $\sigma_{\rm B}$ further decreases. This behavior can be understood from the profile of $a(r)$ and $\lambda(r)$, which are shown in the middle and right panels of Fig. \ref{fig_nonpert_radial} for the mode that is closest to $\numax$. The constant field has a maximal amplitude in the H-burning shell where the sensitivity to the field is the highest. It thus has large values of $a$ (and $\lambda$ values well above 2) in this shell, which produces large non-perturbative effects, and causes mode suppression at low-frequency. When the field becomes more confined in the core, the relative amplitude of the field in the H-burning shell decreases, so that non-perturbative effects weaken. For $\sigma_{\rm B} = 2\times10^{-3}\,R_\star$, the $\lambda$ value is very close to 2 in the H-burning shell. Consequently, the ridges are less distorted than in the constant-field case, and mode suppression occurs at much lower frequency (see right panel of Fig. \ref{fig_echelle_nonpert}). When the field becomes even more confined in the core ($\sigma_{\rm B} = 1.5\times10^{-3}\,R_\star$), its strength $B_0$ increases to maintain the average $\Brcarre$. As a result, the values of $a$ become large at the center, which increases non-perturbative effects and can cause mode suppression. This example demonstrates that measuring the mode frequencies of near-critical fields could give constraints on the radial extent of the core magnetic field, contrary to weaker fields.

\subsection{Latitudinal profile of $B_r$ \label{sect_latitudinal_field}}

{In this study, we found that latitudinal profiles $f(\theta)$ producing the same asymmetry parameter $a_{\rm asym}$ lead to very similar magnetic frequency shifts. To reach this conclusion, }
we considered several different profiles $f(\theta)$. Beside the profile given in Eq. \ref{eq_latitudinal_field}, we explored profiles of the type 
{$f(\theta) = P_1(\cos\theta)  + B P_3(\cos\theta) + C P_5(\cos\theta)$. In this expression, $B$ and $C$ can be determined to produce a given value of $a_{\rm asym}$, although solutions can be found for only a restricted interval (the value $C \approx 2.37$ maximizes the reachable range of $a_{\rm asym} \in [-0.252, 0.832]$).}
The value of $a_{\rm asym}$ can also be ensured by profiles of the type $f(\theta) = \sqrt{1+5a_{\rm asym} P_2(\cos\theta)}$, yielding solutions only over the interval $a_{\rm asym} \in [-0.2, 0.4]$. The analyses presented in the following sections of this paper were repeated with the latitudinal profiles presented above, and we found that our results are statistically unchanged. This suggests that non-perturbative effects {depend on the latitudinal shape of $B_r$ only through the weighted average given by $a_{\rm asym}$, as for the perturbative approach}. In the rest of this study, we used the latitudinal profile given by Eq. \ref{eq_latitudinal_field}.

\section{Application to red giants with near-critical magnetic fields \label{sect_fitdata}} 

We then used the non-perturbative approach described in Sect. \ref{sect_nonpert} to search for strong field configurations that could reproduce the observations for the eight stars of the studied sample. 
For this purpose, we built an asymptotic expression of mixed modes including the effects of magnetic fields following a non-perturbative approach, and we explored the parameter space.

\subsection{Asymptotic model}

To derive asymptotic expressions of magnetic mixed modes, we followed the approach that was first described in \cite{li22} and then used in \cite{li23} and \cite{deheuvels23}. From the work of \cite{shibahashi79} and \cite{unno89}, asymptotic expressions of mixed modes {of azimuthal order $m$} are obtained by solving the equation 
\begin{equation}
\tan \theta_{\textrm{p},m} = q \tan \theta_{\textrm{g},m}
\label{eq_mixmodes}
\end{equation}
where $q$ corresponds to the coupling strength between the p- and g-mode cavities, and $\theta_{\textrm{p},m}$, $\theta_{\textrm{g},m}$ are phase terms that can be expressed as a function of asymptotic frequencies of p and g modes. Following \cite{mosser12a}, we have 
\begin{align}
\theta_{\textrm{p},m} & = \frac{\pi}{\deltanu(n_{\rm p})} (\nu - \nu_{\textrm{p},m}) \\
\theta_{\textrm{g},m} & = \frac{\pi}{\Delta P(n_{\rm g})} \left( \frac{1}{\nu} - P_{\textrm{g},m} \right) - \frac{\pi}{2}
\end{align}
where $\deltanu(n_{\rm p})$ and $\Delta P(n_{\rm g})$ are the local large separation of p modes and period spacing of g modes, and $\nu_{\textrm{p},m}$ and $P_{\textrm{g},m}$ are the frequencies of pure p and periods of pure g modes, respectively\footnote{{As mentioned in \cite{villate26}, the expression of $\theta_{\textrm{g},m}$ was modified compared to \cite{mosser15}, so that we recover the frequencies of pure g modes as solutions of Eq. \ref{eq_mixmodes} when the coupling parameter $q$ tends to zero.}}. 

The effects of the magnetic field on pressure modes can be neglected (\citealt{mathis21}) and we here consider the case of a negligible rotation, so that the asymptotic frequencies $\nu_{\textrm{p},m}$ can be written as
\begin{equation}
\nu_{\textrm{p},m} = \left[ n_{\rm p} + \frac{1}{2} + \varepsilon_{\rm p}  +\frac{\alpha}{2} \left( n - n_{\rm max} \right)^2 \right] \deltanu - d_{01},
\end{equation}
where $\alpha$ measures the second-order effects in the asymptotic development around $\numax$ (as expressed by \citealt{mosser12a}), $\varepsilon_{\rm p}$ is a phase offset, $n_{\rm max} = \numax/\deltanu$, and $d_{01}$ corresponds to the small separation $\delta\nu_{01} = [\nu_{\textrm{p}}(0,n_{\rm p}) + \nu_{\textrm{p}}(0,n_{\rm p}+1) - 2\nu_{\textrm{p}}(1,n_{\rm p})]/2$ when mode frequencies are expressed at first-order. 

For g modes, we include the effects of the magnetic field in a non-perturbative way using Eq. \ref{eq_asympt_mag}. For this purpose, a stellar model is needed. This is an additional difficulty compared to the perturbative approach, for which the magnetic shifts only depend on the average quantity $\omega_{\rm B}$ (Eq. \ref{eq_shift_pert_m0} and \ref{eq_shift_pert_m1}). 
{To explore the parameter space, we computed one reference stellar model for each star} (in Sect. \ref{sect_stellar_model}, we explain the choice of a reference model {and we discuss the validity of this approach}). For this model, we can compute the asymptotic mode frequencies taking into account magnetic fields in a non-perturbative way by solving Eq. \ref{eq_asympt_mag} for each mode. We can also compute non-magnetic asymptotic frequencies $\omega_{\rm g}^{(0)}$ by setting $\lambda=2$ in Eq. \ref{eq_asympt_mag}. We then obtain fractional magnetic frequency shifts as
 \begin{equation}
\frac{\delta\omega_{\textrm{mag},m}}{\omega_{\rm g}^{(0)}} =  \frac{ \int_{\rint}^{\rext}  \frac{N}{r} \sqrt{\lambda_m(r)} \, \hbox{d}r}{  \int_{\rint}^{\rext}  \frac{N}{r} \sqrt{2} \, \hbox{d}r} - 1
\label{eq_magnetic_shift_nonpert}
\end{equation}
When exploring the parameter space, we assume that the fractional frequency shift of a given mode is that of the reference model given by the expression above.
To obtain asymptotic frequencies of pure g modes, we add the magnetic frequency shift given by Eq. \ref{eq_magnetic_shift_nonpert}, leading to the expression
\begin{equation}
P_{\textrm{g},m} = P_{\rm g}^{(0)} \left( 1 + \frac{\delta\omega_{\textrm{mag},m}}{\omega_{\rm g}^{(0)}} \right)^{-1},
\end{equation}
where $P_{\rm g}^{(0)} = (n_{\rm g} + 1/2 + \varepsilon_{\rm g})\dpun$.

Overall, the free parameters of the asymptotic model are the properties of pure p modes ($\deltanu$, $\varepsilon_{\rm p}$, $\alpha$, $d_{01}$), those of pure g modes ($\dpun$, $\varepsilon_{\rm g}$), the coupling strength $q$, and the properties of the magnetic field {given by the parameters $(\delta\omega_0, \sigma_{\rm B}, a_{\rm asym})$}. To characterize the field, we proceed as in Sect. \ref{sect_example}. The free parameters $\delta\omega_0$ and $\sigma_{\rm B}$ are used to build the Gaussian-shaped radial profile $g(r)$ of $B_r$ ($B_0$ is obtained from $\delta\omega_0$ using Eq. \ref{eq_omegaB} and \ref{eq_B0}), and the parameter $a_{\rm asym}$ is used to generate the latitudinal profile $f(\theta)$ from Eq. \ref{eq_latitudinal_field}. For each set of parameters, the asymptotic frequencies of dipole mixed modes for $m=0$ and $m=1$ are calculated by solving Eq. \ref{eq_mixmodes}. Finally, we must remind that asymptotic models are only approximate solutions of the oscillation equations. Departures from asymptotics are taken into account by including {a normal} model error with widths $\sigma_{\rm p}$ and $\sigma_{\rm g}$ for p and g modes, respectively. The quantities $\sigma_{\rm p}$ and $\sigma_{\rm g}$ are considered as free parameters when searching for optimal solutions.

\subsection{Search for reference stellar models \label{sect_stellar_model}}

To search for reference models for the eight stars, we used a grid of stellar models computed with \mesa\ (\citealp{paxton11,paxton13,paxton15,paxton18,paxton19}), with masses varying between 0.9 and 2~$M_\odot$ (step 0.1~$M_\odot$), metallicities between $-0.4$ and 0.4~dex (step 0.1~dex), and a step overshoot was added beyond the convective core with $\aov=0$, 0.1, or 0.2. We used the OPAL equation of state \citep{rogers02} and opacity tables. We used nuclear reaction rates from NACRE \citep{angulo99}. Microscopic diffusion was included by solving the equations of \cite{burgers69} at each time step. The models were evolved until they reach $15\,R_\odot$, keeping track of the internal structure for all models computed along the evolution. For each star, we searched for the model of the grid that best reproduces the asymptotic large separation of p modes $\deltanu$, the frequency of maximum power of the oscillations $\numax$, and the asymptotic period spacing of g modes $\dpun$ by minimizing the function
\begin{equation}
\chi^2 = \frac{(\deltanu - \deltanu^{\rm obs})^2}{\sigma^2_{\deltanu}} + \frac{(\numax - \numax^{\rm obs})^2}{\sigma^2_{\numax}} + \frac{(\dpun - \dpun^{\rm obs})^2}{\sigma^2_{\dpun}} .
\label{eq_chi2}
\end{equation}
We note that $\dpun$ is in fact not known at this stage because the measured period spacing $\dpun^{\rm meas}$ is an underestimated measurement of $\dpun$ (see Sect. \ref{sect_distorted_doublets}). To obtain a first estimate of $\dpun$, we used the results of the perturbative approach obtained in Sect. \ref{sect_pert}. As mentioned above, for the stars with the strongest ridge distortion, this approach cannot fit both ridges simultaneously. In these cases, we used the results of the perturbative approach applied to the ridge showing the weakest distortion alone. We checked a posteriori that it produces estimates of $\dpun$ that are close to the more proper measurements obtained with the non-perturbative approach in Sect. \ref{sect_abim}. 

To check that our conclusions do not critically depend on the choice of a reference model, the fits presented in Sect. \ref{sect_abim} were performed again using the next best models in the grid and we found that the results were statistically indistinguishable. Similarly, for one of the stars (KIC~5700274), the reference model was updated using the asymptotic period spacing $\dpun$ found in Sect. \ref{sect_abim} and a new fit was performed using this model. Again, the posterior distribution of the parameters was found to be nearly identical.

\begin{figure*}
\begin{center}
\includegraphics[width=0.24\linewidth]{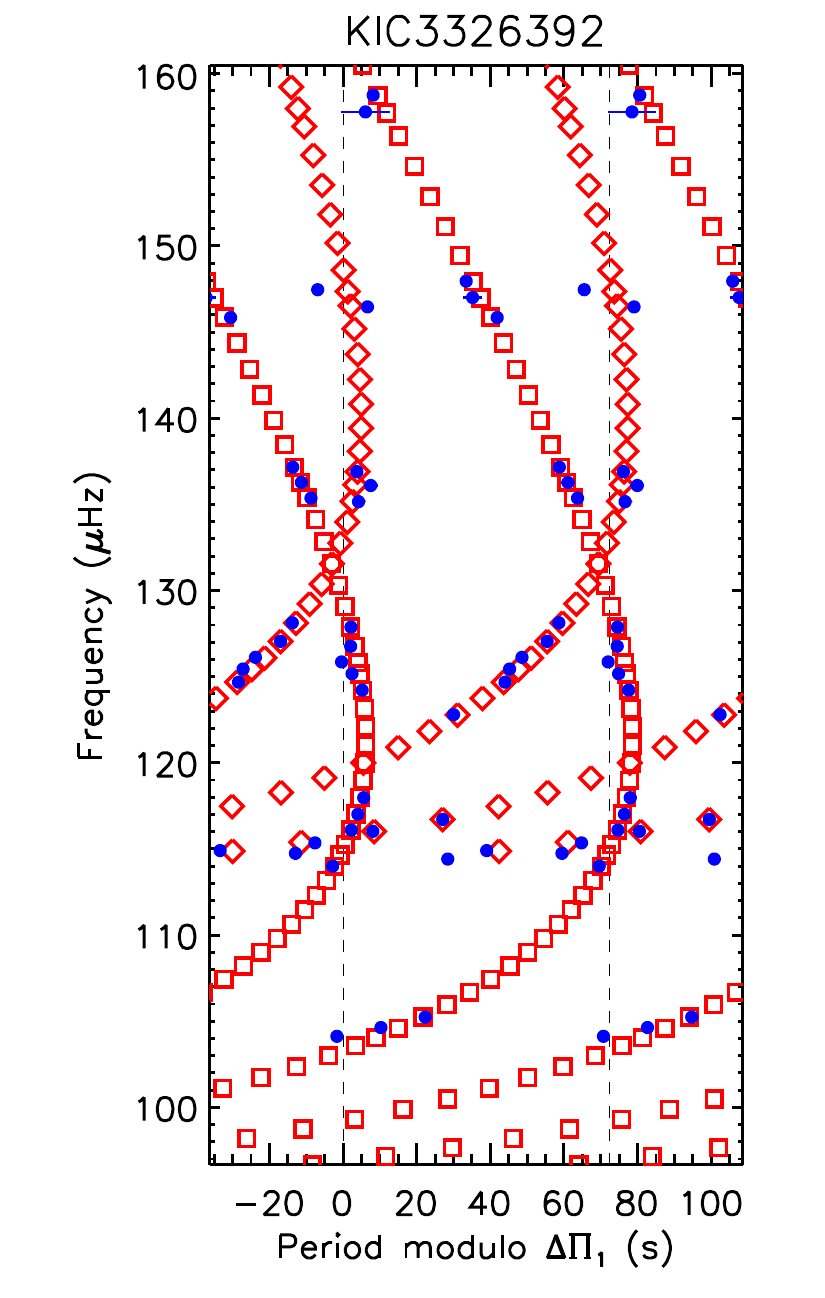}
\includegraphics[width=0.24\linewidth]{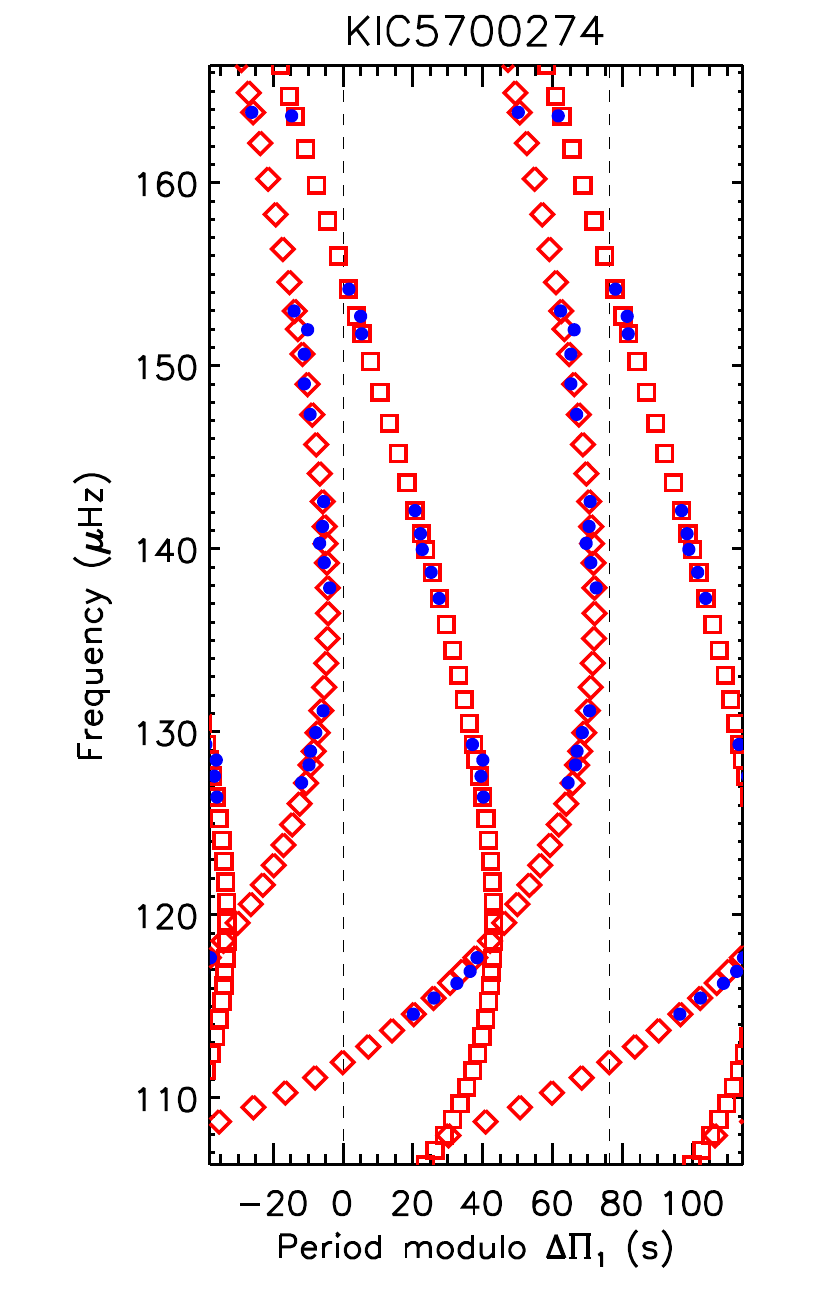}
\includegraphics[width=0.24\linewidth]{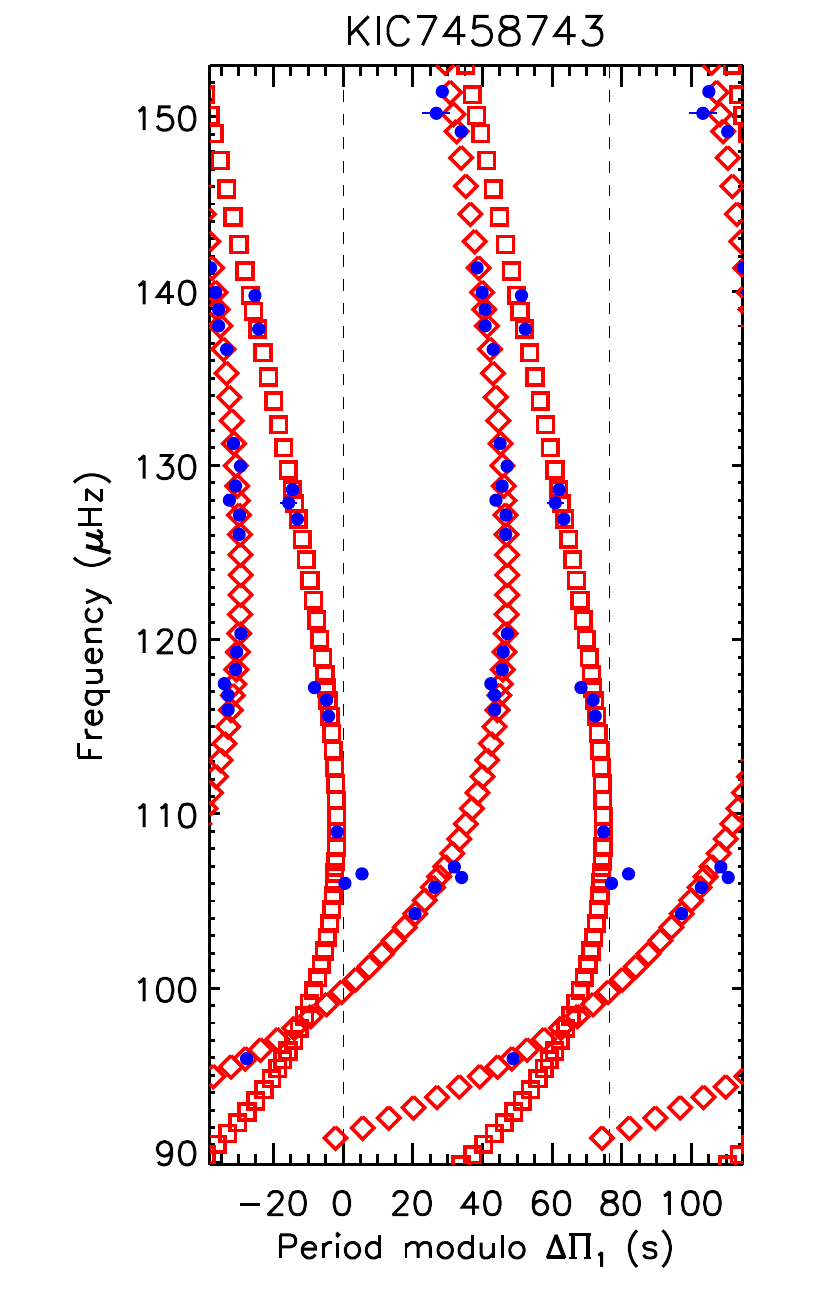}
\includegraphics[width=0.24\linewidth]{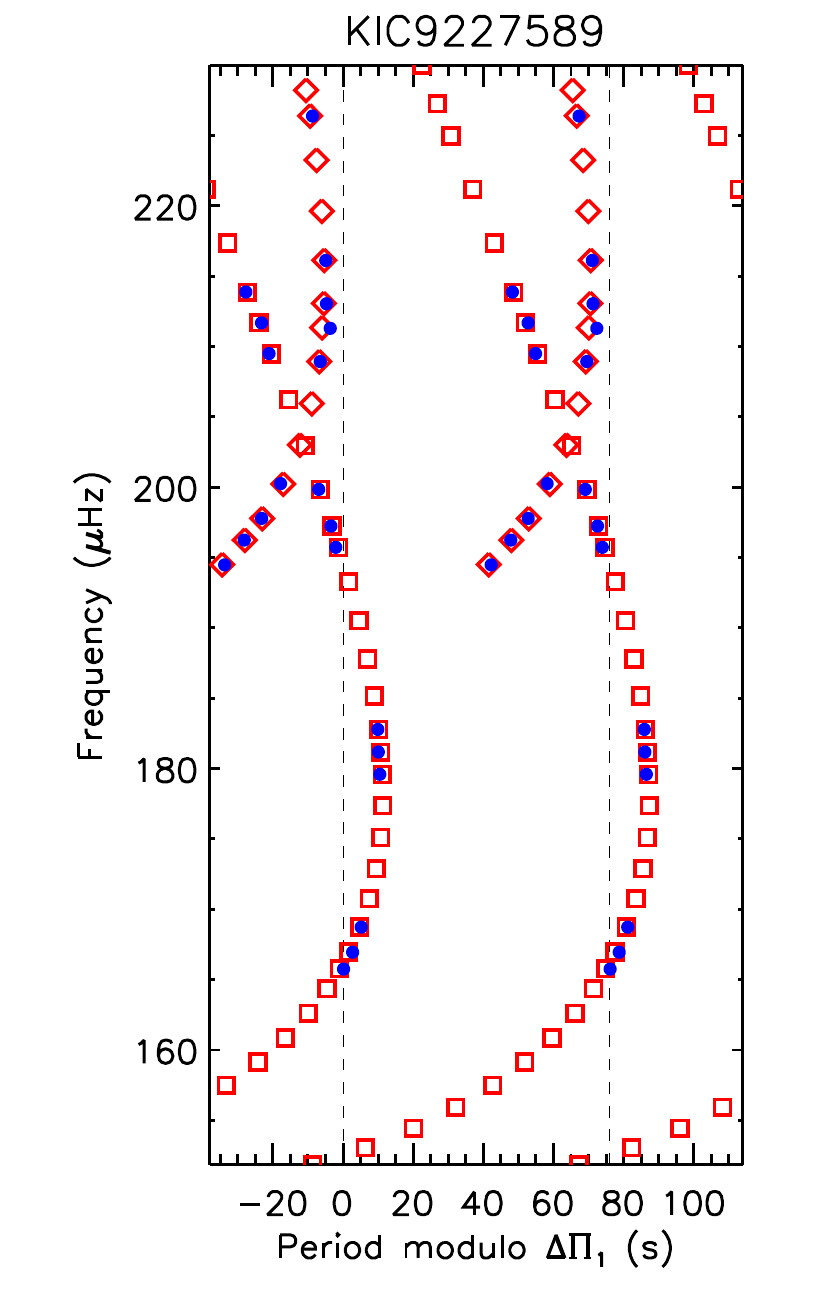}
\includegraphics[width=0.24\linewidth]{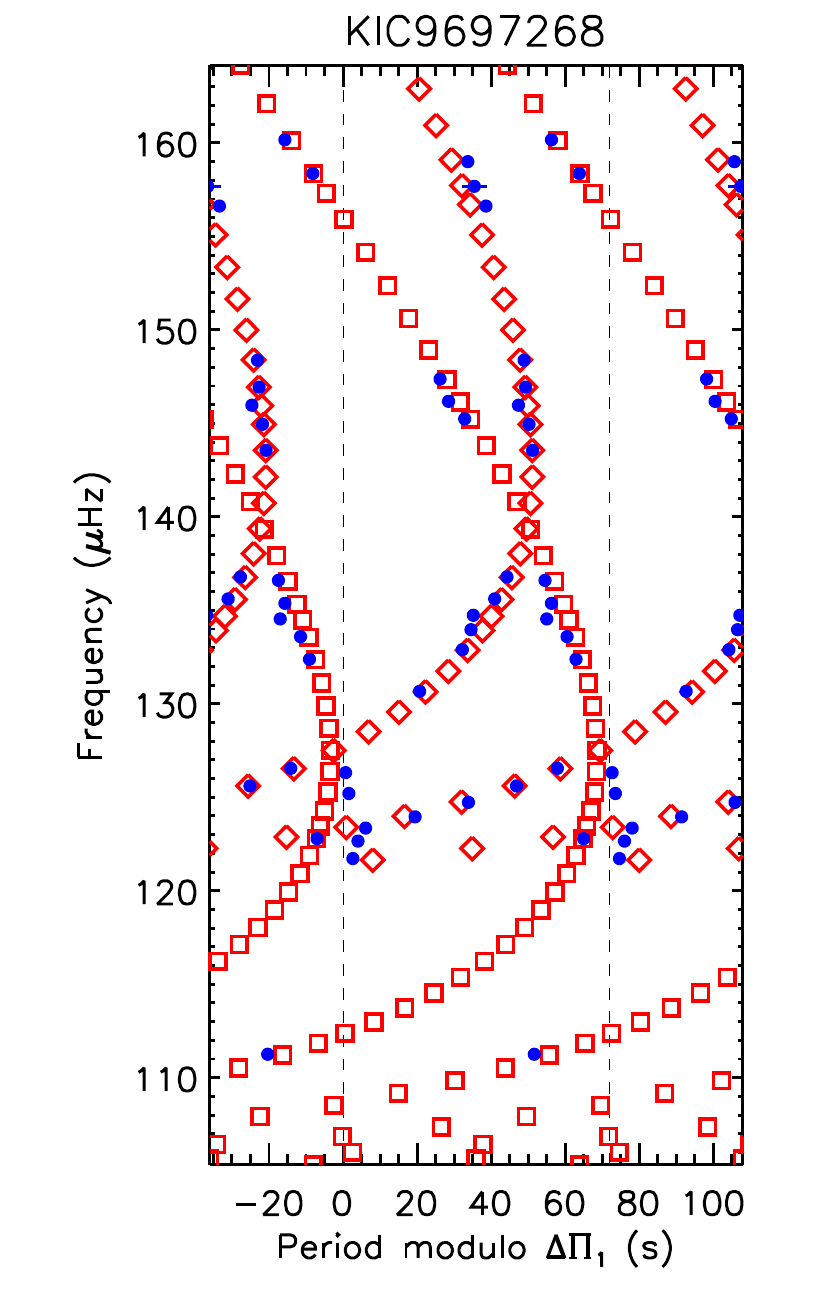}
\includegraphics[width=0.24\linewidth]{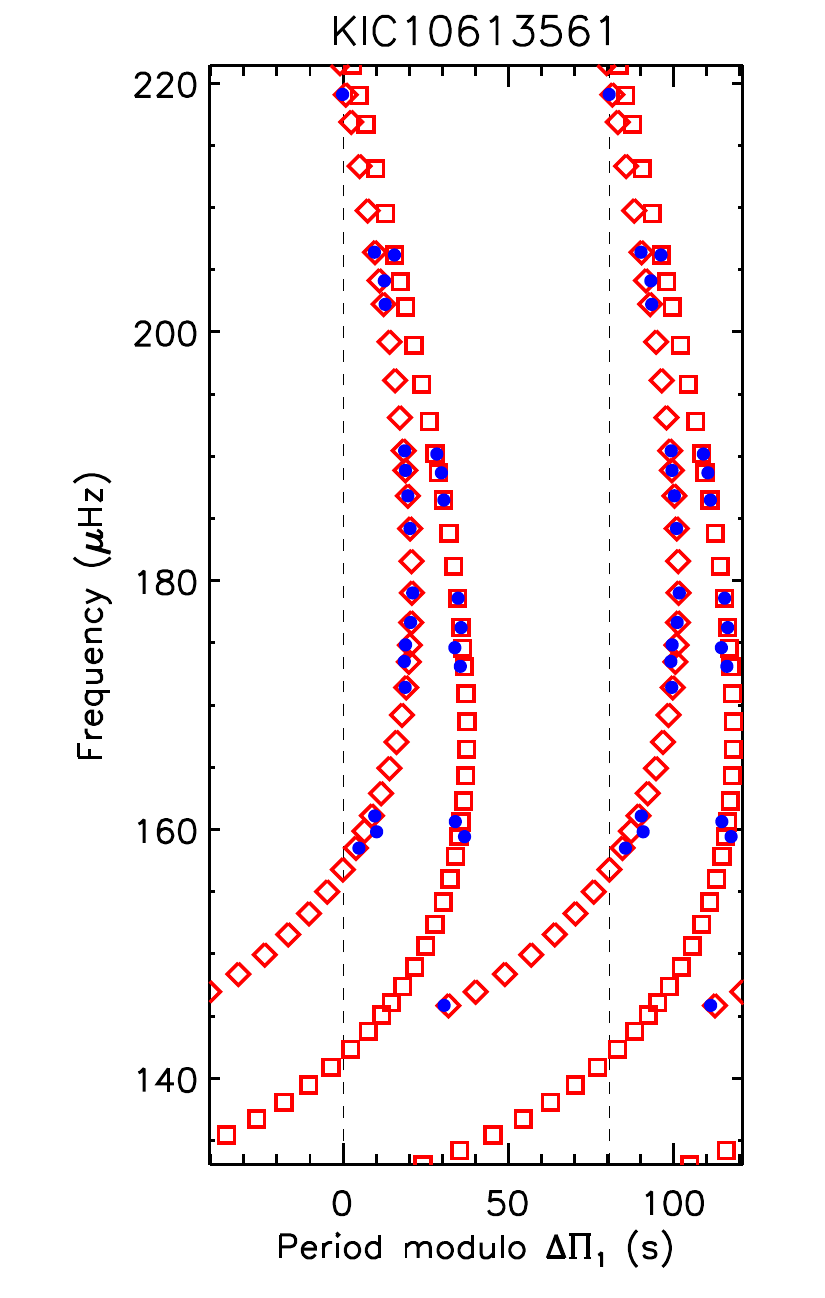}
\includegraphics[width=0.24\linewidth]{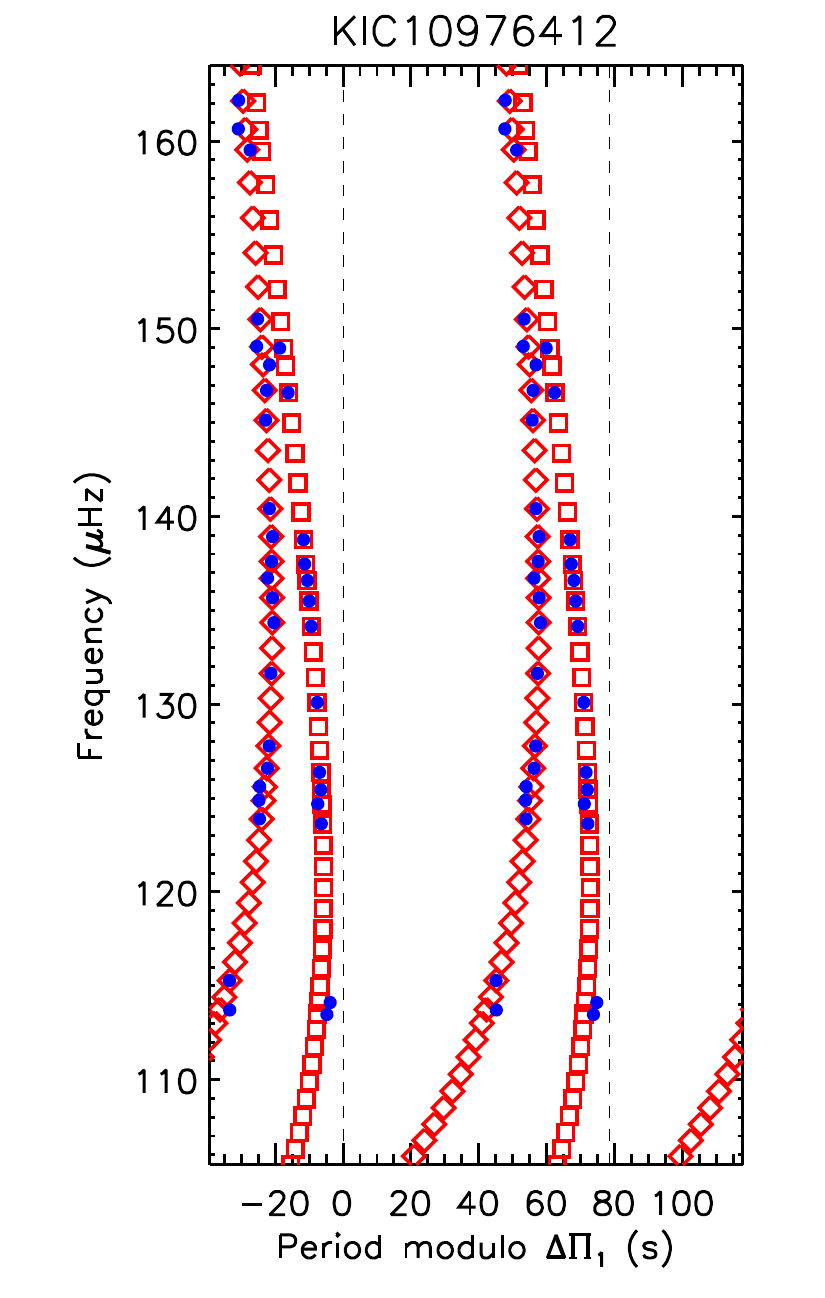}
\includegraphics[width=0.24\linewidth]{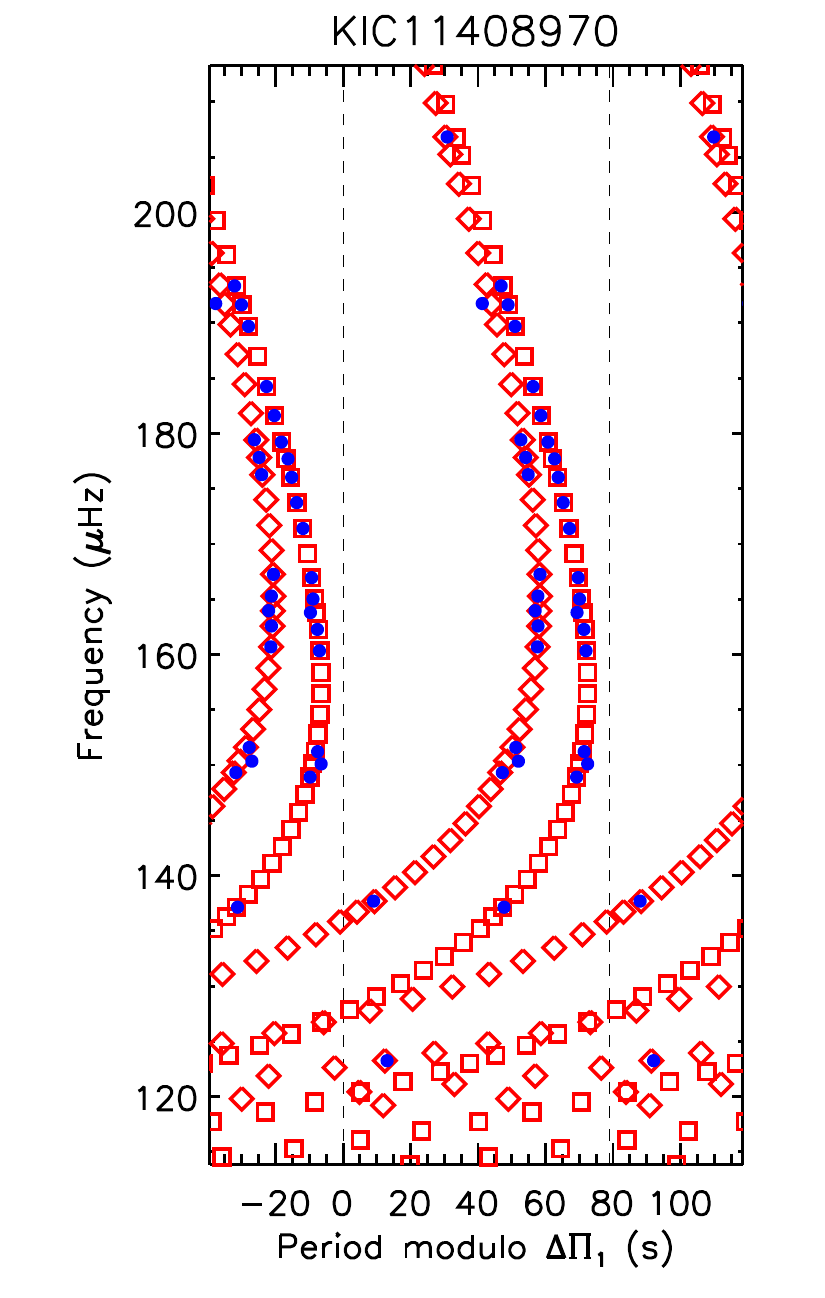}
\end{center}
\caption{Stretched \'echelle diagrams of three stars of our sample. Blue circles and horizontal blue lines correspond to observed modes and their 1-$\sigma$ uncertainties. Red empty circles correspond to the best asymptotic models including magnetic effects in a non-perturbative manner.
\label{fig_results_echelle}}
\end{figure*}

\subsection{Asymptotic fits \label{sect_abim}}
We ﬁt our model to the observed frequencies by following a Bayesian approach. We used the Asteroseismic Bayesian Inference by MCMC (ABIM) code \citep[see][for more details]{villate26}. ABIM is a Fortran code parallelised with OpenMP directives. Sampling is performed with a Markov Chain Monte Carlo method implementing the so-called \emph{stretched move} algorithm proposed by \citet{Goodman_Weare_2010} and including parallel tempering \citep[e.g.,][]{benomar09a}, which improved the exploration of parameter spaces containing numerous local maxima, as is the case here. We typically sample the posterior distributions with 15 parallel chains and 256 walkers for each chain. The initial positions of the walkers are randomly drawn from the prior distributions. The walkers are iterated over 12000 steps, with the 2000 first steps discarded as burn-in to ensure that chains are stationary. The priors for each parameter are provided in Appendix \ref{app_abim}. {To calculate the likelihood, we compare each observed mode to the closest mode in the asymptotic model considered.}

For each step of the MCMC, we need to compute the relation $\lambda(a)$ corresponding to the considered value of $a_{\rm asym}$, which requires to solve Eq. \ref{eq_tarm} using a relaxation scheme, as described in Sect. \ref{sect_horizontal}. In order to reduce the computing time, we pre-computed tables of $\lambda(a)$ relations for all possible asymmetry parameters with a step of $10^{-3}$. Then, during the MCMC procedure, the value of $\lambda$ is obtained by performing a bilinear interpolation for the considered values of $a$ and $a_{\rm asym}$. 

\subsection{Results \label{sect_results}}

\begin{figure*}
\begin{center}
\includegraphics[width=0.49\linewidth]{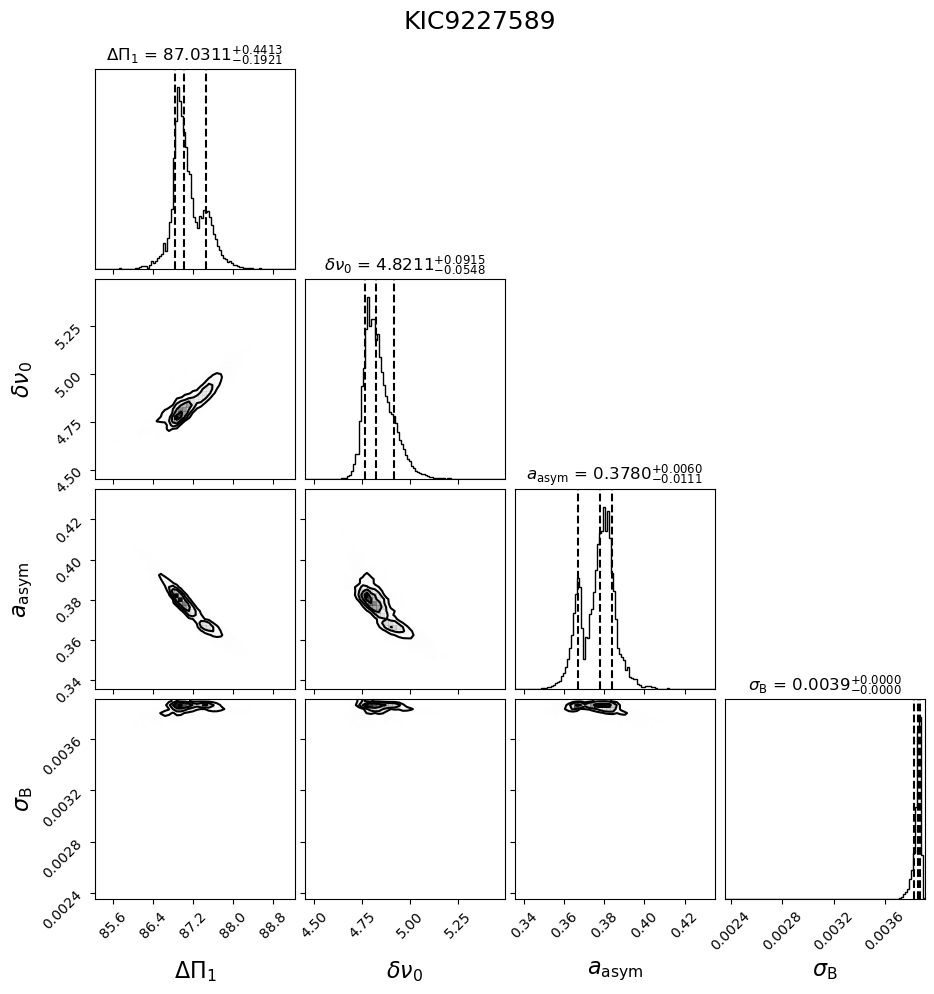}
\includegraphics[width=0.49\linewidth]{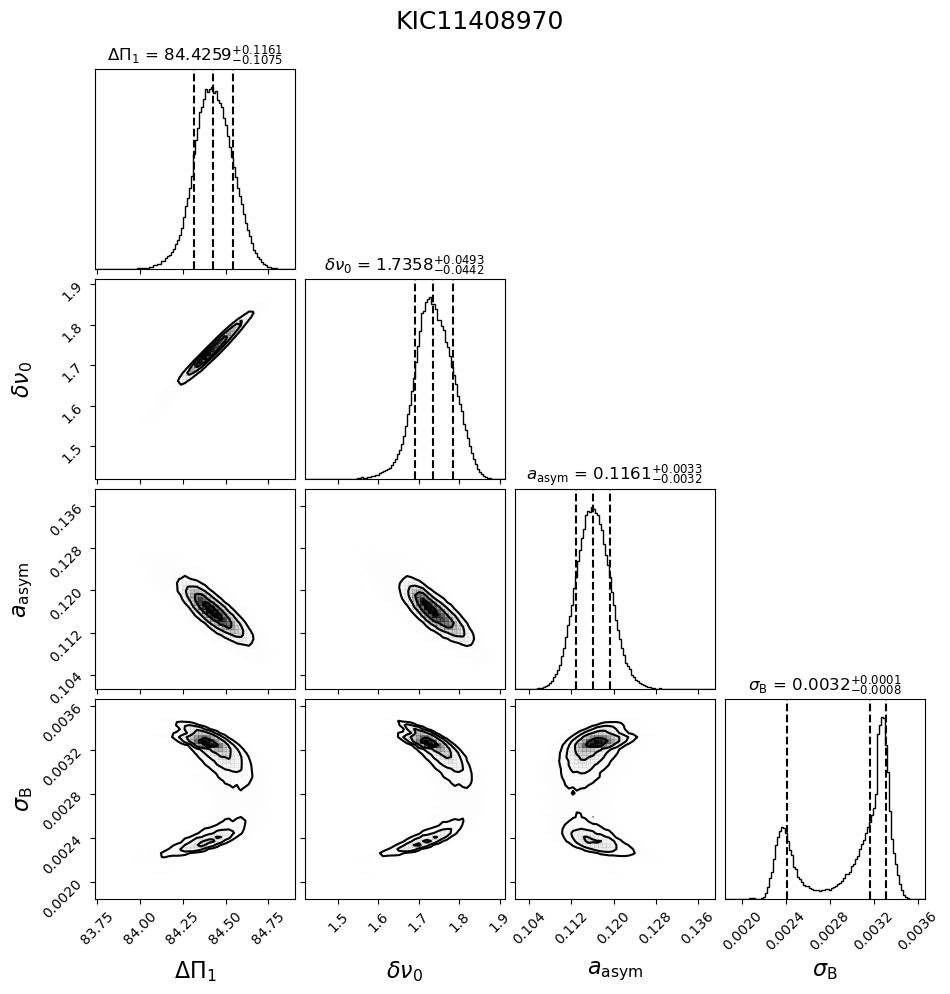}
\end{center}
\caption{Corner plots of \starD\ (left) and \starH\ (right), restricted to the asymptotic period $\dpun$ of dipolar g modes, and the parameters characterizing the core magnetic field: the magnetic shift $\delta\nu_0 = \delta\omega_0/(2\pi)$, $\sigma_B$, and $a_{\rm asym}$. The corner plots for other stars are given in Appendix \ref{app_abim}.
\label{fig_corner}}
\end{figure*}

For all eight stars, we were able to reproduce very well the observed mode frequencies with our asymptotic model. This can be seen in Fig. \ref{fig_results_echelle}, which shows stretched \'echelle diagrams all eight stars 
We recall that no satisfactory fits had been obtained using a perturbative approach, or identifying the modes as $m=\pm1$ components. {We note that the non-perturbative approach as we have applied it in this study involves one more degree of freedom compared to the perturbative approach (the radial extent of the field $\sigma_{\rm B}$). This increase in the degrees of freedom is fully justified because the perturbative approach (i) fails to fit the most distorted stars, and (ii) leads to field measurements that are close to the critical field strength, which makes the perturbative approach not valid (see \citealt{rui24}).}

The optimal parameters of the asymptotic models are given in Table \ref{tab_results} and we show the posterior probability of a selected sample of parameters in Fig. \ref{fig_corner} (see also Fig. \ref{fig_corner_app1} and \ref{fig_corner_app2}). In particular, we obtained measurements of the asymptotic period spacing $\dpun$ that are unbiased by the magnetic field, contrary to the apparent period spacing $\dpun^{\rm meas}$. We thus revised the location of the eight red giants in the $(\deltanu,\dpun)$ plane (red star-shaped symbols in Fig. \ref{fig_dn_dp}). As can be seen in this figure, the stars now clearly belong to the degeneracy sequence, as expected. {This result is quite encouraging since} we did not include any prior stating that $\dpun$ should be such that the stars lie on the degeneracy sequence.

\subsubsection{Field strength}

The measured values of $\delta\omega_0$ could be used to obtain estimates of the measured field strength. We expressed them as the weighted average $\Brcarre^{0.5}$, for consistency with previous core field measurements. To relate $\delta\omega_0$ to $\Brcarre^{0.5}$, we used Eq. \ref{eq_omegaB} and the reference stellar models obtained in Sect. \ref{sect_stellar_model}. We accounted for the uncertainties in the measurements of $\delta\omega_0$ (Table \ref{tab_results}) and uncertainties related to the choice of a reference model were estimated by considering all stellar models from the grid yielding $\chi^2 < \chi^2_{\rm min} + 9$, where $\chi^2_{\rm min}$ is the value of $\chi^2$ (see Eq. \ref{eq_chi2}) for the reference model. We obtained field strengths ranging from about 100 to 700~kG. These measurements are in line with those found by \cite{deheuvels23} for similar stars, except that in this previous study, only lower limits to the field strength were obtained (because only one component per triplet was detected), and a perturbative approach was used even though the fields are near-critical. In the last column of Table \ref{tab_results}, we give the value of the field strength at the center $B_0$ for the median model of each star. We find that $B_0$ is always much larger than $\Brcarre^{0.5}$. This is because the optimal magnetic fields are all confined well below the H-burning shell (see Sect. \ref{sect_radial_profile}). {For the four stars that yielded reasonable fits with the perturbative approach (\starC, \starF, \starG, and \starH), we found that the perturbative approach overestimates the field strength by 5 to 10\%.} 

{In the results presented here, we assumed that the core rotation is too weak to separate $m=\pm1$ components (see Sect. \ref{sect_ID01}). Our alternative explanation for the detection of two components only was that for some as-yet-unknown reason, only one of the $m=\pm1$ components is visible. To check the impact of this alternative scenario on our results, we repeated our analysis of one of our stars (\starA) assuming a core rotation of $\omg = 0.5\,\mu$Hz, which is typical for red giants (\citealt{hatt24}, \citealt{li24}). We found that the results are nearly unchanged, the measured field strength being modified by only 10\,kG, which is within our quoted uncertainties.}

\subsubsection{Mode identification \label{sect_results_ID}}

In this study, we assumed that the two detected ridges correspond to $m=0$ and $m=\pm1$ ridges, but we made no assumption about which ridge corresponds to which $m$-component. With the perturbative approach, the non-rotating case holds an ambiguity. The oscillation spectrum corresponding to $\delta\omega_0$ and $a_{\rm asym}$ is identical to the spectrum obtained with $\delta\omega'_0 = \delta\omega_0 (2-a_{\rm asym})/2$ and $a'_{\rm asym} = 2a_{\rm asym}/(a_{\rm asym}-2)$, the other parameters being unchanged. This degeneracy exists only for $a_{\rm asym} < 0.4$, to ensure that $a'_{\rm asym} \in [-1/2,1]$. For strong fields, using the non-perturbative approach, this degeneracy is lifted. For all eight stars, the preferred solutions with positive asymmetries are clearly preferred, which corresponds to identifying the most distorted ridge of the $m=\pm1$ component. For some stars, the negative asymmetry solutions appear as a small bump in the posterior distribution of $a_{\rm asym}$ (e.g., for \starH), but it remains negligible compared to the positive asymmetry solution. This is in line with the observation made for weaker magnetic fields that stars with positive $a_{\rm asym}$ are more frequent that stars with $a_{\rm asym}<0$.

\begin{table*}
\begin{center}
\caption{Results of fitting asymptotic mode frequencies (including magnetic effects in a non-perturbative way) to the observed frequencies of the red giants in our sample. 
\label{tab_results}}
\begin{tabular}{l c c c c c c}
\hline \hline
\T KIC Id & $\dpun$ & $\delta\omega_0/2\pi$ & $a_{\rm asym}$ & $\sigma_{\rm B}$ & $\Brcarre^{0.5}$ & $B_0$ \\
\B & (s) & ($\mu$Hz) & - & $(\times 10^{-3}\,R_\star)$ & kG & MG \\
\hline
\T \B KIC3326392 & $79.43^{+0.37}_{-0.60}$ & $2.76^{+0.11}_{-0.13}$ & $0.322^{+0.025}_{-0.008}$ & $1.76^{+0.02}_{-0.05}$ \,/\, $2.97^{+0.02}_{-0.09}$ & $215 \pm 7$ & 5.1 \\ 
\T \B KIC5700274 & $81.02^{+0.26}_{-0.31}$ & $1.15^{+0.10}_{-0.11}$ & $0.309^{+0.024}_{-0.018}$ & $3.06^{+0.33}_{-0.50}$ & $179 \pm 11$ & 1.4 \\ 
\T \B KIC7458743 & $79.22^{+0.12}_{-0.15}$ & $0.70^{+0.04}_{-0.06}$ & $0.290^{+0.017}_{-0.013}$ & $2.69^{+0.74}_{-0.74}$ & $111 \pm 4$ & 0.8 \\ 
\T \B KIC9227589 & $87.11^{+0.44}_{-0.19}$ & $4.84^{+0.09}_{-0.05}$ & $0.377^{+0.006}_{-0.011}$ & $3.85^{+0.02}_{-0.03}$ & $681 \pm 20$ & 5.5 \\ 
\T \B KIC9697268 & $82.76^{+0.24}_{-0.41}$ & $3.33^{+0.09}_{-0.11}$ & $0.259^{+0.007}_{-0.008}$ & $2.52^{+0.04}_{-0.04}$ & $331 \pm 36$ & 3.3 \\ 
\T \B KIC10613561 & $85.59^{+0.27}_{-0.22}$ & $1.66^{+0.13}_{-0.10}$ & $0.157^{+0.010}_{-0.010}$ & $2.10^{+0.25}_{-0.23}$ \,/\, $4.47^{+0.20}_{-0.38}$ & $377 \pm 39$ & 1.4 \\ 
\T \B KIC10976412 & $80.38^{+0.17}_{-0.14}$ & $0.54^{+0.06}_{-0.06}$ & $0.273^{+0.027}_{-0.026}$ & $3.59^{+1.23}_{-0.88}$ & $114 \pm 7$ & 0.8 \\ 
\T \B KIC11408970 & $84.43^{+0.12}_{-0.11}$ & $1.74^{+0.05}_{-0.04}$ & $0.116^{+0.003}_{-0.003}$ & $2.39^{+0.12}_{-0.08}$ \,/\, $3.24^{+0.08}_{-0.18}$ & $333 \pm 26$ & 1.9 \\ 
\hline
\end{tabular}
\tablefoot{For \starA, \starF, and \starH, the distribution for $\sigma_{\rm B}$ is bimodal and the possible solutions are given. The last two columns give the average magnetic field strengths in the core $\Brcarre^{0.5}$ measured using a stellar model, and the field strength $B_0$ at the center for the median model (in MG). }
\end{center}
\end{table*}

\subsubsection{Radial extent of the core magnetic field \label{sect_radial_profile}}

As demonstrated in Sect. \ref{sect_radial_field}, we have the possibility to constrain the radial extent of the core magnetic field when the field is near-critical. For all eight stars, we obtained clear contraints on the width $\sigma_{\rm B}$ of the Gaussian function that was used to model the radial profile of $B_r$ (see Table \ref{tab_results}). The posterior distribution of $\sigma_{\rm B}$ can be seen in the corner plots shown in Fig. \ref{fig_corner}, \ref{fig_corner_app1}, and \ref{fig_corner_app2}. For four stars (\starB, \starD, \starE, and \starG), the posterior distribution of $\sigma_{\rm B}$ peaks on a single value. 
For \starA, \starF, \starH, and to a lesser extent \starC, we find a bimodality of the solutions. For instance, for \starA, the observations can be reproduced either with 
$\sigma_{\rm B} = 0.0018\,R_\star$ or with $\sigma_{\rm B} = 0.0029\,R_\star$.
This phenomenon is linked to the behavior described in Sect. \ref{sect_radial_field}. For a given value of $\Brcarre$, non-perturbative effects can be large either when the field is strong near the HBS, or when the field is confined in the central layers. These two possibilities give quite similar oscillation spectra, so that both solutions are possible. In practice, the two solutions lead to similar values of $\Brcarre$ and $a_{\rm asym}$. 

For all eight targets, we found that the radial extent of the core field $\sigma_{\rm B}$ is significantly smaller than the radius of the HBS $r_{\rm HBS}$ (the ratio $\sigma_{\rm B}/r_{\rm HBS}$ ranges from 30\% to 50\% for all eight stars).
Consequently, the field strength in the HBS is less than 10\% of its maximal intensity for all stars. If we consider magnetic fields that extend closer to the HBS, non-perturbative effects strongly increase, the ridges become too distorted, and mode suppression occurs at too-high frequency compared to the observations, so that no satisfactory solution can be found.

\subsubsection{Mode suppression}

For \starA, \starD, and \starE, seismic observations have shown that one ridge undergoes mode suppression at low frequency. Even though we did not include as a constraint in our fit that one ridge is suppressed, our best models indeed show mode suppression on the same ridge (the $m=\pm1$ ridge according to Sect. \ref{sect_results_ID}) for all three targets. Moreover, we find that the suppression of the ridge in our optimal models occurs at a frequency that is {very} close to the lowest-frequency mode detected in the ridge. This can be seen in the stretched \'echelle diagrams shown in Fig. \ref{fig_results_echelle}. {For \starD\ and \starE, the lowest-frequency detected mode coincides exactly with the last non-suppressed $m=1$ mode in our optimal model. For \starA, mode suppression occurs at a frequency of $114.83\,\mu$Hz, which is only slightly higher than the lowest-frequency observed modes ($114.41\,\mu$Hz).}
We find this agreement quite encouraging, although more thorough work will be needed to determine to what extent the observed frequency at which modes become suppressed can be used to infer magnetic field properties. We also checked the reason for mode suppression in the three stars. For \starD\ and \starE, the solution for $\sigma_{\rm B}$ is unimodal and mode suppression occurs because $a(r)$ exceeds the maximal value of $a_{\rm max}$ in the core. For \starA, the solutions are bimodal, and as could be expected, we find that $a(r)$ exceeds $a_{\rm max}$ in the HBS for the solution with higher $\sigma_{\rm B}$, and near the core for the solution with lower $\sigma_{\rm B}$.

\section{Discussion \label{sect_discussion}}

\subsection{Origin of the detected fields}

One of the main scenarios is that the magnetic fields detected in red giants originate from dynamo action in main-sequence convective cores. Numerical simulations of convective cores in A-type stars show the presence of such fields, with intensities of a few tens of kG (\citealt{brun05}). After the end of the main sequence, these fields could relax into stable configurations (\citealt{becerra22}), although they might be reshaped by the core contraction (\citealt{gouhier22}). The ohmic diffusion timescale being longer than the star's lifetime (\citealt{cantiello16}), we expect these fields to survive in red giant cores and to remain mostly confined in previously convective layers. As stars evolve in the post-main sequence phase, the HBS progresses outwards in mass and eventually outreaches the maximal extent of the convective core during the MS. In this case, the core magnetic field does not extend to the HBS. This reduces the seismically-measured field strength $\Brcarre$, which has maximal sensitivity near the HBS. This also reduces non-perturbative effects for near-critical fields, as shown in Sect. \ref{sect_radial_field}, and it corresponds to the configuration that gives best agreement with the seismic observations for our eight targets (Sect. \ref{sect_results}). The moment when the HBS leaves the previously-convective core strongly depends on the stellar mass: it occurs sooner for lower-mass stars ($M\sim 1.2 M_\odot$), which have small MS convective cores. 

Interestingly, we observe that red giants with magnetically distorted doublets tend to have lower masses than other magnetic red giants. Indeed, seven stars (out of eight) have seismic masses that are between 1.12 and 1.24~$M_\odot$, in agreement with the masses of our best-fit models (lying between 1.1 and 1.2~$M_\odot$), while the average seismic mass of magnetic giants detected so far is $1.31\pm0.15\,M_\odot$ (\citealt{li22}, \citealt{li23}, \citealt{deheuvels23}, \citealt{hatt24}). The last star, \starE, has an estimated mass of $1.43\pm0.09\,M_\odot$ from seismic scaling relations (\citealt{yu18}), {but the period spacing $\dpun$ we measure for this star (see Table \ref{tab_results}) makes it more compatible with lower-mass models (our best-fit model has a mass of 1.2~$M_\odot$).}
For all eight stars, we used our best-fit models to estimate the radius $r_{\rm cc}$ of the layer corresponding to the maximal extent of the convective core during the MS. We found that $r_{\rm cc}$ is below the HBS for all stars ($r_{\rm cc}/r_{\rm HBS}$ is between 40\% and 60\%). This observation holds regardless of the amount of core overshooting that is added beyond the MS core ($r_{\rm cc}$ varies by only $\sim10$\% when $\aov$ is increased from 0 to 0.2). In fact, we find that the radius $r_{\rm cc}$ is remarkably similar to the radial extent of the magnetic field $\sigma_{\rm B}$ that we measured in Sect. \ref{sect_results} (the ratio $r_{\rm cc}/\sigma_{\rm B}$ is between 1 and 1.3 for all eight stars). The magnetic fields that we measure for the stars of our sample are thus consistent with the hypothesis that they originate from convective core dynamo.

We note that if our stars were more massive $M\gtrsim1.4\,M_\odot$, we would be in the inverse situation, the HBS would still be within the layers that were previously convective during the MS. In this case, we have shown that non-perturbative effects are much more severe, and mode suppression occurs at higher frequency. It is therefore tempting to suggest that we are detecting strongly distorted doublets in the stars of our sample precisely because they are low-mass stars, and that more massive stars at the same evolutionary state and with similar field intensity might have lost one or both components to mode suppression. A more thorough search for red giants with near-critical fields could help test this hypothesis.

\subsection{Assumptions}

We now discuss some of the assumptions that were made in this study. 

\subsubsection{Radial JWKB approximation in non-perturbative approach}

Our non-perturbative approach relies on a number of assumptions, which were already made by \cite{rui23} and \cite{rui24} in their development. The use of a radial JWKB approximation in the unseparated problem would deserve to be validated, ideally using a complete non-perturbative treatment that includes the effects of the magnetic field. This is beyond the scope of this work, but we note that developments pursuing this goal are currently under way (\citealt{fort24}). One of the underlying assumptions is that the radial wavenumber $k_r$ introduced in Eq. \ref{eq_phase_function} depends only on $r$. We investigated this assumption using the ray tracing approach proposed by \cite{loi20a}. The latter study integrates Hamilton's equations in spherical coordinates using a stellar model and assuming a particular magnetic configuration (a Prendergast field). This procedure gives access to the profile of $k_r$ for magnetic fields with arbitrary intensities. We reproduced the approach of \cite{loi20a} and found that $k_r$ remains approximately a function of $r$ only, even for field intensities that approach $B_{\rm c}$. This would need to be shown for arbitrary magnetic configurations.

\subsubsection{Profile of the magnetic field}

In this study, we have assumed that the magnetic field is axisymmetric, mainly because only two components per dipole multiplet are visible. We note that if the core rotation of our stars is indeed very slow, as proposed in Sect. \ref{sect_ID01}, then the ratio between the magnetic splitting and the rotational splitting is large. In this case, the alignment of the pulsations is expected to be dictated by the magnetic field rather than the rotation. If the magnetic field possesses an axis of symmetry, then we would indeed be in an axisymmetric configuration along the magnetic axis. So far, non-perturbative treatments of magnetic effects on oscillation mode frequencies were developed under the assumption of an axisymmetric field (\citealt{rui23}). When complete non-perturbative calculations are available, it will be interesting to investigate the non-axisymmetric case as well. In this study, we also separated the radial field component into $B_r(r,\theta) = B_0 f(\theta) g(r)$. This assumption seems reasonable because the oscillation modes are sensitive to the magnetic field only in the g-mode cavity, which has a very small extension in the star (typically $\sim1$\% of the stellar radius) and most of their sensitivity is in the very thin H-burning shell. 

\section{Conclusion \label{sect_conclusion}}

In this paper, we detected strong deviations from the regular period spacing pattern of g modes in eight \kepler\ red giants exhibiting $l=1$ doublets. The measured period spacings vary with frequency and they appear abnormally low compared to what is expected for red giant branch stars with strong electron-degeneracy in the core (\citealt{deheuvels22}). We ruled out buoyancy glitches as the cause of these features, and investigated the magnetic origin. To produce such distortions in the g-mode pattern, the core magnetic field needs to be close to the critical strength $B_{\rm c}$ above which the propagation of magneto-gravity waves is inhibited (\citealt{fuller15}). 

We took into account the effects of magnetic fields on oscillation modes using the non-perturbative approach developed by \cite{rui23} and \cite{rui24}. We showed that the radial dependence of the radial component of the field $B_r$ has a large impact on the intensity of the magnetic frequency shift. This is in contrast with results from the perturbative approach, whereby magnetic frequency shifts only depend on a weighted average of $B_r^2$ (\citealt{li22}). This demonstrates that information on the radial profile of $B_r$ can be obtained for red giants that host near-critical core magnetic fields. On the contrary, the latitudinal dependence of $B_r$ seems to impact magnetic frequency shifts only through the weighted average $a_{\rm asym}$ (Eq. \ref{eq_pert_asym}), like in the case of the perturbative approach.

One critical question in our work regards the identification of the two components detected per $l=1$ multiplet. If these components are identified as $m=\pm1$, as would be most natural, no satisfactory fit to the data can be found for any of the eight stars.
Also, this identification fails to explain the difference of amplitude between the two components in \starC, or the suppression of one of the two ridges in \starA, \starD, and \starE. As an alternative, we explored the identification of the two components as $m=0$ and $m=1$. This assumption enabled us to satisfactorily account for all observational features. First, using a non-perturbative approach, we obtained very good fits to the two distorted ridges in all eight stars. Secondly, the period spacings $\dpun$ obtained under this assumption bring the eight stars back to the degeneracy sequence where they belong in the $(\deltanu, \dpun)$ plane (see Fig. \ref{fig_dn_dp}). Finally, our best-fit solutions for \starA, \starD, and \starE\ have a partially suppressed $m=1$ ridge, in agreement with the observations, and the frequencies of mode suppression are very close to the observed ones. Despite these successes, it should be reminded that we currently cannot satisfactorily explain that only two ridges (and not three) are observed. In this paper, we have worked under the assumption that the core rotation is too slow to separate the $m=\pm1$ ridges. So far, magnetic red giants were found to have similar core rotation compared to regular red giants (\citealt{li22}, \citealt{hatt24}, \citealt{villate26}). However, the stars studied here have stronger fields than the stars from these previous works, so it is conceivable that in the strong-field regime, the transport of angular momentum could be more efficient. Finally, we note that a toroidal field could act differently on the $m=\pm1$ components (\citealt{dhouib22}), so it would be interesting in the future to investigate the impact of a strong, near-critical toroidal field on the oscillations modes of our eight stars.

To reproduce the observations, we had to invoke core magnetic fields $B_r$ that are confined well below the H-burning shell (Gaussian-shaped profiles for $B_r$ with widths $\sigma_{\rm B}$ corresponding to 30-50\% of the radius of the HBS). Fields extending all the way to the HBS induce stronger non-perturbative effects, so that g-mode period distortion is too large, and mode suppression occurs at too high frequency to reproduce the observations. If our interpretation is correct, this is the first time we can place constraints on the radial extent of the core magnetic field.

The eight stars under study have a relative low mass compared to other red giants. Our best-fit stellar models have masses between 1.1 and 1.2\,$M_\odot$. They thus had a small convective core during the main sequence. Interestingly, the maximal size of the main-sequence convective core in our models approximately corresponds to our measurements of the radial extent $\sigma_{\rm B}$ of the core field. This means that the detected fields could be the remnant of fields generated by dynamo effect in the main-sequence convective core. More massive stars have larger convective cores, so that the HBS is still within the previously-convective layers at the bottom of the RGB. As we have shown in this paper, non-perturbative effects are stronger in this case and lead to more mode suppression. This might partially explain why the mass distribution of red giants with direct detection of core magnetic fields peaks at a lower mass than the distribution of red giants with suppressed dipolar modes  (see \citealt{villate26}). This calls for further investigation.

\begin{acknowledgements}
{The authors thank the anonymous referee for comments and suggestions that improved the clarity of the manuscript.} S.D. and J.B. acknowledge support from from the project BEAMING ANR-18-CE31-0001 of the French National Research Agency (ANR).
This work has been supported by CNES, focused on the preparation of the PLATO mission.
This paper includes data collected by the \kepler\ mission and obtained from the MAST data archive at the Space Telescope Science Institute (STScI). Funding
for the \kepler\ mission is provided by the NASA Science Mission Directorate.
STScI is operated by the Association of Universities for Research in Astronomy,
Inc., under NASA contract NAS 5–26555.
\end{acknowledgements}

\bibliographystyle{aa.bst} 
\bibliography{biblio}

\begin{appendix}

\section{Mode frequencies \label{app_freq}}

We give the measured mode frequencies for all eight stars of the sample as obtained from our Maximum-Likelihood Estimation technique (see Sect. \ref{sect_MLE}) in Tables \ref{tab_freq} and \ref{tab_freq2}.

\begin{table*}
\begin{center}
\caption{Frequencies of $l=0$ and $l=1$ detected oscillation modes for \starA, \starB, \starC, and \starD. \label{tab_freq}}
\begin{minipage}[t]{0.24\textwidth}
\centering
\vspace{0pt}
\begin{tabular}{l c}
\multicolumn{2}{c}{{KIC~3326392}} \\
\hline \hline
\T \B $l$ & $\nu_{n,l} (\mu$Hz) \\
\hline
\T 0 & $ 98.915 \pm 0.014 $ \\
0 & $ 109.413 \pm 0.009 $ \\
0 & $ 120.098 \pm 0.010 $ \\
0 & $ 130.714 \pm 0.011 $ \\
0 & $ 141.527 \pm 0.020 $ \\
0 & $ 152.470 \pm 0.029 $ \\
1 & $ 104.123 \pm 0.009 $ \\
1 & $ 104.632 \pm 0.011 $ \\
1 & $ 105.238 \pm 0.003 $ \\
1 & $ 114.002 \pm 0.005 $ \\
1 & $ 114.412 \pm 0.008 $ \\
1 & $ 114.743 \pm 0.007 $ \\
1 & $ 114.904 \pm 0.007 $ \\
1 & $ 115.349 \pm 0.006 $ \\
1 & $ 116.021 \pm 0.013 $ \\
1 & $ 116.095 \pm 0.004 $ \\
1 & $ 116.710 \pm 0.007 $ \\
1 & $ 117.015 \pm 0.006 $ \\
1 & $ 117.972 \pm 0.009 $ \\
1 & $ 122.794 \pm 0.009 $ \\
1 & $ 124.214 \pm 0.008 $ \\
1 & $ 124.677 \pm 0.007 $ \\
1 & $ 125.170 \pm 0.006 $ \\
1 & $ 125.442 \pm 0.007 $ \\
1 & $ 125.854 \pm 0.006 $ \\
1 & $ 126.122 \pm 0.007 $ \\
1 & $ 126.762 \pm 0.006 $ \\
1 & $ 127.047 \pm 0.006 $ \\
1 & $ 127.875 \pm 0.011 $ \\
1 & $ 128.128 \pm 0.006 $ \\
1 & $ 135.155 \pm 0.009 $ \\
1 & $ 135.361 \pm 0.008 $ \\
1 & $ 136.094 \pm 0.019 $ \\
1 & $ 136.273 \pm 0.017 $ \\
1 & $ 136.901 \pm 0.009 $ \\
1 & $ 137.160 \pm 0.009 $ \\
1 & $ 145.842 \pm 0.011 $ \\
1 & $ 146.460 \pm 0.019 $ \\
1 & $ 147.006 \pm 0.026 $ \\
1 & $ 147.458 \pm 0.015 $ \\
1 & $ 147.957 \pm 0.009 $ \\
1 & $ 157.778 \pm 0.071 $ \\
\B 1 & $ 158.753 \pm 0.016 $ \\
\hline
\end{tabular}
\end{minipage}
\hfill
\begin{minipage}[t]{0.24\textwidth}
\centering
\vspace{0pt}
\begin{tabular}{l c}
\multicolumn{2}{c}{{KIC~5700274}} \\
\hline \hline
\T \B $l$ & $\nu_{n,l} (\mu$Hz) \\
\hline
\T 0 & $ 110.774 \pm 0.012 $ \\
0 & $ 122.344 \pm 0.011 $ \\
0 & $ 134.303 \pm 0.015 $ \\
0 & $ 146.125 \pm 0.015 $ \\
0 & $ 158.162 \pm 0.021 $ \\
1 & $ 114.573 \pm 0.009 $ \\
1 & $ 115.455 \pm 0.002 $ \\
1 & $ 116.259 \pm 0.013 $ \\
1 & $ 116.911 \pm 0.014 $ \\
1 & $ 117.658 \pm 0.004 $ \\
1 & $ 126.436 \pm 0.004 $ \\
1 & $ 127.218 \pm 0.005 $ \\
1 & $ 127.575 \pm 0.007 $ \\
1 & $ 128.190 \pm 0.005 $ \\
1 & $ 128.458 \pm 0.005 $ \\
1 & $ 128.945 \pm 0.008 $ \\
1 & $ 129.310 \pm 0.004 $ \\
1 & $ 129.957 \pm 0.005 $ \\
1 & $ 131.149 \pm 0.006 $ \\
1 & $ 137.289 \pm 0.007 $ \\
1 & $ 137.861 \pm 0.003 $ \\
1 & $ 138.713 \pm 0.005 $ \\
1 & $ 139.242 \pm 0.014 $ \\
1 & $ 139.962 \pm 0.006 $ \\
1 & $ 140.298 \pm 0.006 $ \\
1 & $ 140.823 \pm 0.004 $ \\
1 & $ 141.222 \pm 0.006 $ \\
1 & $ 142.086 \pm 0.006 $ \\
1 & $ 142.580 \pm 0.006 $ \\
1 & $ 147.342 \pm 0.012 $ \\
1 & $ 149.011 \pm 0.012 $ \\
1 & $ 150.631 \pm 0.006 $ \\
1 & $ 151.744 \pm 0.007 $ \\
1 & $ 151.965 \pm 0.015 $ \\
1 & $ 152.703 \pm 0.009 $ \\
1 & $ 152.991 \pm 0.008 $ \\
1 & $ 154.186 \pm 0.002 $ \\
1 & $ 163.654 \pm 0.007 $ \\
\B 1 & $ 163.847 \pm 0.017 $ \\
\hline
\end{tabular}
\end{minipage}
\hfill
\begin{minipage}[t]{0.24\textwidth}
\centering
\vspace{0pt}
\begin{tabular}{l c}
\multicolumn{2}{c}{{KIC~7458743}} \\
\hline \hline
\T \B $l$ & $\nu_{n,l} (\mu$Hz) \\
\hline
\T 0 & $ 90.397 \pm 0.023 $ \\
0 & $ 100.743 \pm 0.021 $ \\
0 & $ 111.541 \pm 0.015 $ \\
0 & $ 122.389 \pm 0.016 $ \\
0 & $ 133.229 \pm 0.015 $ \\
0 & $ 144.288 \pm 0.022 $ \\
1 & $ 95.950 \pm 0.011 $ \\
1 & $ 104.271 \pm 0.002 $ \\
1 & $ 105.784 \pm 0.012 $ \\
1 & $ 106.015 \pm 0.006 $ \\
1 & $ 106.351 \pm 0.009 $ \\
1 & $ 106.555 \pm 0.005 $ \\
1 & $ 106.956 \pm 0.005 $ \\
1 & $ 108.962 \pm 0.005 $ \\
1 & $ 115.609 \pm 0.002 $ \\
1 & $ 115.968 \pm 0.003 $ \\
1 & $ 116.530 \pm 0.007 $ \\
1 & $ 116.815 \pm 0.005 $ \\
1 & $ 117.249 \pm 0.007 $ \\
1 & $ 117.471 \pm 0.007 $ \\
1 & $ 118.285 \pm 0.005 $ \\
1 & $ 119.294 \pm 0.013 $ \\
1 & $ 120.344 \pm 0.002 $ \\
1 & $ 126.037 \pm 0.002 $ \\
1 & $ 126.919 \pm 0.005 $ \\
1 & $ 127.147 \pm 0.006 $ \\
1 & $ 127.849 \pm 0.023 $ \\
1 & $ 128.002 \pm 0.008 $ \\
1 & $ 128.608 \pm 0.004 $ \\
1 & $ 128.823 \pm 0.007 $ \\
1 & $ 129.959 \pm 0.010 $ \\
1 & $ 131.252 \pm 0.005 $ \\
1 & $ 136.655 \pm 0.005 $ \\
1 & $ 137.822 \pm 0.009 $ \\
1 & $ 138.006 \pm 0.006 $ \\
1 & $ 138.941 \pm 0.012 $ \\
1 & $ 139.746 \pm 0.007 $ \\
1 & $ 139.927 \pm 0.009 $ \\
1 & $ 141.338 \pm 0.004 $ \\
1 & $ 149.147 \pm 0.024 $ \\
1 & $ 150.205 \pm 0.047 $ \\
\B 1 & $ 151.449 \pm 0.019 $ \\
\hline
\end{tabular}
\end{minipage}
\hfill
\begin{minipage}[t]{0.24\textwidth}
\centering
\vspace{0pt}
\begin{tabular}{l c}
\multicolumn{2}{c}{{KIC~9227589}} \\
\hline \hline
\T \B $l$ & $\nu_{n,l} (\mu$Hz) \\
\hline
\T 0 & $ 158.332 \pm 0.018 $ \\
0 & $ 173.423 \pm 0.008 $ \\
0 & $ 188.759 \pm 0.006 $ \\
0 & $ 203.972 \pm 0.008 $ \\
0 & $ 219.393 \pm 0.014 $ \\
0 & $ 235.015 \pm 0.024 $ \\
1 & $ 165.750 \pm 0.002 $ \\
1 & $ 166.949 \pm 0.011 $ \\
1 & $ 168.730 \pm 0.002 $ \\
1 & $ 179.595 \pm 0.006 $ \\
1 & $ 181.176 \pm 0.008 $ \\
1 & $ 182.783 \pm 0.004 $ \\
1 & $ 194.476 \pm 0.007 $ \\
1 & $ 195.724 \pm 0.006 $ \\
1 & $ 196.240 \pm 0.005 $ \\
1 & $ 197.236 \pm 0.004 $ \\
1 & $ 197.779 \pm 0.007 $ \\
1 & $ 199.842 \pm 0.003 $ \\
1 & $ 200.251 \pm 0.002 $ \\
1 & $ 208.925 \pm 0.005 $ \\
1 & $ 209.491 \pm 0.012 $ \\
1 & $ 211.288 \pm 0.011 $ \\
1 & $ 211.672 \pm 0.008 $ \\
1 & $ 213.035 \pm 0.008 $ \\
1 & $ 213.868 \pm 0.007 $ \\
1 & $ 216.109 \pm 0.002 $ \\
\B 1 & $ 226.371 \pm 0.017 $ \\
\hline
\end{tabular}
\end{minipage}
\hfill
\end{center}
\end{table*}

\begin{table*}
\begin{center}
\caption{Frequencies of $l=0$ and $l=1$ detected oscillation modes for \starE, \starF, \starG, and \starH. \label{tab_freq2}}
\begin{minipage}[t]{0.24\textwidth}
\centering
\vspace{0pt}
\begin{tabular}{l c}
\multicolumn{2}{c}{{KIC~9697268}} \\
\hline \hline
\T \B $l$ & $\nu_{n,l} (\mu$Hz) \\
\hline
\T 0 & $ 95.551 \pm 0.017 $ \\
0 & $ 106.141 \pm 0.011 $ \\
0 & $ 117.282 \pm 0.015 $ \\
0 & $ 128.739 \pm 0.008 $ \\
0 & $ 140.086 \pm 0.011 $ \\
0 & $ 151.627 \pm 0.017 $ \\
1 & $ 111.253 \pm 0.002 $ \\
1 & $ 121.720 \pm 0.005 $ \\
1 & $ 122.655 \pm 0.005 $ \\
1 & $ 122.783 \pm 0.006 $ \\
1 & $ 123.351 \pm 0.007 $ \\
1 & $ 123.951 \pm 0.004 $ \\
1 & $ 124.727 \pm 0.004 $ \\
1 & $ 125.201 \pm 0.005 $ \\
1 & $ 125.604 \pm 0.008 $ \\
1 & $ 126.316 \pm 0.002 $ \\
1 & $ 126.547 \pm 0.009 $ \\
1 & $ 130.668 \pm 0.004 $ \\
1 & $ 132.381 \pm 0.003 $ \\
1 & $ 132.893 \pm 0.004 $ \\
1 & $ 133.591 \pm 0.004 $ \\
1 & $ 133.964 \pm 0.003 $ \\
1 & $ 134.538 \pm 0.008 $ \\
1 & $ 134.735 \pm 0.005 $ \\
1 & $ 135.375 \pm 0.004 $ \\
1 & $ 135.611 \pm 0.004 $ \\
1 & $ 136.603 \pm 0.005 $ \\
1 & $ 136.786 \pm 0.004 $ \\
1 & $ 143.559 \pm 0.017 $ \\
1 & $ 144.950 \pm 0.010 $ \\
1 & $ 145.242 \pm 0.015 $ \\
1 & $ 145.970 \pm 0.009 $ \\
1 & $ 146.178 \pm 0.006 $ \\
1 & $ 146.938 \pm 0.005 $ \\
1 & $ 147.371 \pm 0.009 $ \\
1 & $ 148.376 \pm 0.007 $ \\
1 & $ 156.624 \pm 0.016 $ \\
1 & $ 157.681 \pm 0.042 $ \\
1 & $ 158.363 \pm 0.014 $ \\
1 & $ 158.999 \pm 0.006 $ \\
\B 1 & $ 160.168 \pm 0.002 $ \\
\hline
\end{tabular}
\end{minipage}
\hfill
\begin{minipage}[t]{0.24\textwidth}
\centering
\vspace{0pt}
\begin{tabular}{l c}
\multicolumn{2}{c}{{KIC~10613561}} \\
\hline \hline
\T \B $l$ & $\nu_{n,l} (\mu$Hz) \\
\hline
\T 0 & $ 138.175 \pm 0.004 $ \\
0 & $ 152.311 \pm 0.017 $ \\
0 & $ 166.962 \pm 0.012 $ \\
0 & $ 181.714 \pm 0.007 $ \\
0 & $ 196.390 \pm 0.012 $ \\
0 & $ 211.367 \pm 0.034 $ \\
1 & $ 145.866 \pm 0.013 $ \\
1 & $ 158.502 \pm 0.018 $ \\
1 & $ 159.428 \pm 0.009 $ \\
1 & $ 159.818 \pm 0.010 $ \\
1 & $ 160.635 \pm 0.031 $ \\
1 & $ 161.090 \pm 0.008 $ \\
1 & $ 171.427 \pm 0.007 $ \\
1 & $ 173.117 \pm 0.007 $ \\
1 & $ 173.507 \pm 0.007 $ \\
1 & $ 174.614 \pm 0.012 $ \\
1 & $ 174.845 \pm 0.006 $ \\
1 & $ 176.229 \pm 0.003 $ \\
1 & $ 176.644 \pm 0.022 $ \\
1 & $ 178.593 \pm 0.010 $ \\
1 & $ 179.014 \pm 0.011 $ \\
1 & $ 184.195 \pm 0.005 $ \\
1 & $ 186.480 \pm 0.003 $ \\
1 & $ 186.823 \pm 0.004 $ \\
1 & $ 188.670 \pm 0.007 $ \\
1 & $ 188.865 \pm 0.008 $ \\
1 & $ 190.182 \pm 0.008 $ \\
1 & $ 190.438 \pm 0.006 $ \\
1 & $ 202.211 \pm 0.009 $ \\
1 & $ 204.102 \pm 0.015 $ \\
1 & $ 206.177 \pm 0.004 $ \\
1 & $ 206.401 \pm 0.007 $ \\
\B 1 & $ 219.085 \pm 0.014 $ \\
\hline
\end{tabular}
\end{minipage}
\hfill
\begin{minipage}[t]{0.24\textwidth}
\centering
\vspace{0pt}
\begin{tabular}{l c}
\multicolumn{2}{c}{{KIC~10976412}} \\
\hline \hline
\T \B $l$ & $\nu_{n,l} (\mu$Hz) \\
\hline
\T 0 & $ 96.661 \pm 0.011 $ \\
0 & $ 107.790 \pm 0.017 $ \\
0 & $ 119.247 \pm 0.009 $ \\
0 & $ 130.851 \pm 0.011 $ \\
0 & $ 142.415 \pm 0.008 $ \\
0 & $ 154.252 \pm 0.024 $ \\
1 & $ 113.467 \pm 0.008 $ \\
1 & $ 113.711 \pm 0.009 $ \\
1 & $ 114.108 \pm 0.007 $ \\
1 & $ 115.282 \pm 0.005 $ \\
1 & $ 123.631 \pm 0.003 $ \\
1 & $ 123.889 \pm 0.003 $ \\
1 & $ 124.688 \pm 0.004 $ \\
1 & $ 124.879 \pm 0.004 $ \\
1 & $ 125.440 \pm 0.007 $ \\
1 & $ 125.618 \pm 0.006 $ \\
1 & $ 126.369 \pm 0.002 $ \\
1 & $ 126.587 \pm 0.004 $ \\
1 & $ 127.774 \pm 0.004 $ \\
1 & $ 130.093 \pm 0.002 $ \\
1 & $ 131.645 \pm 0.002 $ \\
1 & $ 134.149 \pm 0.002 $ \\
1 & $ 134.340 \pm 0.005 $ \\
1 & $ 135.496 \pm 0.005 $ \\
1 & $ 135.670 \pm 0.004 $ \\
1 & $ 136.591 \pm 0.011 $ \\
1 & $ 136.712 \pm 0.004 $ \\
1 & $ 137.465 \pm 0.004 $ \\
1 & $ 137.604 \pm 0.005 $ \\
1 & $ 138.760 \pm 0.002 $ \\
1 & $ 138.925 \pm 0.002 $ \\
1 & $ 140.416 \pm 0.031 $ \\
1 & $ 145.133 \pm 0.006 $ \\
1 & $ 146.591 \pm 0.006 $ \\
1 & $ 146.717 \pm 0.007 $ \\
1 & $ 148.075 \pm 0.009 $ \\
1 & $ 148.965 \pm 0.007 $ \\
1 & $ 149.061 \pm 0.006 $ \\
1 & $ 150.517 \pm 0.008 $ \\
1 & $ 159.518 \pm 0.016 $ \\
1 & $ 160.647 \pm 0.016 $ \\
\B 1 & $ 162.169 \pm 0.043 $ \\
\hline
\end{tabular}
\end{minipage}
\hfill
\begin{minipage}[t]{0.24\textwidth}
\centering
\vspace{0pt}
\begin{tabular}{l c}
\multicolumn{2}{c}{{KIC~11408970}} \\
\hline \hline
\T \B $l$ & $\nu_{n,l} (\mu$Hz) \\
\hline
\T 0 & $ 129.812 \pm 0.031 $ \\
0 & $ 143.078 \pm 0.013 $ \\
0 & $ 156.947 \pm 0.007 $ \\
0 & $ 170.771 \pm 0.007 $ \\
0 & $ 184.625 \pm 0.011 $ \\
0 & $ 198.778 \pm 0.031 $ \\
1 & $ 123.244 \pm 0.015 $ \\
1 & $ 137.133 \pm 0.016 $ \\
1 & $ 137.684 \pm 0.002 $ \\
1 & $ 148.907 \pm 0.003 $ \\
1 & $ 149.310 \pm 0.005 $ \\
1 & $ 150.099 \pm 0.007 $ \\
1 & $ 150.347 \pm 0.009 $ \\
1 & $ 151.211 \pm 0.006 $ \\
1 & $ 151.590 \pm 0.007 $ \\
1 & $ 160.335 \pm 0.002 $ \\
1 & $ 160.691 \pm 0.004 $ \\
1 & $ 162.270 \pm 0.003 $ \\
1 & $ 162.580 \pm 0.006 $ \\
1 & $ 163.797 \pm 0.007 $ \\
1 & $ 163.962 \pm 0.007 $ \\
1 & $ 165.022 \pm 0.005 $ \\
1 & $ 165.289 \pm 0.004 $ \\
1 & $ 166.954 \pm 0.003 $ \\
1 & $ 167.253 \pm 0.005 $ \\
1 & $ 171.405 \pm 0.002 $ \\
1 & $ 173.743 \pm 0.004 $ \\
1 & $ 176.036 \pm 0.002 $ \\
1 & $ 176.272 \pm 0.007 $ \\
1 & $ 177.706 \pm 0.007 $ \\
1 & $ 177.829 \pm 0.006 $ \\
1 & $ 179.223 \pm 0.008 $ \\
1 & $ 179.437 \pm 0.003 $ \\
1 & $ 181.619 \pm 0.004 $ \\
1 & $ 184.249 \pm 0.002 $ \\
1 & $ 189.692 \pm 0.012 $ \\
1 & $ 191.641 \pm 0.019 $ \\
1 & $ 191.758 \pm 0.002 $ \\
1 & $ 193.353 \pm 0.015 $ \\
\B 1 & $ 206.830 \pm 0.007 $ \\
\hline
\end{tabular}
\end{minipage}
\hfill
\end{center}
\end{table*}

\section{Oscillation equations in non-perturbative framework \label{app_nonpert}}

{We here detail the main steps that lead to the expressions for g mode frequencies with magnetic field taken into account in a non-perturbative manner. The development essentially follows that of \cite{rui23}.} With the assumptions made in Sect. \ref{sect_nonpert_assumptions}, the oscillation equations can be written as follows. The continuity equation reads:
\begin{equation}
\frac{1}{r^2}\frac{\partial}{\partial r} (r^2\xi_r) + \frac{1}{r \sin \theta} \frac{\partial}{\partial \theta} (\sin \theta \xi_\theta) +  \frac{1}{r \sin \theta} \frac{\partial \xi_\varphi}{\partial \varphi} = 0 
\end{equation}
The momentum equations are written as:
\begin{align}
& -\rho_0 \omega^2 \xi_r = - \frac{\partial p'}{\partial r} - \rho' g_0 \\
& -\rho_0 \omega^2 \xi_\theta = - \frac{1}{r} \frac{\partial p'}{\partial \theta} + \frac{1}{4\pi} B_r^2 \frac{\partial^2 \xi_\theta}{\partial r^2} \\
& -\rho_0 \omega^2 \xi_\varphi = - \frac{1}{r\sin\theta} \frac{\partial p'}{\partial \varphi} + \frac{1}{4\pi} B_r^2 \frac{\partial^2 \xi_\varphi}{\partial r^2} 
\end{align}
We assume the oscillations are adiabatic, so the energy equation can be replaced by the adiabaticity relation 
\begin{equation}
 \frac{\rho'}{\rho_0} = \frac{1}{\Gamma_{1,0}} \frac{p'}{p_0} + \frac{N^2 \xi_r}{g_0},
\label{eq_adiab}
\end{equation}
where $\Gamma_{1,0}$ is the adiabatic index. For gravity waves, the first term in the right-hand-side of Eq. \ref{eq_adiab} can be neglected, so that $\rho' \simeq \rho_0 N^2 \xi_r /g_0$.
 
All the eigenfunctions are written as the product of a plane-like wave function $E(r) = \exp\left( - i \int k_r \,\hbox{d}r \right)$ and an envelope function (for instance $\hat{p}(r,\theta)$ for the pressure perturbation) that is a slowly varying function of the radius. The continuity equation then becomes
\begin{equation}
 -i k_r \widehat{\xi_r}  - \frac{1}{r} \frac{\partial}{\partial \mu} \left( \sqrt{1-\mu^2} \widehat{\xi_\theta} \right) +  \frac{1}{r \sqrt{1-\mu^2}} im \widehat{\xi_\varphi} = 0 
\end{equation}
Combining the radial equation of motion with the adiabaticity relation, we obtain
\begin{equation}
- \rho_0 \omega^2 \widehat{\xi_r} = i k_r \hat{p} - \rho_0 N^2 \widehat{\xi_r}
\end{equation}
Since $\omega \ll N$, the left-hand term of this equation can be neglected. The horizontal equations of motion yield
\begin{align}
\widehat{\xi_\theta}  & = - \frac{1}{\rho_0  r \omega^2} \frac{\sqrt{1-\mu^2}}{1-b^2 f^2} \frac{\partial \hat{p}}{\partial \mu} \\
\widehat{\xi_\varphi}  & = \frac{1}{\rho_0  r \omega^2} \frac{1}{\sqrt{1-\mu^2}(1-b^2 f^2)}  im \hat{p}
\end{align}
We insert these expressions into the continuity equation, yielding
\begin{equation}
\lambda \hat{p} +  \frac{\partial}{\partial \mu} \left( \frac{1-\mu^2}{1-b^2 f^2} \frac{\partial \hat{p}}{\partial \mu} \right) -  \frac{m^2}{ (1-\mu^2)(1-b^2 f^2)} \hat{p} = 0,
\end{equation}
where
\begin{equation}
\lambda =  \left( \frac{r \omega k_r}{ N} \right)^2 = \left( \frac{b}{a} \right)^2.
\end{equation}

\section{Magnetic field profile \label{app_field_profile}}

\subsection{Latitudinal profile of $B_r$ \label{app_latitudinal_profile}}

To build the latitudinal profile $f(\theta)$ introduced in Eq. \ref{eq_latitudinal_field}, we need to determine the value of $\delta$ corresponding to a given {$a_{\rm asym} = \int_{-1}^1 f^2 P_2 \hbox{d}\mu/\int_{-1}^1 f^2  \hbox{d}\mu$. The denominator can be expressed as $I_1 e^{-\delta}$, where \begin{equation}
I_n = \int_{-1}^1 \mu^{2n} e^{3\delta\mu^2} \,\hbox{d}\mu
\end{equation}
To calculate these integrals numerically, we express them as:
\begin{equation}
I_n = 2 \sum_{k=0}^{+\infty} \frac{(3\delta)^k}{k! (2k+2n+1)}.
\end{equation}
and terms in the summation are added until they become smaller than $10^{-15}$. The numerator can be written as $e^{-\delta}(3I_2-I_1)/2$, so that 
\begin{equation}
a_{\rm asym} = \frac{3}{2} \frac{I_2}{I_1} - \frac{1}{2}.
\end{equation}}
The value of $\delta$ corresponding to a given $a_{\rm asym}$ is obtained by using a Newton-Raphson algorithm and making use of the following expression for the derivative $\hbox{d}a_{\rm asym}/\hbox{d}\delta$:
{\begin{equation}
\frac{\hbox{d}a_{\rm asym}}{\hbox{d}\delta} = \frac{9}{2} \frac{I_3}{I_1} - 2\left(a+\frac{1}{2}\right)^2.
\end{equation}}

\subsection{Alternative expression for $f(\theta)$ \label{app_alt_latitudinal_profile}}

We can also build a latitudinal profile $f(\theta)$ for the radial field $B_r$ in the shape 
{$f(\theta) = P_1(\cos\theta) + B P_3(\cos\theta) + C P_5(\cos\theta)$.}
For a given value of the asymmetry parameter $a_{\rm asym}$, the field must satisfy Eq. \ref{eq_pert_asym}. This is achieved provided
{\begin{equation}
a_{\rm asym} = \frac{\frac{4}{15} + \frac{8}{105}B^2 + \frac{20}{429}C^2 + \frac{12}{35}B + \frac{40}{231}BC}{\frac{2}{3} + \frac{2}{7}B^2 + \frac{2}{11}C^2}
\end{equation}
We find that the value $C\approx 2.37$ maximizes the interval of asymmetry parameters that are reachable.
Solutions to the previous equation can be found for 
$-0.252 \lesssim a_{\rm asym} \lesssim 0.832$.} This does not cover the full range of possible values for $a_{\rm asym}$, but gives satisfactory coverage for this study at least for the solutions with positive asymmetries.
We then divide the obtained profile $f(\theta)$ by is maximum absolute value to satisfy the condition $\max(|f|) = 1$.

\subsection{Making the field divergence-free \label{app_divergence_free}}

We need to ensure that the magnetic field that we build satisfies the condition $\nabla \cdot \vec B = 0$. The field is axisymmetric, so that the azimuthal component $B_\varphi(r,\theta)$ can be arbitrarily chosen without affecting the divergence of $\vec B$. We add a $\theta$-component $B_\theta(r,\theta)$ in such a way that {
\begin{equation}
\frac{1}{r^2}\frac{\partial}{\partial r}(r^2B_r) + \frac{1}{r \sin\theta}    \frac{\partial}{\partial \theta} (\sin\theta B_\theta) = 0 
\end{equation}
This is achieved if $B_\theta$ takes the form
\begin{equation}
B_{\theta}(r,\theta) = -  \frac{B_0}{r}  \frac{\hbox{d}}{\hbox{d}r}\left[ r^2 g(r) \right]  \frac{\int_0^\theta f(\theta')\sin\theta' \,\hbox{d}\theta'}{\sin\theta}
\end{equation}}
This solution is regular at the poles ($\mu = \cos\theta \rightarrow \pm 1$). Indeed, by expending $\mu = \pm( 1-\varepsilon)$, one finds that the numerator is equivalent to $e^{\delta}\varepsilon$ while the denominator varies as $\sqrt{2\varepsilon}$.

\section{Priors and results for ABIM \label{app_abim}}

\subsection{Priors}

When using the ABIM code to fit asymptotic expressions of mixed modes (including effects from rotation and magnetic fields) to the observations (Sect. \ref{sect_abim}), we impose a set of priors on the free parameters. The priors are chosen either uniform (U) over a range $[x_{\rm min}, x_{\rm max}]$, ``uniform-periodic'' (UP), which is a uniform distribution for a periodic parameter (the steps of the walkers are calculated modulo the period), or ``uniform-normal'' (UG), which is a uniform distribution with Gaussian tails at the edges ($\sigma_{\rm min}$ and $\sigma_{\rm max}$ are the widths of the tails at both ends).

Our choices for the priors are summarized in Table \ref{tab_priors}. For the large separation $\deltanu$, we used a uniform prior centered on the value measured by \cite{yu18} with an interval of $\pm1\,\mu$Hz. The phase terms $\varepsilon_{\rm p}$ and $\varepsilon_{\rm g}$ have a periodicity of one. The priors on $\alpha$ and $d_{01}$ were designed based on measurements of these parameters for other similar red giants. The uniform prior on the period spacing $\dpun$ was chosen to include the apparent period spacing $\dpun^{\rm meas}$ (which is lower than the expected value) and extend 6~s above the degeneracy sequence (\citealt{deheuvels22}). For the coupling strength $q$, we chose a uniform distribution ranging up to 0.2, which covers the expected values on the RGB. The uniform prior on the magnetic frequency shift $\delta\nu_0$ (which is related to the quantity $\Brcarre$) was chosen to extend well beyond the values measured by \cite{deheuvels23}. The asymmetry parameter $a_{\rm asym}$ takes values between $-0.5$ and 1, which are in principle all reachable by the latitudinal profile that we chose (Eq. \ref{eq_latitudinal_field}). However, we must have $\delta\rightarrow\pm\infty$ to reach the edges of this range, which causes numerical problems. For this reason, we restricted the prior distribution to $[-0.45, 0.9]$. The radial extension of the field is parametrized by $\sigma_{\rm B}$, for which we chose a uniform prior distribution in the range $[0.001, 0.009]\,R_\star$. The upper limit of this interval corresponds to a field that is nearly constant within the g-mode cavity. Finally, the priors on $\sigma_{\rm p}$ and $\sigma_{\rm g}$, which characterize the model errors, were chosen based on model errors computed for the red giants studied in \cite{li24}.

\FloatBarrier
\begin{figure*}
\begin{center}
\includegraphics[width=0.49\linewidth]{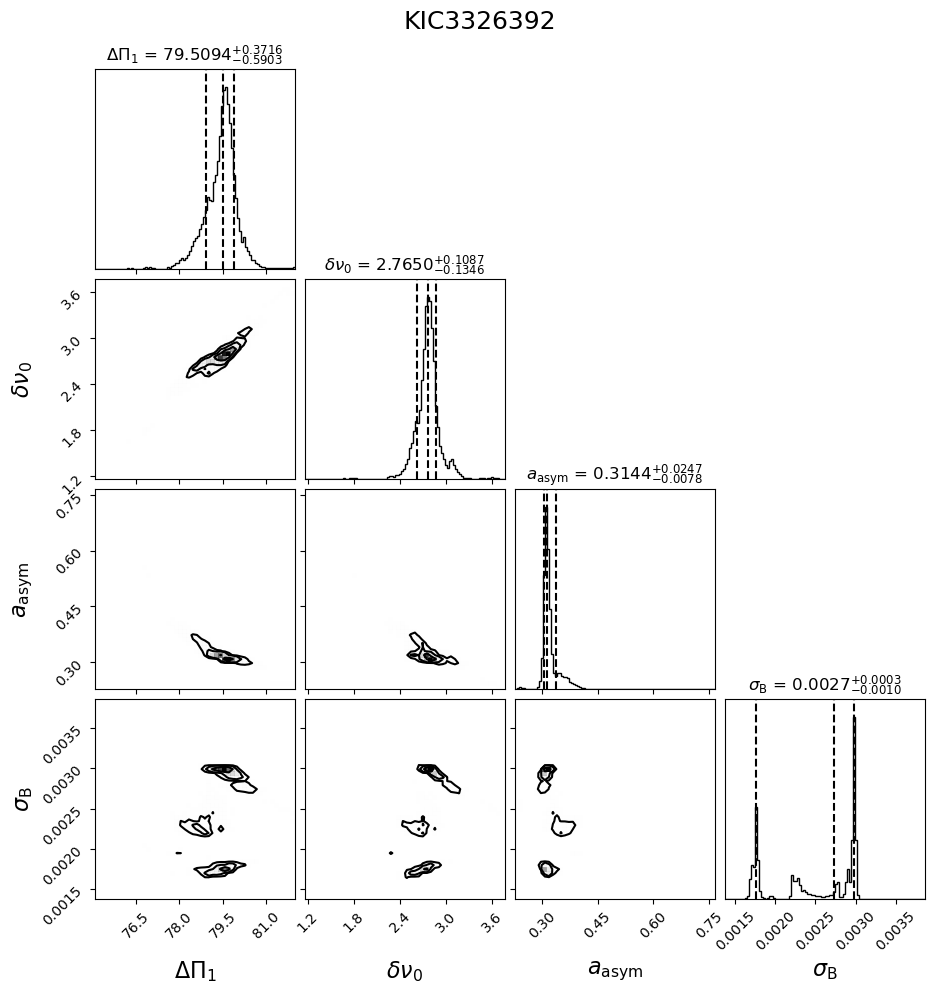}
\includegraphics[width=0.49\linewidth]{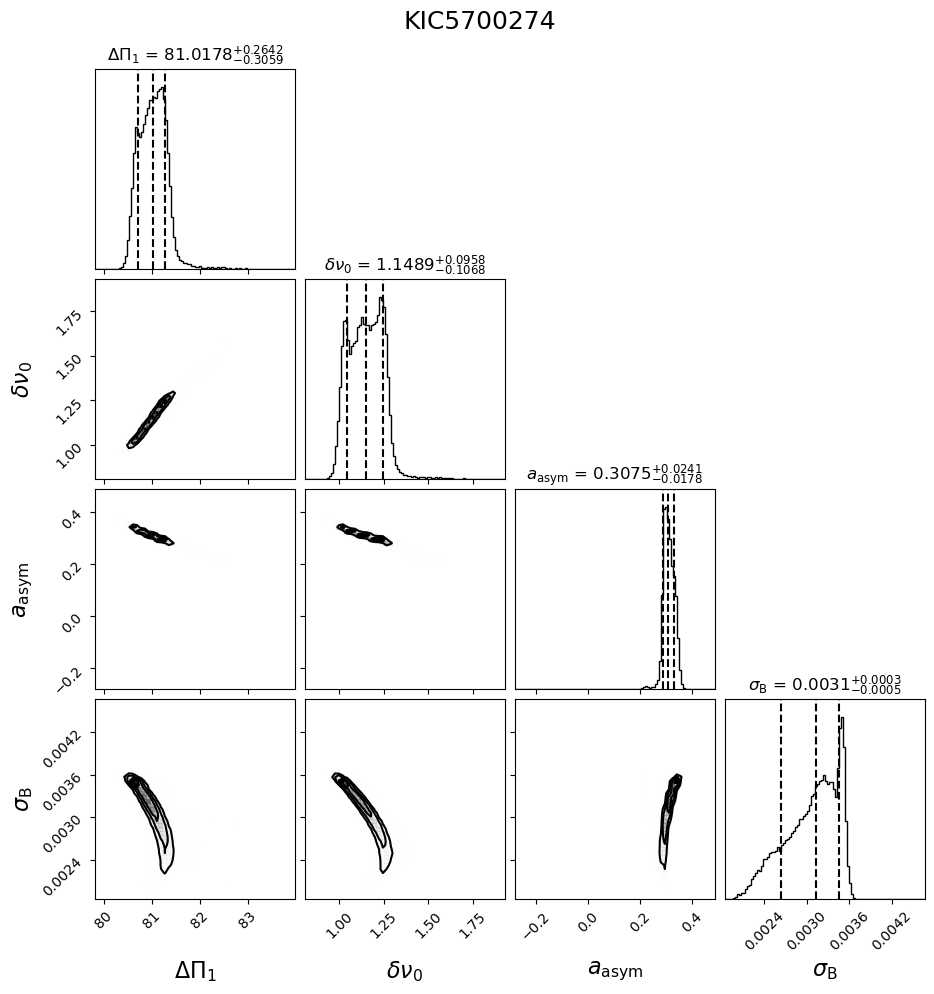}
\end{center}
\caption{Same as Fig. \ref{fig_corner} for \starA\ (left) and \starB\ (right).
\label{fig_corner_app1}}
\end{figure*}

\begin{table}[!htbp]
\caption{Choice of priors for ABIM.} 
\label{tab_priors}      
\centering
\small
\begin{tabular}{c c c c c c}
    \hline\hline
        Parameter & {$x_\mathrm{min}$} &{$x_\mathrm{max}$} & $\sigma_\mathrm{min}$ & $\sigma_\mathrm{max}$ & Law \\
        \hline
         
         $\Delta\nu$ ($\mu$Hz) & $\Delta\nu_0-1$&$\Delta\nu_0+1$ & & & U\\
         $\epsilon_p$ & $0.5$ & $1.5$ & & & UP\\
         $\alpha$ & $-0.05$ & $0.05$ & $0.005$ & $0.005$& UG \\
         $d_{01}$ ($\mu$Hz) & $-0.2$ & $0.2$ & $0.1$ & $0.1$ & UG \\
         
         $\epsilon_g$ & $-0.2$ &$0.8$& & & UP \\
         $\Delta\Pi_1$ (s) & $\dpun^{\rm meas} -2$ & $\dpun^{\rm deg}+6$ & & & U \\
         
         $q$ & $0$ & $0.2$ & & & U \\
         
         $\delta\nu_0$ ($\mu$Hz)& 0 & 10 &  &  & U \\
         $a_{\rm asym}$ & $-0.45$ & $0.9$ & & & U\\  
         $\sigma_{\rm B}$ & 0.001 & 0.009 & & & U \\
         
         $\sigma_g$ ($\mu$Hz)& $0$& $0.1$& & & U \\
         $ \sigma_p$ ($\mu$Hz)&$0$&$0.3$&$0$ & $0.1$ & UG\\
         \hline
    \end{tabular}
    \tablefoot{In the last column, ``U'' stands for uniform distribution, ``UP'' for uniform periodic, and ``UG'' for normal (or Gaussian) uniform.}
\end{table}

\subsection{Corner plots}

In Fig. \ref{fig_corner_app1} and \ref{fig_corner_app1}, we show corner plots similar to those of Fig. \ref{fig_corner} for \starA, \starB, \starC, \starE, \starG, and \starH. For clarity, the corner plots are restricted to the asymptotic period $\dpun$ of dipolar g modes, and the parameters characterizing the core magnetic field: the magnetic shift $\delta\nu_0 = \delta\omega_0/(2\pi)$, $\sigma_B$, and $a_{\rm asym}$.

\FloatBarrier
\begin{figure*}
\begin{center}
\includegraphics[width=0.49\linewidth]{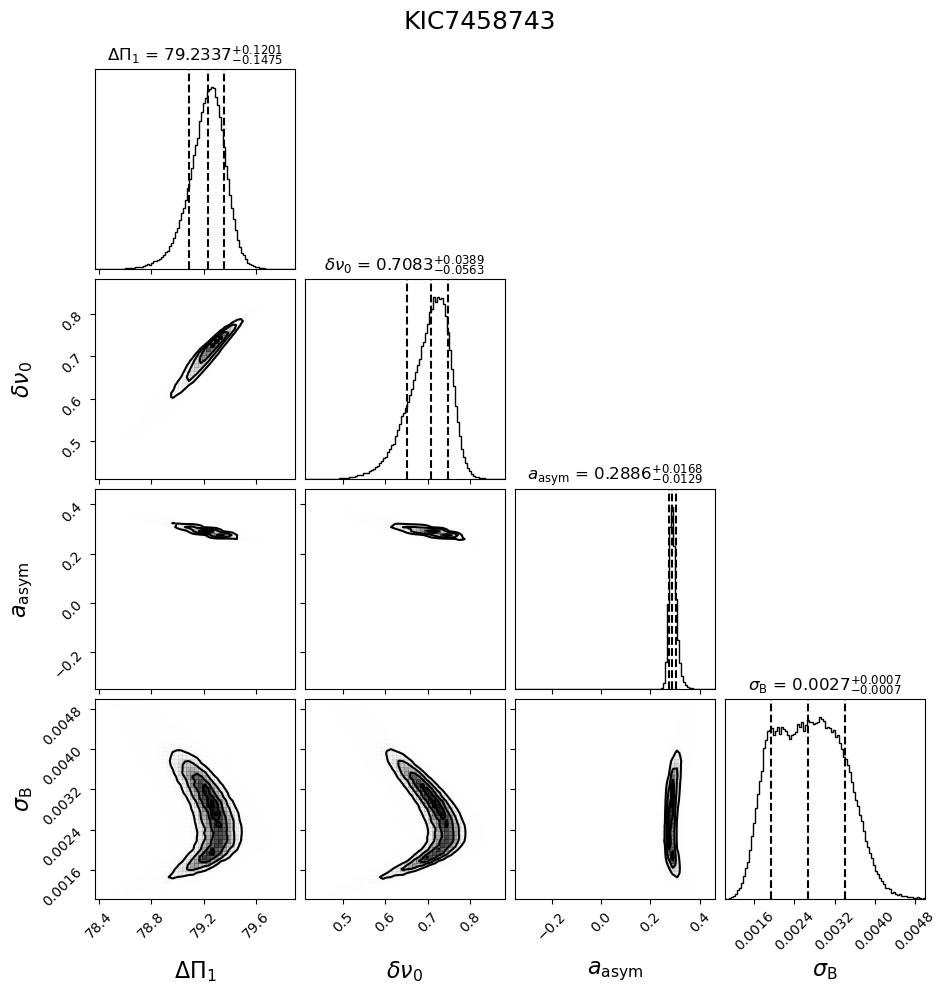}
\includegraphics[width=0.49\linewidth]{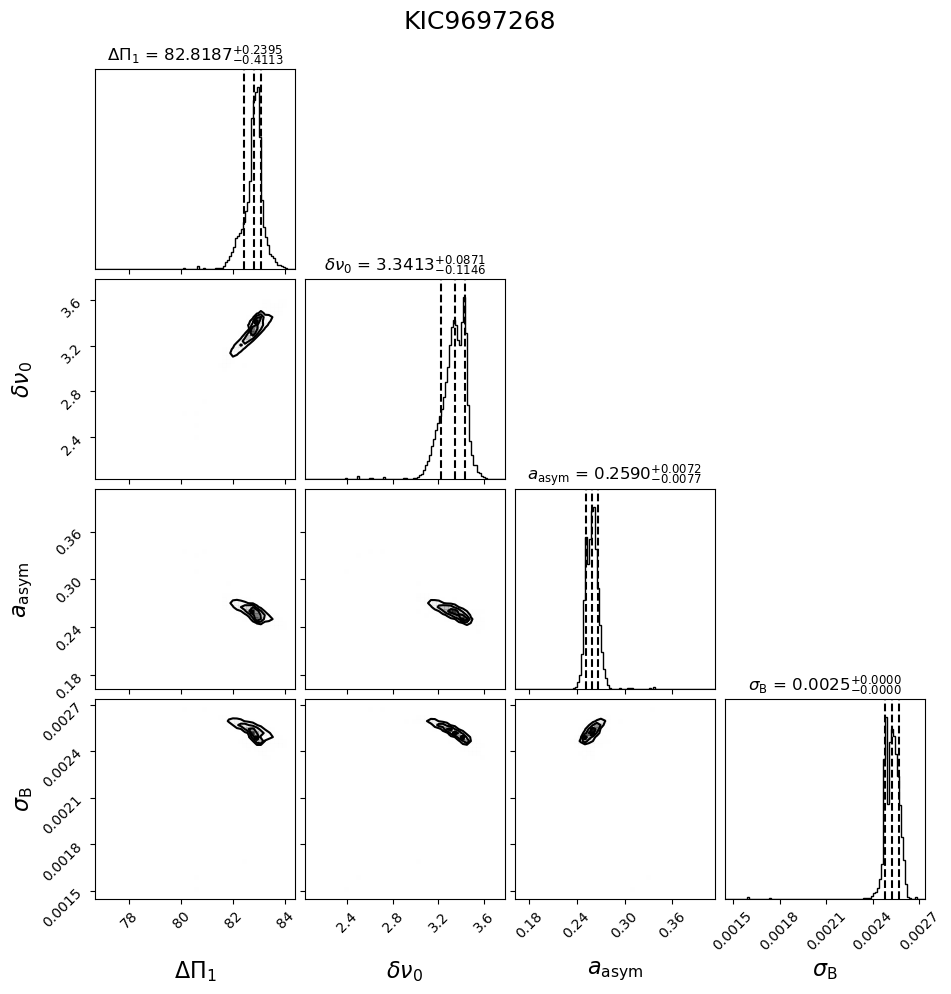}
\includegraphics[width=0.49\linewidth]{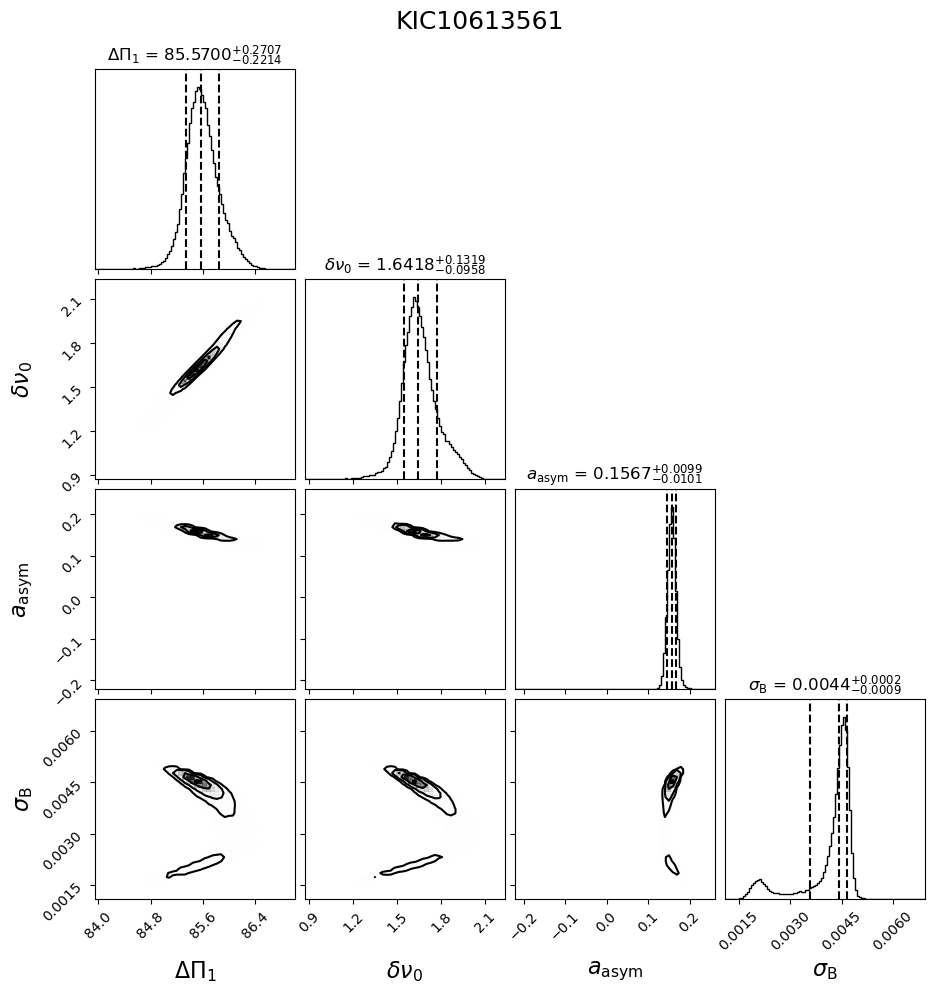}
\includegraphics[width=0.49\linewidth]{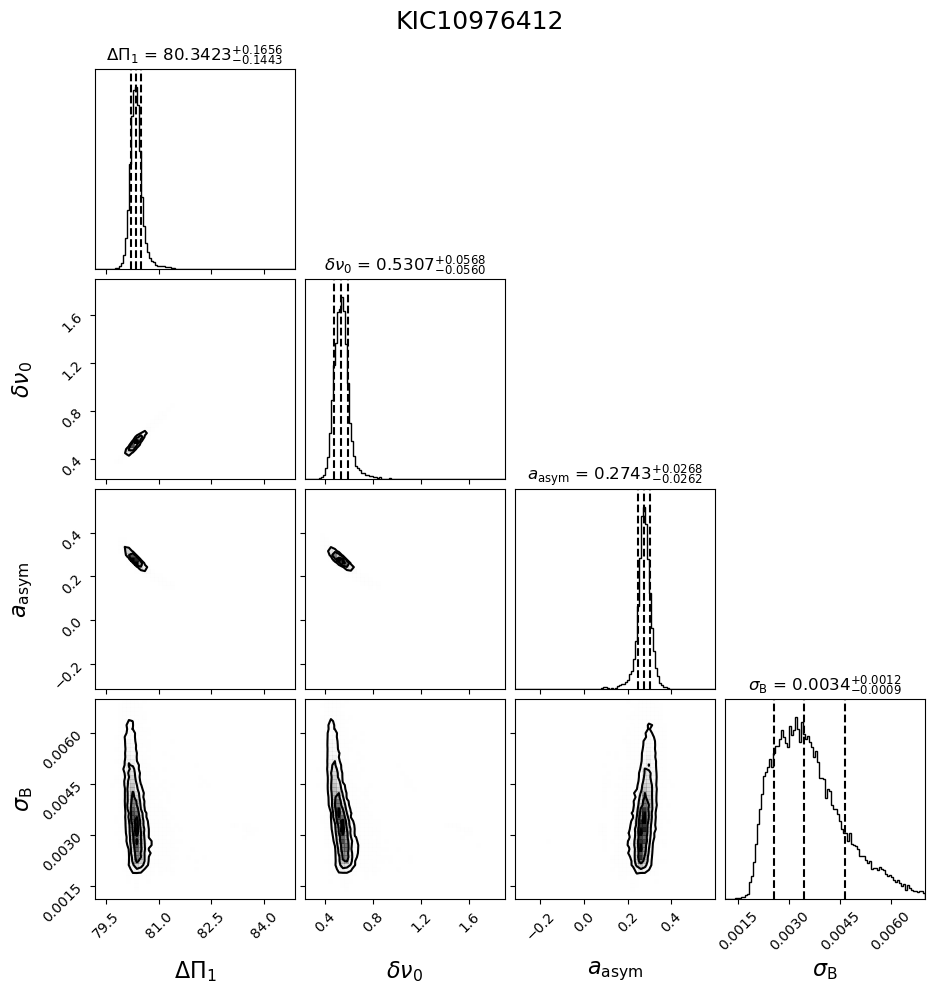}
\end{center}
\caption{Same as Fig. \ref{fig_corner} for \starC\ (top left), and \starE\ (top right), \starF\ (bottom left) and \starG\ (bottom right).
\label{fig_corner_app2}}
\end{figure*}

\end{appendix}

\end{document}